\documentclass[prd,twocolumn,nofootinbib,superscriptaddress,fleqn,amssymb,amsfonts,amsmath]{revtex4-2}
\usepackage{graphicx,slashed,xcolor,multirow,bbold,mathtools,sidecap,tikz,bm,ulem,enumitem,booktabs,array}
\usepackage[colorlinks=true,linktoc=page,linkcolor=purple,citecolor=teal,urlcolor=magenta]{hyperref}
\usepackage[compat=1.1.0]{tikz-feynman}
\graphicspath{ {figures/} }

\definecolor{lime}{HTML}{A6CE39}
\DeclareRobustCommand{\orcidicon}{\hspace{-2.1mm}
\begin{tikzpicture}
\draw[lime,fill=lime] (0,0.0) circle [radius=0.13] node[white] {{\fontfamily{qag}\selectfont \tiny ID}}; \draw[white,fill=white] (-0.0525,0.095) circle [radius=0.007]; 
\end{tikzpicture} \hspace{-3.7mm} }
\foreach \x in {A, ..., Z}
 {\expandafter\xdef\csname orcid\x\endcsname{\noexpand\href{https://orcid.org/\csname orcidauthor\x\endcsname} {\noexpand\orcidicon}}}

\let\emph\textit

\begin{document}

\title{Probing compressed mass spectra in the type-II seesaw model at the LHC}

\author{Saiyad Ashanujjaman\orcidA{}}
\email{saiyad.a@iopb.res.in}
\affiliation{Department of Physics, SGTB Khalsa College, Delhi 110007, India}
\affiliation{Department of Physics and Astrophysics, University of Delhi, Delhi 110007, India}

\author{Siddharth P. Maharathy\orcidB{}}
\email{siddharth.m@iopb.res.in}
\affiliation{Institute of Physics, Bhubaneswar, Sachivalaya Marg, Sainik School, Bhubaneswar 751005, India}                                                           
\affiliation{Homi Bhabha National Institute, Training School Complex, Anushakti Nagar, Mumbai 400094, India}
                                                                                                                                                                                      
\begin{abstract}
Despite a great deal of effort in searching for the triplet-like Higgses in the type-II seesaw model, evidence for their production is yet to be found  at the LHC. As such, one might be in the balance regarding this model's relevance at the electroweak scale. In this work, we peruse a scenario, akin to compressed mass spectra in Supersymmetry, which might have eluded the experimental searches thus far. We perform a multivariate analysis to distinguish signals with a pair of same-sign leptons with low invariant mass from the SM processes, including those accruing from {\it fake} leptons and electron charge misidentification, and find that a significant part of the hitherto unconstrained parameter space could be probed with the already collected Run 2 13 TeV LHC and future HL-LHC data.
\end{abstract}

\maketitle                                                                                           

\section{\label{sec:intro} Introduction}
The type-II seesaw model \cite{Konetschny:1977bn,Cheng:1980qt,Lazarides:1980nt,Schechter:1980gr,Mohapatra:1980yp,Magg:1980ut}, extending the Standard Model (SM) with an $SU(2)_L$ triplet scalar field with hypercharge $Y=1$, offers a tenable explanation for the observed neutrino masses and mixings, as such it is arguably the most widely-studied neutrino mass model \cite{Huitu:1996su,Gunion:1996pq,Chakrabarti:1998qy,Muhlleitner:2003me,Akeroyd:2005gt,Akeroyd:2007zv,Garayoa:2007fw,Han:2007bk,delAguila:2008cj,FileviezPerez:2008jbu,Perez:2008ha,Akeroyd:2009hb,Akeroyd:2010ip,Melfo:2011nx,Aoki:2011pz,Akeroyd:2011zza,Arbabifar:2012bd,Chiang:2012dk,Akeroyd:2012nd,Chun:2012zu,delAguila:2013mia,Chun:2013vma,Kanemura:2013vxa,Dev:2013ff,Kanemura:2014goa,Kanemura:2014ipa,kang:2014jia,Han:2015hba,Han:2015sca,Mitra:2016wpr,Ghosh:2017pxl,Antusch:2018svb,BhupalDev:2018tox,deMelo:2019asm,Primulando:2019evb,Chun:2019hce,Padhan:2019jlc,Gluza:2020qrt,Ashanujjaman:2021txz,Ashanujjaman:2022ofg,Rodejohann:2010bv,Blunier:2016peh,Nomura:2017abh,Crivellin:2018ahj,Agrawal:2018pci,Rahili:2019ixf,Bandyopadhyay:2020mnp,Ashanujjaman:2022tdn,Rodejohann:2010jh,Li:2023ksw,Maharathy:2023dtp,Fridell:2023gjx,Dev:2019hev,Yang:2021skb,Deppisch:2015qwa,Cai:2017mow,Chun:2003ej,Kadastik:2007yd,Arhrib:2011uy,Chun:2012jw,Das:2016bir}. This model accommodates, in addition to the 125 GeV Higgs, several other physical states: doubly charged scalar ($H^{\pm\pm}$), singly charged scalar ($H^\pm$) and CP-even and CP-odd neutral scalars ($H^0$ and $A^0$). The phenomenology of these states, particularly $H^{\pm\pm}$, has been studied extensively at the Large Hadron Collider (LHC) \cite{Huitu:1996su,Gunion:1996pq,Chakrabarti:1998qy,Muhlleitner:2003me,Akeroyd:2005gt,Akeroyd:2007zv,Garayoa:2007fw,Han:2007bk,delAguila:2008cj,FileviezPerez:2008jbu,Perez:2008ha,Akeroyd:2009hb,Akeroyd:2010ip,Melfo:2011nx,Aoki:2011pz,Akeroyd:2011zza,Arbabifar:2012bd,Chiang:2012dk,Akeroyd:2012nd,Chun:2012zu,delAguila:2013mia,Chun:2013vma,Kanemura:2013vxa,Dev:2013ff,Kanemura:2014goa,Kanemura:2014ipa,kang:2014jia,Han:2015hba,Han:2015sca,Mitra:2016wpr,Ghosh:2017pxl,Antusch:2018svb,BhupalDev:2018tox,deMelo:2019asm,Primulando:2019evb,Chun:2019hce,Padhan:2019jlc,Gluza:2020qrt,Ashanujjaman:2021txz,Ashanujjaman:2022ofg}, electron colliders \cite{Rodejohann:2010bv,Blunier:2016peh,Nomura:2017abh,Crivellin:2018ahj,Agrawal:2018pci,Rahili:2019ixf,Bandyopadhyay:2020mnp,Ashanujjaman:2022tdn}, muon colliders \cite{Rodejohann:2010jh,Li:2023ksw,Maharathy:2023dtp,Fridell:2023gjx} and electron-proton colliders \cite{Dev:2019hev,Yang:2021skb}; see Refs.~\cite{Deppisch:2015qwa,Cai:2017mow,Ashanujjaman:2021txz} for comprehensive reviews. Experimental collaborations have performed several searches for $H^{\pm \pm}$ \cite{ATLAS:2012hi,Chatrchyan:2012ya,ATLAS:2014kca,Khachatryan:2014sta,CMS:2016cpz,CMS:2017pet,Aaboud:2017qph,CMS:2017fhs,Aaboud:2018qcu,Aad:2021lzu,ATLAS:2021jol,ATLAS:2022yzd}, and non-observations of any significant excess over the SM expectations have led to stringent limits on them. While the ATLAS collaboration has set a lower limit of 1020 GeV for $H^{\pm \pm}$ decaying into $\ell^\pm \ell^\pm$ (with $\ell = e,\mu$) \cite{ATLAS:2022yzd}, the CMS collaboration has set a lower limit of 535 GeV for those decaying into $\tau^\pm \tau^\pm$ \cite{CMS:2017pet}. For $H^{\pm \pm}$ decaying into $W^\pm W^\pm$, the ATLAS collaboration has excluded them within the mass range 200--350 GeV \cite{ATLAS:2021jol}. 

Recently, Ref.~\cite{Ashanujjaman:2021txz} has estimated exclusion limits on $m_{H^{\pm \pm}}$ for a vast model parameter space, characterised by the mass-splitting $\Delta m = m_{H^{\pm \pm}} - m_{H^\pm}$ and the triplet vacuum expectation value (VEV) $v_t$, by recasting several searches performed by the CMS and ATLAS collaborations.\footnote{For $H^{\pm \pm}$ decaying into $W^\pm W^\pm$, Ref.~\cite{Ashanujjaman:2021txz} estimates an improved exclusion range of 200--400 GeV compared to 200--350 GeV obtained by the ATLAS collaboration in Ref.~\cite{ATLAS:2021jol}.} The regions corresponding to $\Delta m = 0$ and $\Delta m < 0$, being dominated with the so-called ``golden decays'' of $H^{\pm \pm}$ to $\ell^\pm \ell^\pm$ and $W^\pm W^\pm$, are highly constrained by the LHC searches, as such $H^{\pm \pm}$ up to 1115(420) GeV masses are excluded for small (large) $v_t$. On the other hand, a significantly large part of the model parameter space characterised by $\Delta m \sim \mathcal{O}(10)$ GeV and $v_t \sim \mathcal{O}(10^{-7})$--$\mathcal{O}(10^{-3})$ is largely unconstrained by the existing LHC searches. A detailed analysis for probing this region at future $e^-e^+$ colliders has been performed in Ref.~\cite{Ashanujjaman:2022tdn}. This region is dominated by the exclusive decays to one or more off-shell $W^\pm$-bosons and $H^0/A^0$. While the latter decays to $\nu \nu$ for $v_t \leq \mathcal{O}(10^{-4})$ and to $b\bar{b},t\bar{t},ZZ,Zh,hh$ for $v_t \geq \mathcal{O}(10^{-4})$, the former results in leptons and jets with relatively low transverse momentum ($p_T$), termed as soft leptons and soft jets. 

The present work concerns part of the abovementioned unconstrained region where $H^0/A^0$ exclusively decays to $\nu \nu$. Therefore, the resulting final state is populated with soft leptons and jets and low missing transverse momentum ($p_T^{\rm miss}$). Typically, these are beset with oversized background contributions from quantum chromodynamics (QCD) multijet, multitop and Drell-Yan processes, thus are not highly sensitive to new physics searches. Such signals, akin to those appearing in compressed mass spectra in Supersymmetry, in principle, can be distinguished from the SM background by requiring either a jet with large $p_T$ from initial state radiation that leads to a high boost of the decaying particle pair and thus large $p_T^{\rm miss}$ \cite{ATLAS:2021kxv,CMS:2021far}, or two or more soft leptons along with large $p_T^{\rm miss}$ \cite{ATLAS:2019lng,CMS:2021edw}. However, as it turns out, the requirement of an energetic jet or large $p_T^{\rm miss}$ exceedingly reduces the signal rate, thereby making such searches insensitive to the present model \cite{Ashanujjaman:2021txz}. Keeping this in mind, we relax the aforementioned requirement to retain signal acceptance, and require only a pair of same-sign leptons with low invariant mass. We perform a multivariate analysis to distinguish such characteristic signal from the SM background, including those arising from {\it fake} leptons and electron charge misidentification.

The rest of this work is structured as follows. We briefly outline the type-II seesaw model in Sec.~\ref{sec:model}, followed by a detailed collider analysis in Sec.~\ref{sec:collider} and a summary in Sec.~\ref{sec:summary}.

\section{\label{sec:model} The Higgs triplet}
In addition to the SM field content, the type-II seesaw model employs an $SU(2)_L$ triplet scalar field with $Y=1$:
\[
\Delta = \begin{pmatrix} \Delta^+/\sqrt{2} & \Delta^{++} \\ \Delta^0 & -\Delta^+/\sqrt{2} \end{pmatrix}.
\]
The scalar potential involving $\Delta$ and the SM Higgs doublet $\Phi = \begin{pmatrix} \Phi^+ & \Phi^0 \end{pmatrix}^T $ is given by \cite{Arhrib:2011uy}
\begin{align*}
V(\Phi,\Delta) =& -m_\Phi^2{\Phi^\dagger \Phi} + \frac{\lambda}{4}(\Phi^\dagger \Phi)^2 + m_\Delta^2{\rm Tr}(\Delta^{\dagger}{\Delta}) 
\\
& + [\mu(\Phi^T{i}\sigma^2\Delta^\dagger \Phi)+{\rm h.c.}] + \lambda_1(\Phi^\dagger \Phi){\rm Tr}(\Delta^{\dagger}{\Delta})
\\
&  + \lambda_2[{\rm Tr}(\Delta^{\dagger}{\Delta})]^2 + \lambda_3{\rm Tr}[(\Delta^{\dagger}{\Delta})^2] + \lambda_4{\Phi^\dagger \Delta \Delta^\dagger \Phi},
\end{align*}
where $m_\Phi^2, m_\Delta^2$ and $\mu$ are the mass parameters, $\lambda$ and $\lambda_i$ ($i=1,\dots,4$) are the dimensionless quartic couplings. The neutral components $\Phi^0$ and $\Delta^0$ procures respective VEVs $v_d$ and $v_t$ such that $\sqrt{v_d^2+2v_t^2}=246$ GeV. For a detailed description of the main dynamical features of the scalar potential, see Ref.~\cite{Arhrib:2011uy}. After the electroweak symmetry breaking, mixing of the identically charged states results in several physical states: 
\begin{enumerate}[label=$(\roman*)$]
\item $\Phi^0$ and $\Delta^0$ mix into two CP-even states ($h$ and $H^0$) and two CP-odd states ($G^0$ and $A^0$), 
\item $\Phi^\pm$ and $\Delta^\pm$ mix into two mass states $G^\pm$ and $H^\pm$, 
\item $\Delta^{\pm \pm}$ is aligned with its mass state $H^{\pm \pm}$.
\end{enumerate}
$G^0$ and $G^\pm$ are the {\it would-be} Nambu-Goldstone bosons, $h$ is identified as the 125 GeV Higgs observed at the LHC, and the rest follows the sum rule
\[
m_{H^{\pm\pm}}^2-m_{H^\pm}^2 \approx m_{H^\pm}^2 - m_{H^0/A^0}^2 \approx -\frac{\lambda_4}{4}v_d^2.
\]

The Yukawa interaction $Y^{\nu}_{ij} L^T_i C i \sigma^2 \Delta L_j$ ($L_i$ stands for the SM lepton doublet with $i\in e,\mu,\tau$, and $C$ the charge-conjugation operator) induces masses for the neutrinos:
\[
m_\nu=\sqrt{2}Y^\nu v_t.
\]

\begin{figure}[htb!]
\centering
\includegraphics[width=0.75\columnwidth]{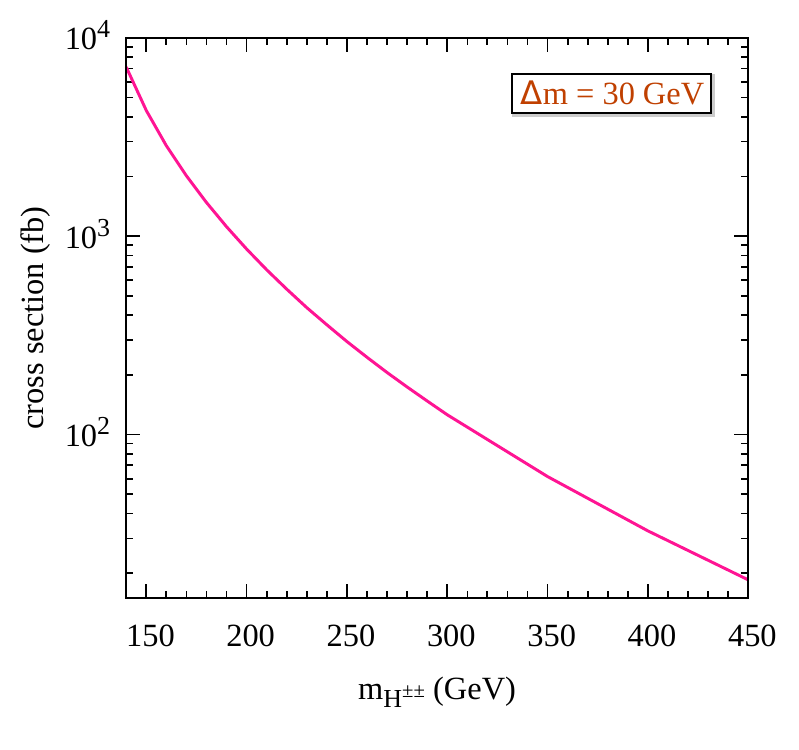}
\caption{\label{fig:xsec} LO production cross section for the triplet-like Higgses for $\Delta m = 30$ GeV at the 13 TeV LHC.}
\end{figure}

The triple-like Higgses are pair produced at the LHC via the neutral and charged current Drell-Yan mechanisms:\footnote{They are also produced via $t/u$-channel photon fusion as well as vector-boson fusion processes. However, such processes are rather sub-dominant.}
\begin{align*}
&q \bar{q} \to \gamma^*/Z^* \to H^{++} H^{--}, H^+ H^-, H^0 A^0
\\
&q q^\prime \to W^{\pm *} \to H^{\pm\pm} H^\pm, H^\pm H^0, H^\pm A^0
\end{align*}
We evaluate the leading order (LO) cross sections using the \texttt{SARAH 4.14.4} \cite{Staub:2013tta,Staub:2015kfa} generated \texttt{UFO} \cite{Degrande:2011ua} modules in \texttt{MadGraph5\_aMC\_v2.7.3} \cite{Alwall:2011uj,Alwall:2014hca} with the \texttt{NNPDF23\_lo\_as\_0130\_qed} parton distribution function \cite{Ball:2013hta,NNPDF:2014otw}. Fig.~\ref{fig:xsec} shows the LO production cross section for the triplet-like Higgses at the 13 TeV LHC as a function of $m_{H^{\pm\pm}}$ for $\Delta m = 30$ GeV. Following the QCD corrections estimated in Ref.~\cite{Fuks:2019clu}, we naively scale the LO cross section by a next-to-leading order (NLO) $K$-factor of 1.15.

\section{\label{sec:collider} Collider analysis}
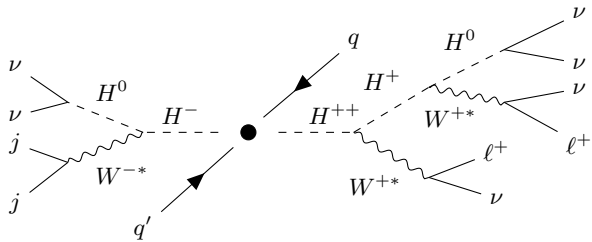
\begin{figure}[htb!]
\begin{tikzpicture}
\begin{feynman}
\vertex[large, dot] (c) {};
\vertex[right=0.4cm of c] (cr);
\vertex[left=0.4cm of c] (cl);
\vertex[right=1.0cm of cr] (r1);
\vertex[right=1.0cm of r1] (r20);
\vertex[above=0.6cm of r20] (r2);
\vertex[right=1.0cm of r2] (r30);
\vertex[above=0.5cm of r30] (r3);
\vertex[right=1.0cm of r3] (r50);
\vertex[above=0.3cm of r50] (r5) {$\nu$};
\vertex[below=0.0cm of r50] (r6) {$\nu$};
\vertex[below=0.7cm of r3] (r4);
\vertex[right=1.0cm of r4] (r70);
\vertex[above=0.0cm of r70] (r7) {$\nu$};
\vertex[below=0.3cm of r70] (r8) {$\ell^+$};
\vertex[below=0.6cm of r20] (r9);
\vertex[right=0.9cm of r9] (r100);
\vertex[above=0.1cm of r100] (r10) {$\ell^+$};
\vertex[below=0.1cm of r100] (r11) {$\nu$};
\vertex[left=1.0cm of cl] (l1);
\vertex[left=1.0cm of l1] (l20);
\vertex[above=0.4cm of l20] (l2);
\vertex[below=0.4cm of l20] (l3);
\vertex[left=0.7cm of l2] (l40);
\vertex[above=0.3cm of l40] (l4) {$\nu$};
\vertex[below=0.0cm of l40] (l5) {$\nu$};
\vertex[left=0.7cm of l3] (l60);
\vertex[above=0.0cm of l60] (l6) {$j$};
\vertex[below=0.3cm of l60] (l7) {$j$};
\vertex[right=0.2cm of c] (u10);
\vertex[above=0.2cm of u10] (u1);
\vertex[above=1.0cm of r1] (u2) {$q$};
\vertex[left=0.2cm of c] (d10);
\vertex[below=0.2cm of d10] (d1);
\vertex[below=1.0cm of l1] (d2) {$q^\prime$};
\diagram*{
(cr) -- [scalar, edge label=\(H^{++}\), near end] (r1) -- [scalar, edge label=\(H^+\), near end] (r2) -- [scalar, edge label=\(H^0\), near end] (r3),  
(r2) -- [boson, edge label'=\(W^{+*}\), near end] (r4),
(r3) -- (r5), (r3) -- (r6),
(r4) -- (r7), (r4) -- (r8),
(r1) -- [boson, edge label'=\(W^{+*}\), near end] (r9),
(r9) -- (r10), (r9) -- (r11),
(cl) -- [scalar, edge label'=\(H^-\)] (l1) -- [scalar, edge label'=\(H^0\), near end] (l2),
(l1) -- [boson, edge label=\(W^{-*}\), near end] (l3),
(l2) -- (l4), (l2) -- (l5),
(l3) -- (l6), (l3) -- (l7),
(u2) -- [fermion] (u1),
(d2) -- [fermion] (d1)
};
\end{feynman}
\end{tikzpicture}
\caption{\label{fig:Feynman} Schematic diagram for $qq^\prime \to H^{++}H^-$ leading to a pair of same-sign leptons and $p_T^{\rm miss}$ in the final state.}
\end{figure}

In broad terms, the phenomenology of this model depends on three parameters, namely $m_{H^{\pm \pm}}$, $v_t$ and $\Delta m=m_{H^{\pm \pm}}-m_{H^\pm}$, see Refs.~\cite{FileviezPerez:2008jbu,Aoki:2011pz,Ashanujjaman:2021txz} for detailed discussions. As mentioned earlier, the present work concerns the unconstrained parameters space characterised by $\Delta m \sim \mathcal{O}(10)$ GeV and $v_t \lesssim \mathcal{O}(10^{-4})$, where $H^{\pm\pm}$ and $H^\pm$ exclusively decay into off-shell $W^\pm$-bosons and $H^0/A^0$, with the latter further decaying into $\nu \nu$.\footnote{The analysis presented in this work is largely insensitive to the value of $v_t$ as long as $v_t \lesssim \mathcal{O}(10^{-4})$, thus we do not commit to a fixed value for $v_t$.} Therefore, the resulting final state is populated with soft leptons and jets and low $p_T^{\rm miss}$. A pair of same-sign leptons constitute the final state signature of the search presented in this work, see Fig.~\ref{fig:Feynman}.

In the following, we briefly outline the relevant SM background processes, the selection of various physics objects, and event selection, then perform a multivariate analysis to distinguish the signal from background.

\subsection{\label{sec:backg} SM backgrounds}
The final state with a pair of same-sign leptons is beset with less SM background contributions compared to other final states with leptons and jets. However, for the present analysis, we consider numerous SM processes such as diboson, triboson and tetraboson processes, Higgsstrahlung processes, single and multi-top productions in association with/without gauge bosons, and Drell-Yan processes. All these processes are generated using {\tt MadGraph5\_aMC\_v2.7.3} \cite{Alwall:2011uj,Alwall:2014hca} at the LO precision in perturbative QCD, with the {\it MLM merging} scheme to consolidate additional partons using {\tt PYTHIA~8.2} \cite{Sjostrand:2014zea}, and then naively scaled by appropriate NLO (or higher, whichever is available in the literature) $K$-factors \cite{Catani:2009sm,Balossini:2009sa,Campbell:2011bn,Cascioli:2014yka,Campbell:2016jau,LHCHiggsCrossSectionWorkingGroup:2016ypw,Shen:2016ape,Nhung:2013jta,Shen:2015cwj,Wang:2016fvj,Alwall:2014hca,Frederix:2014hta,Kidonakis:2015nna,Muselli:2015kba,Broggio:2019ewu,Frederix:2017wme}.

The relevant backgrounds can be broadly divided into three categories:
\begin{enumerate}[label=$(\roman*)$]
\item Prompt background: Particles originating from (very close vicinity of) the primary interaction point constitute this background, with the dominant contributions coming from the diboson and top-pair production processes.
\item Non-prompt and {\it fake} background: Processes where a jet is misidentified as a lepton or additional leptons originate from initial and final state radiation (ISR and FSR) photon conversions and in-flight heavy-flavour decays constitute this background. Though the lepton isolation requirements (mentioned in Section~\ref{sec:object}) significantly subdue this contribution, a considerable fraction passes the object selection. Estimating this contribution requires a data-driven approach, the so-called {\it fake factor} method, which is beyond the scope of this work. We adopt a conservative approach, assuming a $p_T$-dependent probability of 0.1–0.3\% for a jet to be misidentified as a lepton \cite{ATLAS:2016iqc}.
\item Electron charge misidentification: Bremsstrahlung interaction of the electrons with the inner detector material triggering trident events and sniff tracks could lead to charge misidentification. Therefore, a small fraction of prompt background events with a pair of opposite-sign leptons (mainly from Drell-Yan and top-pair production processes) can lead to same-sign leptons in the final states. To account for this effect, all prompt electrons are naively corrected with a $p_T$- and $\eta$-dependent charge misidentification probability: $P(p_T,\eta)=\sigma(p_T) \times f(\eta)$, where $\sigma(p_T)$ and $f(\eta)$ ranges from 0.02 to 0.1 and 0.03 to 1, respectively \cite{ATLAS:2017xqs}.
\end{enumerate}

\subsection{\label{sec:object} Object selection}
We pass the \texttt{MadGraph5\_aMC\_v2.7.3} generated parton-level events into {\tt PYTHIA~8.2} \cite{Sjostrand:2014zea} to simulate subsequent decays for the unstable particles, ISR and FSR, showering, fragmentation and hadronisation, and then into {\tt Delphes~3.4.2} with the default CMS card \cite{deFavereau:2013fsa} for simulating detector effects as well as reconstructing various physics objects, {\it viz.} photons, electrons, muons and jets.

The jet constituents are clustered using the {\it anti-k$_T$ algorithm} \cite{Cacciari:2008gp} with a jet radius $R=0.4$ as implemented in {\tt FastJet 3.3.2} \cite{Cacciari:2011ma}. While the jets are required to be within the pseudorapidity range $|\eta| < 2.4$ and have a transverse momentum $p_T > 20$ GeV, the leptons (electrons and muons) are required to have $|\eta|<2.5$ and $p_T > 10$ GeV. Further, to ensure that the leptons are isolated, we demand the scalar sum of the $p_T$s of all other objects lying within a cone of radius 0.3(0.4) around an electron (a muon) to be smaller than 10\%(15\%) of its $p_T$. Finally, $p_T^{\rm miss}$ is estimated from the momentum imbalance in the transverse direction associated to all reconstructed objects (including photons) in an event.

\subsection{\label{sec:event} Event selection}
The present analysis requires events with a pair of same-sign isolated leptons. For such events, we apply the following preselection requirements.
\begin{enumerate}[label=$(\roman*)$]
\item We reject events with $m_{\ell\ell} \in [3,3.2]$ GeV to lessen the contribution from $J/\psi$ resonance. No veto is applied around other resonances like $\Upsilon$ or $\Psi$ as these contributions are rather sub-dominant.
\item We require $m_{\ell\ell} > 1$ GeV and the angular separation between the leptons $\Delta R_{\ell\ell} > 0.05$ to suppress the nearly collinear lepton pairs resulting from ISR/FSR photon conversions or spurious pairs of tracks with shared hits from muon bremsstrahlung interactions.
\item Leptons are required to separated from the reconstructed jets by $\Delta R_{\ell,j} > 0.4$. This, along with the lepton isolation requirements, suppress non-prompt leptons from in-flight heavy-flavour decays.
\item Both the leptons are required to have $p_T > 15$ GeV. Though this is at par with the dilepton triggers used in 2015 in the Run 2 LHC \cite{ATLAS:2016wtr}, a little higher $p_T$ thresholds have been used \cite{ATLAS:2019dpa,CMS:2020cmk} in the later years. A trigger with higher $p_T$ thresholds would reduce the signal acceptance. Therefore, to retain the signal rate, a `combined' trigger with $p_T(\ell_{1,2}) > 15$ GeV and the dilepton invariant mass $m_{\ell\ell} < 60$ GeV can be used.
\end{enumerate}

\begin{figure}[htb!]
\centering
\includegraphics[width=0.75\columnwidth]{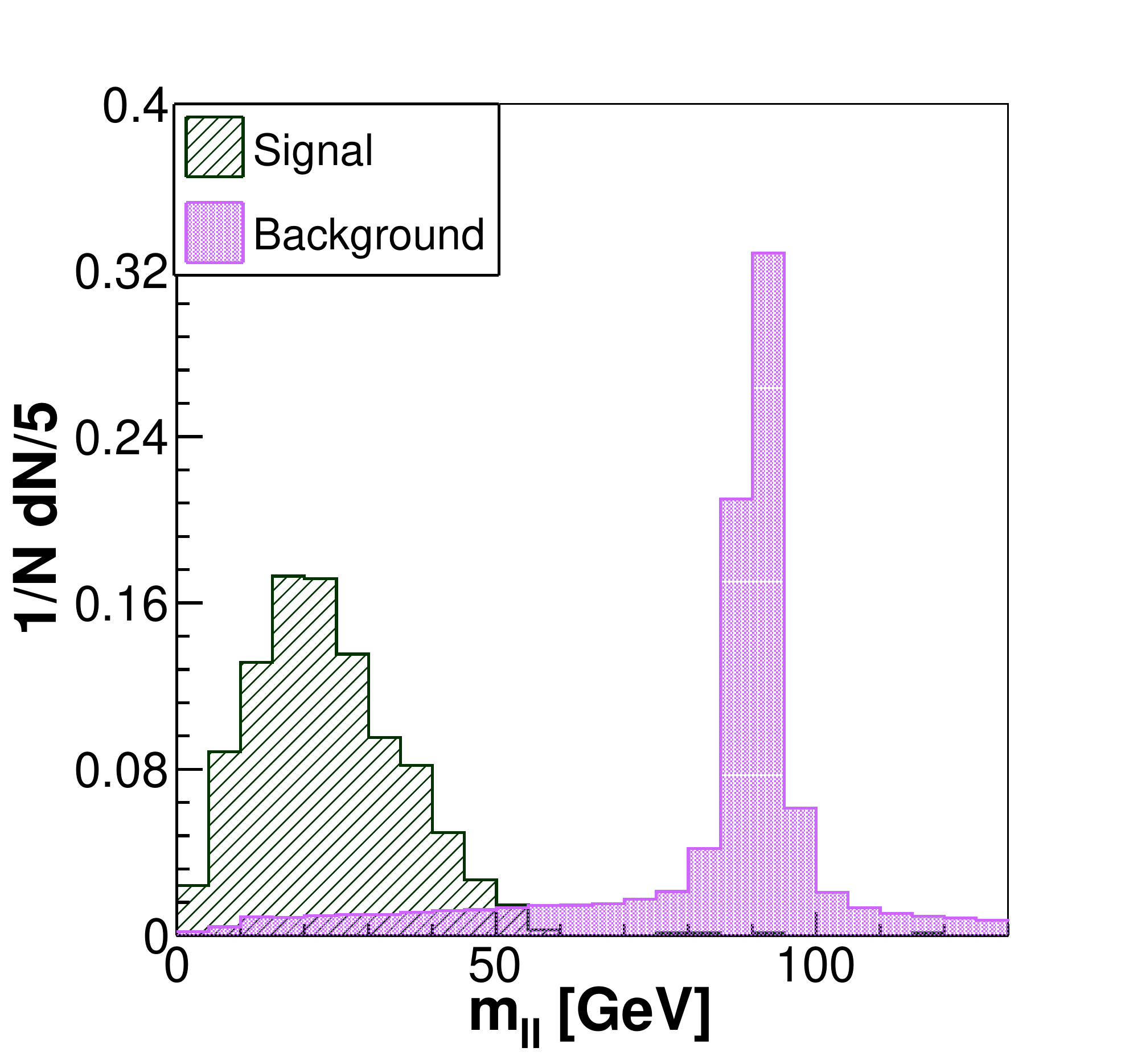}
\caption{\label{fig:mll} Normalised distributions of $m_{\ell\ell}$ for the signal (BP1) and background events after the $p_T(\ell_{1,2}) > 15$ GeV selection.}
\end{figure}

Shown in Fig.~\ref{fig:mll} is the normalised distributions of $m_{\ell\ell}$ for the signal and background events after the $p_T(\ell_{1,2}) > 15$ GeV selection. The signal events are shown for a benchmark defined by
\[
{\rm BP1}: m_{H^{\pm\pm}} = 200 {\rm ~GeV},~ \Delta m = 30 {\rm ~GeV}.
\]
For the signal, it falls rapidly with an end point near 60 GeV as occasioned by the compressed mass spectrum considered in BP1. On the contrary, the background boasts a peak at the $Z$-boson mass with the lion's share of the contributions accruing from $Z\to e^-e^+$ due to electron charge misidentification. Not only does the selection $m_{\ell\ell} < 60$ GeV help with the trigger, but it also vanquishes the oversized background contribution.

\subsection{\label{sec:bdt} Multivariate analysis}
We now perform a multivariate analysis with the BDT classifier implemented in the {\tt TMVA~4.3} toolkit \cite{Hocker:2007ht} integrated into {\tt ROOT~6.24} \cite{Brun:2000es} to distinguish the signals from backgrounds. For training and testing the classifier, $10^6$ signal events for each $m_{H^{\pm\pm}}$ within the [150,400] GeV range in steps of 50 GeV, and at least of worth 3000 fb$^{-1}$ luminosity of background events are fed. Of these, 50\% are picked randomly for training, and the rest are used for testing. The classifier is trained with the {\it adaptive boost} algorithm with a learning rate of 0.1, 500 decision trees with 5\% minimum node size and a depth of 5 layers per tree into a forest, and the {\it Gini index} being used for node splitting. The relevant BDT hyperparameters are summarised in Table~\ref{table:bdtParametrs}.

\begin{table}[htb!]
\centering
\begin{tabular}{ll}
\toprule
BDT hyperparameter & Optimised choice \\
\midrule
NTrees & 500 \\ 
MinNodeSize  & 5\% \\ 
MaxDepth & 5 \\ 
BoostType & AdaBoost\\ 
AdaBoostBeta & 0.1  \\ 
UseBaggedBoost & True \\ 
BaggedSampleFraction & 0.5 \\
SeparationType & GiniIndex \\ 
nCuts & 20 \\ 
\bottomrule
\end{tabular} 
\caption{\label{table:bdtParametrs} Summary of optimised BDT hyperparameters.}
\end{table}

\begin{figure}[htb!]
\centering
\includegraphics[width=0.48\columnwidth]{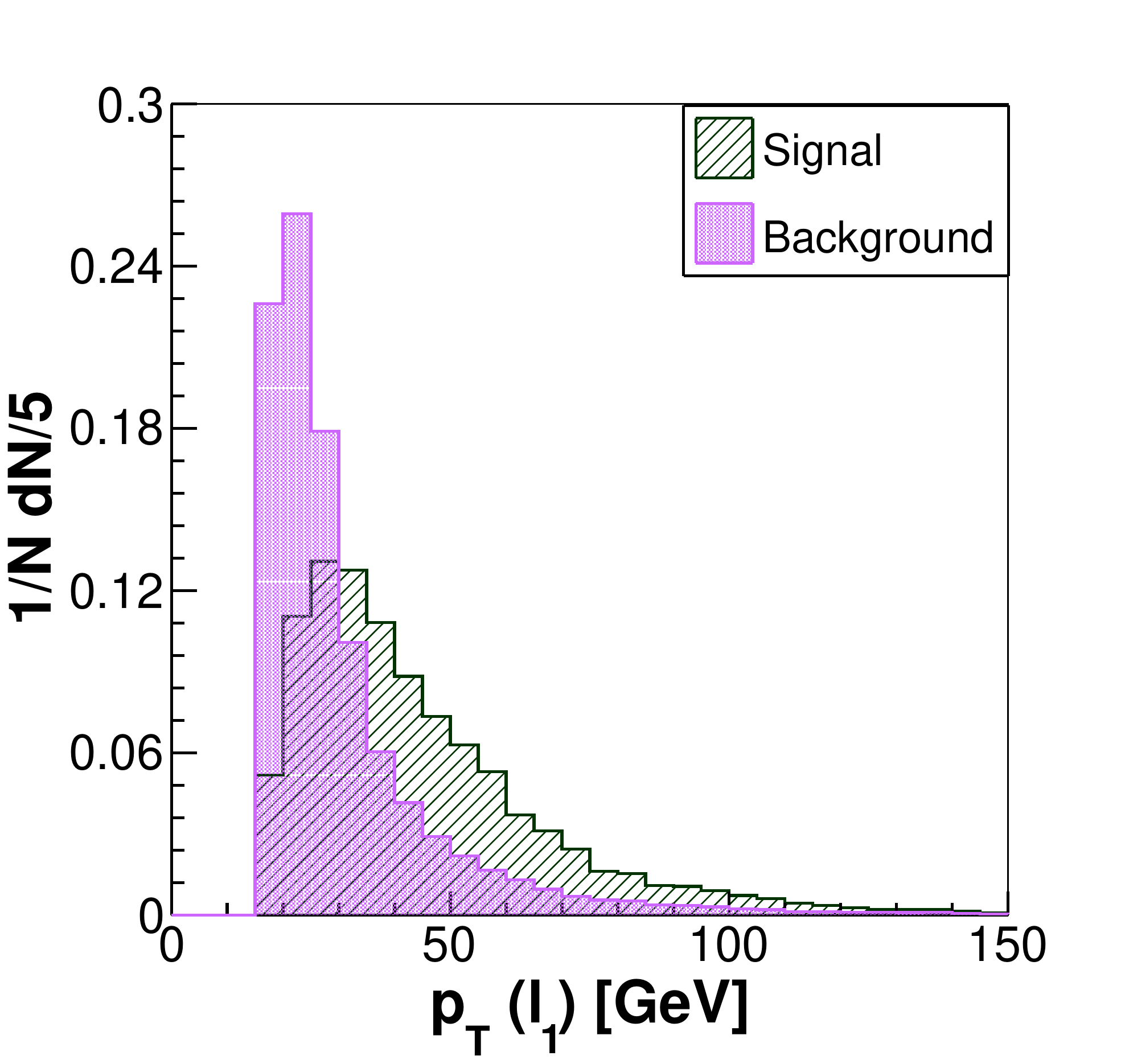}
\includegraphics[width=0.48\columnwidth]{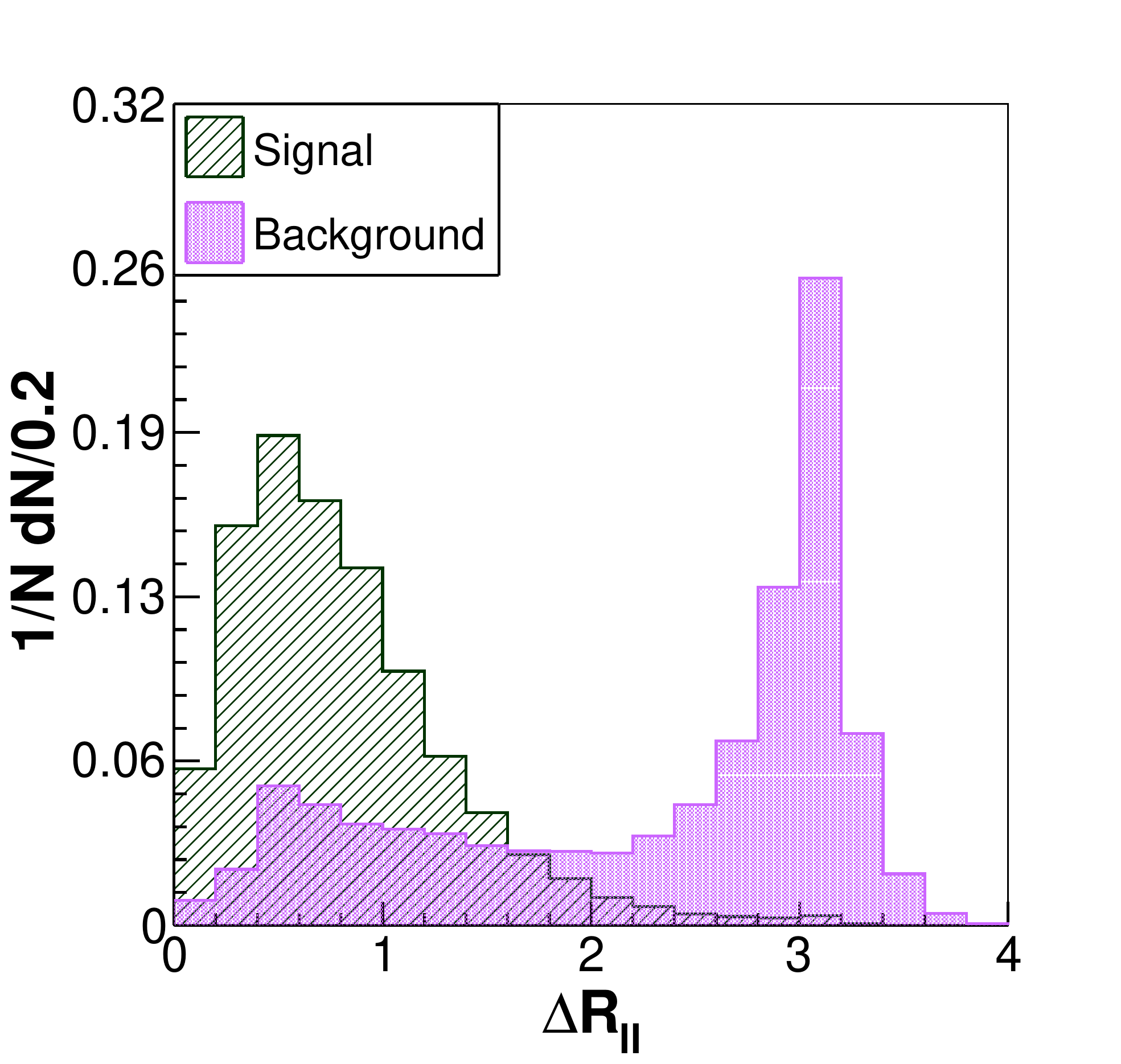}
\includegraphics[width=0.48\columnwidth]{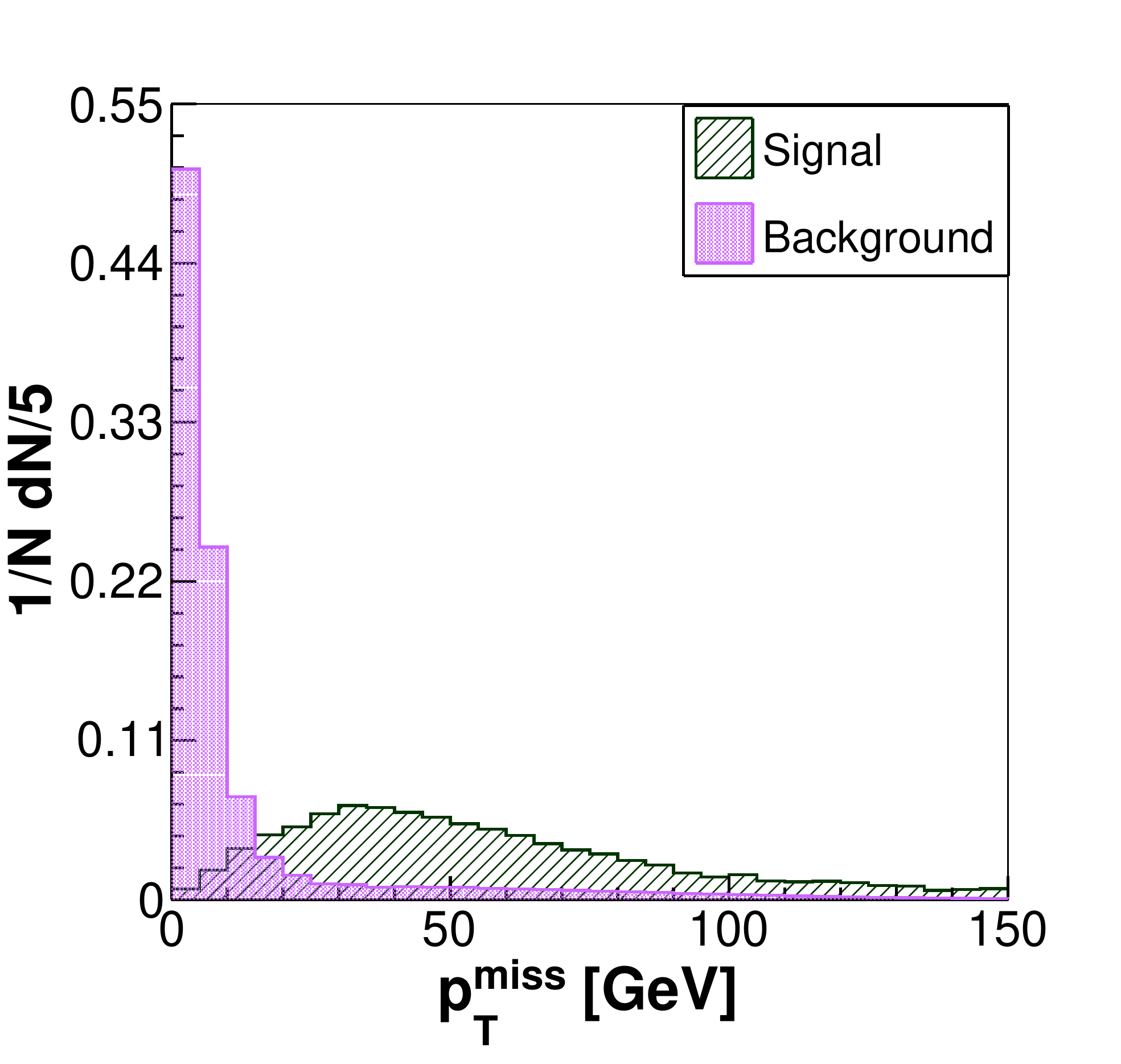}
\includegraphics[width=0.48\columnwidth]{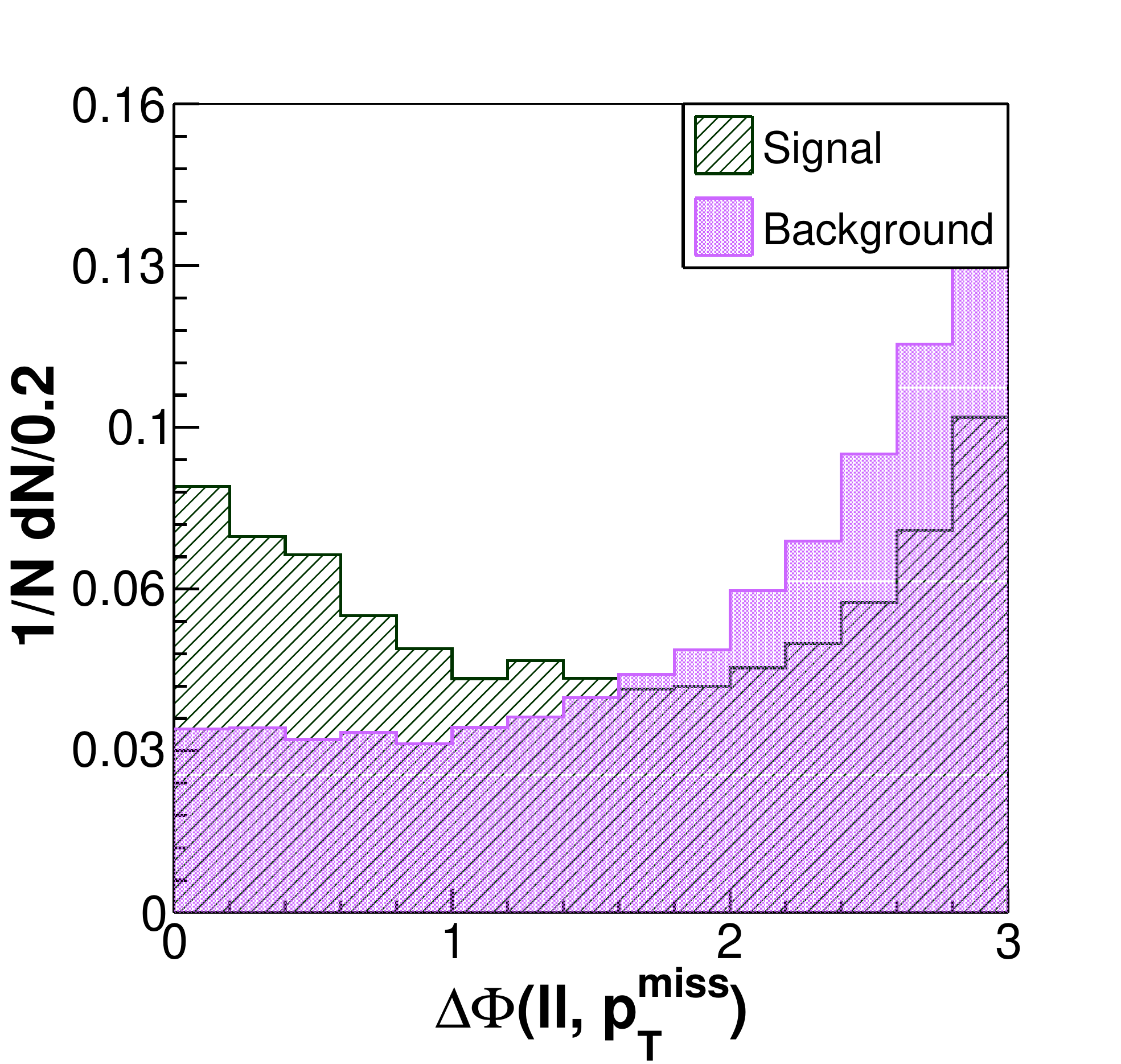}
\caption{\label{fig:bdtInput} Normalised distributions for some of the input features. The signal distributions are for BP1.}
\end{figure}

\begin{figure}[htb!]
\centering
\includegraphics[width=0.55\columnwidth]{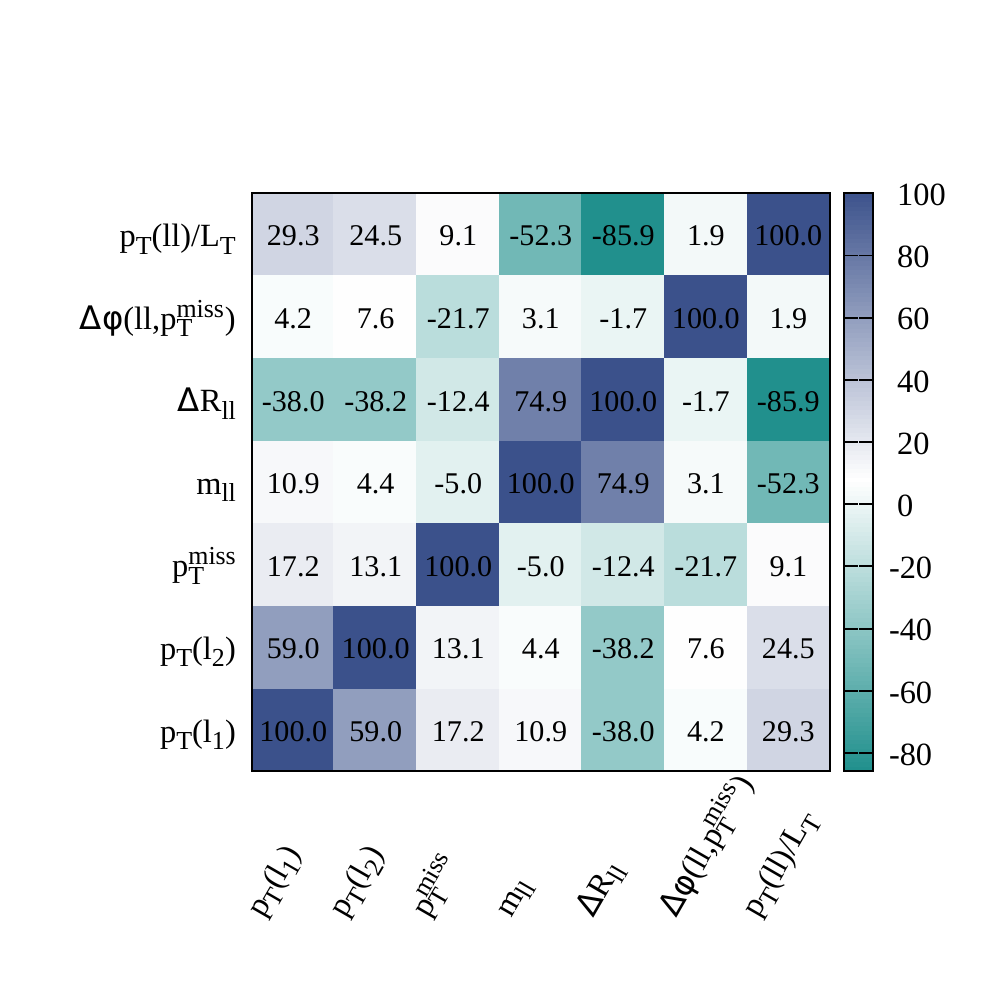}
\includegraphics[width=0.43\columnwidth]{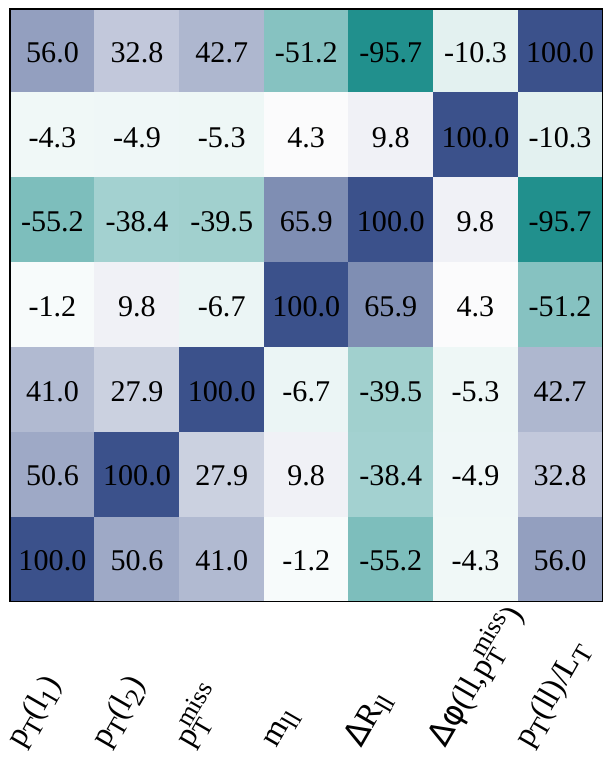}
\caption{\label{fig:bdtCorr} Correlations in \% among the input features for the signal (left) and background (right).}
\end{figure}

\begin{figure}[htb!]
\centering
\includegraphics[width=0.8\columnwidth]{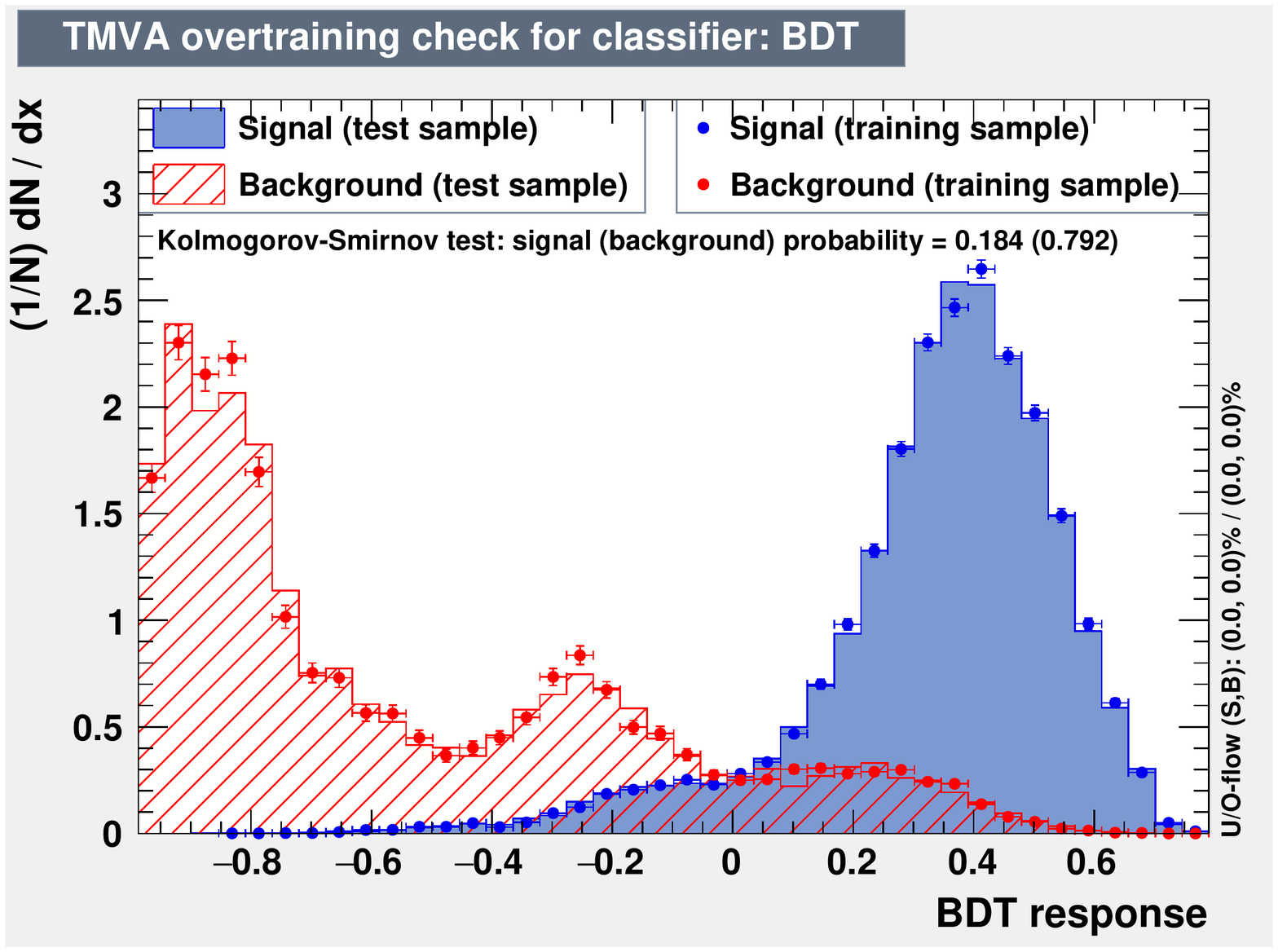}
\caption{\label{fig:bdtKS} BDT response curves for the training and testing subsamples.}
\end{figure}

We use the following kinematic variables as input features to the BDT classifier:\footnote{Just like $\Delta R_{\ell\ell}$, the azimuthal separation between the leptons $\Delta \phi_{\ell\ell}$ can be used as an input feature. However, these two variables are highly correlated, thus we drop the latter.}.
\[
p_T(\ell_{1,2}), ~p_T^{\rm miss}, ~m_{\ell\ell}, ~\Delta R_{\ell\ell}, ~\frac{p_T(\ell\ell)}{L_T} {\rm ~and~} \Delta\phi(\ell\ell,p_T^{\rm miss}),
\]
where $p_T(\ell\ell)$ is the dilepton system's $p_T$, and $L_T$ is the scalar sum of all jets and leptons' $p_T$. Normalised distributions for some of these features are shown in Fig.~\ref{fig:bdtInput}, the rest are not shown for brevity. These features constitute a minimal set with a good separation power between the signal and background, which is usually measured in terms of the method-unspecific separation and method-specific ranking. For a given feature $x$, the former is defined as
\[
\langle S^2 \rangle = \frac{1}{2} \int \frac{\left[\hat{x}_S(x)-\hat{x}_B(x)\right]^2}{\hat{x}_S(x)+\hat{x}_B(x)} dx
\]
where $\hat{x}_S(x)$ and $\hat{x}_B(x)$ are the probability density functions of $x$ for the signal and background, respectively. The method-specific ranking demonstrates the relative importance of the input features in separating the signal from background. Both these measures are shown in Table~\ref{table:bdtSeparationRanking}. As it turns out, $\Delta R_{\ell\ell}$ is the best separating variable, while $p_T(\ell_{1,2})$ are the ones with least separating power. Shown in Fig.~\ref{fig:bdtCorr} is the Pearson’s linear correlation matrix, with the coefficients defined as
\[
\rho(x,y) = \frac{\langle xy \rangle - \langle x \rangle \langle y \rangle}{\sigma_x \sigma_y},
\]
where $\langle x \rangle$ and $\sigma_x$, respectively, are the expectation value and standard deviation of $x$. These input features are not highly correlated, and thus constitute a minimal set. Lastly, the classifier is checked for overtraining by performing the Kolmogorov-Smirnov (KS) test which compares the BDT response curves for the training and testing subsamples. The response curves shown in Fig.~\ref{fig:bdtKS} exhibit no considerable overtraining.

\begin{table}[htb!]
\centering
\begin{tabular}{lll}
\toprule
 & Method-unspecific & Method-specific \\ 
Feature & separation & ranking \\ 
\midrule
$p_T^{\rm miss}$ & 0.5945 & 0.214 \\
$\Delta R_{\ell\ell}$ & 0.4777 & 0.3015 \\
$\frac{p_T(\ell\ell)}{L_T}$ & 0.4586 & 0.2086 \\
$m_{\ell\ell}$ & 0.368 & 0.08534 \\
$p_T(\ell_1)$ & 0.1481 & 0.04751 \\
$p_T(\ell_2)$ & 0.08474 & 0.05806 \\
$\Delta \phi(\ell\ell, p_T^{\rm miss})$ & 0.03997 & 0.08499 \\
\bottomrule
\end{tabular}
\caption{\label{table:bdtSeparationRanking} Method-unspecific separation and method-specific ranking of the input features.}
\end{table}

\begin{figure}[htb!]
\centering
\includegraphics[width=0.99\columnwidth]{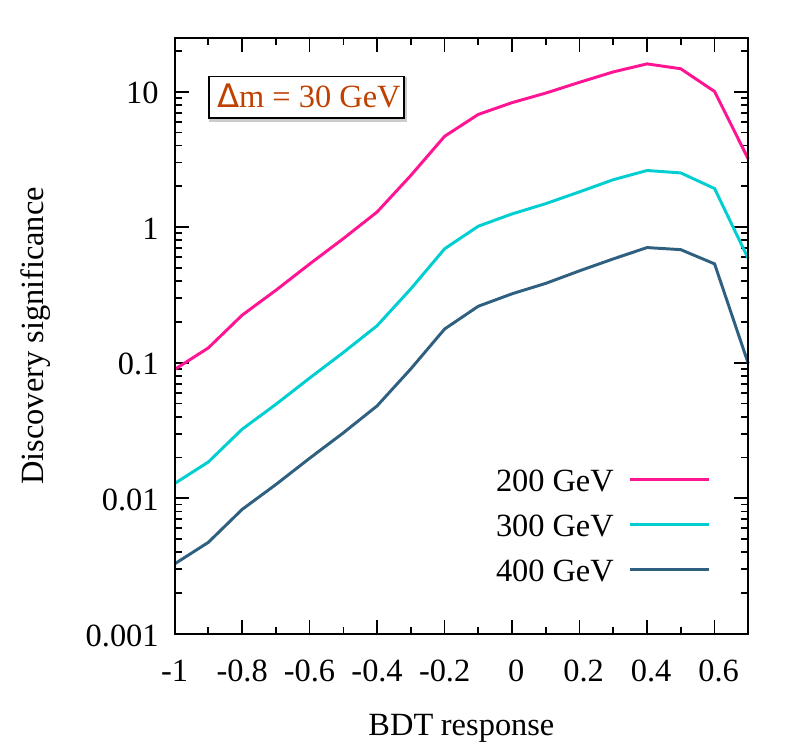}
\caption{\label{fig:SigDis139} Discovery significance as a function of the BDT response for $m_{H^{\pm\pm}}$ = 200, 300 and 400 GeV with $\Delta m=30$ GeV.}
\end{figure}

For brevity, the receiver-operator-characteristic (ROC) curve quantifying the BDT performance is not shown. Instead, we show the variation of the discovery significance (estimated using the formula given in Section.~\ref{sec:result}) as a function of the BDT response for $m_{H^{\pm\pm}}$ = 200, 300 and 400 GeV with $\Delta m=30$ GeV in Fig.~\ref{fig:SigDis139}. As apparent from this, the discovery significance reaches it's maximum for the BDT response of 0.4, albeit irrespective of $m_{H^{\pm\pm}}$, demonstrating the robustness of the BDT classifier's training. Therefore, to maximise the sensitivity of the search, we require
\[
{\rm BDT~response} > 0.4.
\]
Table~\ref{table:cut-flow} shows the remaining background and signal (for $m_{H^{\pm\pm}}=200,300$ and 400 GeV with $\Delta m=30$ GeV) cross sections after the above selection.

\begin{table}[htb!]
\centering %
\begin{tabular}{p{4cm} l}
\toprule
Event sample & Cross section (fb) \\
\midrule
$\gamma^*/Z^*$ & 2.6344\\
$W^\pm Z$ & 1.4435\\  
$W^\pm W^\pm V$ ($V=W,Z$) & 0.7913\\
$t\bar{t}V$ & 0.7313\\
$t\bar{t}$ & 0.5535\\ 
Others & 1.0618 \\
\hline
Total background & 7.2158\\
\hline
Signal: $m_{H^{\pm\pm}} = 200$ GeV & 4.6235\\
Signal: $m_{H^{\pm\pm}} = 300$ GeV & 0.6906\\
Signal: $m_{H^{\pm\pm}} = 400$ GeV & 0.1841\\
\bottomrule
\end{tabular} 
\caption{\label{table:cut-flow} Background and signal cross sections (fb) after the selection: BDT response $> 0.4$.}
\end{table}

\subsection{\label{sec:result} Discovery and exclusion projection}
\begin{figure}[htb!]
\centering
\includegraphics[width=0.99\columnwidth]{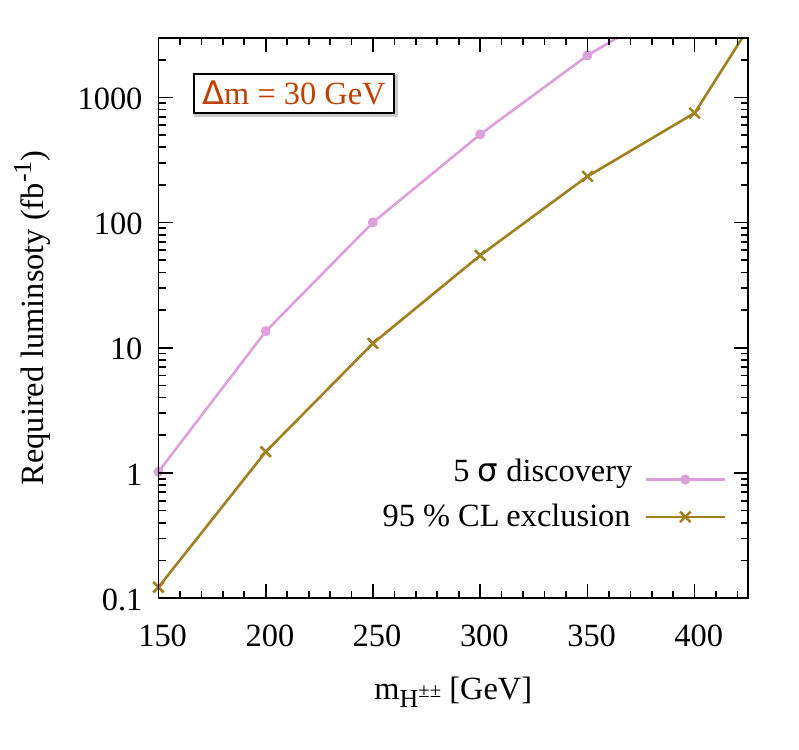}
\caption{\label{fig:lumi} Required luminosity (fb$^{-1}$) for the $5\sigma$ discovery and 95\% CL exclusion for different $m_{H^{\pm\pm}}$ with $\Delta m =30$ GeV.}
\end{figure}

Finally, we estimate the discovery and exclusion projection for different $m_{H^{\pm\pm}}$. The median expected discovery and exclusion significances are estimated as \cite{Cowan:2010js,Li:1983fv,Cousins:2007yta}
\begin{align*}
Z_{\rm dis} =& \Bigg[ 2\Bigg( (s+b) \ln \left[ \frac{(s+b)(b+\delta_b^2)}{b^2+(s+b)\delta_b^2} \right]
\\
&-\frac{b^2}{\delta_b^2} \ln\left [1+ \frac{\delta_b^2 s}{b(b+\delta_b^2)} \right] \Bigg) \Bigg]^{1/2},
\end{align*}

\begin{align*}
Z_{\rm exc} =& \Bigg[ 2 \left\{ s-b \ln \left( \frac{b+s+x}{2b} \right) - \frac{b^2}{\delta_b^2} \ln \left( \frac{b-s+x}{2b} \right) \right\} 
\\
&- (b+s-x)(1+b/\delta_b^2) \Bigg]^{1/2}
\end{align*}
where $x = \sqrt{(s+b)^2 - 4sb\delta_b^2/(b+\delta_b^2)}$, $s$ and $b$ are the numbers of signal and background events, respectively, and $\delta_b$ is the background uncertainty. Without going into the intricacy of estimating the latter, we assume it to be 20\%, albeit conservatively. In Fig.~\ref{fig:lumi}, we show the required luminosities (in fb$^{-1}$) needed to achieve 95\% CL ($1.645 \sigma$) exclusion and $5\sigma$ discovery for different $m_{H^{\pm\pm}}$ with $\Delta m =30$ GeV. For high $m_{H^{\pm\pm}}$, the signal cross section falls quickly, and thus the sensitivity. 

We find that, for $\Delta m =30$ GeV, this model could be probed up to $m_{H^{\pm\pm}} = 260 (330)$ GeV with $5\sigma$ discovery (95\% CL exclusion) significance with the already collected Run 2 LHC data ($\sim 139$ fb$^{-1}$). Whereas, for the high-luminosity LHC (HL-LHC) with 3000 fb$^{-1}$ data, this limit extends to 360 (420) GeV.

\section{\label{sec:summary} Summary}
Non-observations of any significant excess over the SM expectation in all the experimental searches looking for doubly-charged Higgses thus far have led to stringent limits on them. While interpreted in the context of the type-II seesaw model, the triplet-like Higgses are already excluded up to a few hundred GeV masses for a vast region of the model parameter space \cite{Ashanujjaman:2021txz}. In this work, we peruse a scenario, akin to compressed mass spectra in Supersymmetry, which these experimental searches are not sensitive to and consequently, is largely unconstrained. We perform a multivariate analysis to distinguish signals with a pair of same-sign leptons with low invariant mass from the SM processes, including those concerning {\it fake} leptons and electron charge misidentification, and find that a significant part of the hitherto unconstrained parameter space could be probed with the already collected Run2 LHC and future HL-LHC data.

In closing, we mention that the search strategy presented here is also applicable to other scalar extensions of the SM with compressed mass spectra where the lightest scalar state decays invisibly.

\acknowledgments SA acknowledges support from the SERB Core Research Grants CRG/2018/004889. The simulations were supported in part by the SAMKHYA: High Performance Computing Facility provided by Institute of Physics, Bhubaneswar.

\bibliography{v0}

\begin{thebibliography}{120}%
\makeatletter
\providecommand \@ifxundefined [1]{%
 \@ifx{#1\undefined}
}%
\providecommand \@ifnum [1]{%
 \ifnum #1\expandafter \@firstoftwo
 \else \expandafter \@secondoftwo
 \fi
}%
\providecommand \@ifx [1]{%
 \ifx #1\expandafter \@firstoftwo
 \else \expandafter \@secondoftwo
 \fi
}%
\providecommand \natexlab [1]{#1}%
\providecommand \enquote  [1]{``#1''}%
\providecommand \bibnamefont  [1]{#1}%
\providecommand \bibfnamefont [1]{#1}%
\providecommand \citenamefont [1]{#1}%
\providecommand \href@noop [0]{\@secondoftwo}%
\providecommand \href [0]{\begingroup \@sanitize@url \@href}%
\providecommand \@href[1]{\@@startlink{#1}\@@href}%
\providecommand \@@href[1]{\endgroup#1\@@endlink}%
\providecommand \@sanitize@url [0]{\catcode `\\12\catcode `\$12\catcode
  `\&12\catcode `\#12\catcode `\^12\catcode `\_12\catcode `\%12\relax}%
\providecommand \@@startlink[1]{}%
\providecommand \@@endlink[0]{}%
\providecommand \url  [0]{\begingroup\@sanitize@url \@url }%
\providecommand \@url [1]{\endgroup\@href {#1}{\urlprefix }}%
\providecommand \urlprefix  [0]{URL }%
\providecommand \Eprint [0]{\href }%
\providecommand \doibase [0]{https://doi.org/}%
\providecommand \selectlanguage [0]{\@gobble}%
\providecommand \bibinfo  [0]{\@secondoftwo}%
\providecommand \bibfield  [0]{\@secondoftwo}%
\providecommand \translation [1]{[#1]}%
\providecommand \BibitemOpen [0]{}%
\providecommand \bibitemStop [0]{}%
\providecommand \bibitemNoStop [0]{.\EOS\space}%
\providecommand \EOS [0]{\spacefactor3000\relax}%
\providecommand \BibitemShut  [1]{\csname bibitem#1\endcsname}%
\let\auto@bib@innerbib\@empty
\bibitem [{\citenamefont {Konetschny}\ and\ \citenamefont
  {Kummer}(1977)}]{Konetschny:1977bn}%
  \BibitemOpen
  \bibfield  {author} {\bibinfo {author} {\bibfnamefont {W.}~\bibnamefont
  {Konetschny}}\ and\ \bibinfo {author} {\bibfnamefont {W.}~\bibnamefont
  {Kummer}},\ }\bibfield  {title} {\bibinfo {title} {{Nonconservation of Total
  Lepton Number with Scalar Bosons}},\ }\href
  {https://doi.org/10.1016/0370-2693(77)90407-5} {\bibfield  {journal}
  {\bibinfo  {journal} {Phys. Lett. B}\ }\textbf {\bibinfo {volume} {70}},\
  \bibinfo {pages} {433} (\bibinfo {year} {1977})}\BibitemShut {NoStop}%
\bibitem [{\citenamefont {Cheng}\ and\ \citenamefont
  {Li}(1980)}]{Cheng:1980qt}%
  \BibitemOpen
  \bibfield  {author} {\bibinfo {author} {\bibfnamefont {T.~P.}\ \bibnamefont
  {Cheng}}\ and\ \bibinfo {author} {\bibfnamefont {L.-F.}\ \bibnamefont {Li}},\
  }\bibfield  {title} {\bibinfo {title} {{Neutrino Masses, Mixings and
  Oscillations in SU(2) x U(1) Models of Electroweak Interactions}},\ }\href
  {https://doi.org/10.1103/PhysRevD.22.2860} {\bibfield  {journal} {\bibinfo
  {journal} {Phys. Rev. D}\ }\textbf {\bibinfo {volume} {22}},\ \bibinfo
  {pages} {2860} (\bibinfo {year} {1980})}\BibitemShut {NoStop}%
\bibitem [{\citenamefont {Lazarides}\ \emph {et~al.}(1981)\citenamefont
  {Lazarides}, \citenamefont {Shafi},\ and\ \citenamefont
  {Wetterich}}]{Lazarides:1980nt}%
  \BibitemOpen
  \bibfield  {author} {\bibinfo {author} {\bibfnamefont {G.}~\bibnamefont
  {Lazarides}}, \bibinfo {author} {\bibfnamefont {Q.}~\bibnamefont {Shafi}},\
  and\ \bibinfo {author} {\bibfnamefont {C.}~\bibnamefont {Wetterich}},\
  }\bibfield  {title} {\bibinfo {title} {{Proton Lifetime and Fermion Masses in
  an SO(10) Model}},\ }\href {https://doi.org/10.1016/0550-3213(81)90354-0}
  {\bibfield  {journal} {\bibinfo  {journal} {Nucl. Phys. B}\ }\textbf
  {\bibinfo {volume} {181}},\ \bibinfo {pages} {287} (\bibinfo {year}
  {1981})}\BibitemShut {NoStop}%
\bibitem [{\citenamefont {Schechter}\ and\ \citenamefont
  {Valle}(1980)}]{Schechter:1980gr}%
  \BibitemOpen
  \bibfield  {author} {\bibinfo {author} {\bibfnamefont {J.}~\bibnamefont
  {Schechter}}\ and\ \bibinfo {author} {\bibfnamefont {J.~W.~F.}\ \bibnamefont
  {Valle}},\ }\bibfield  {title} {\bibinfo {title} {{Neutrino Masses in SU(2) x
  U(1) Theories}},\ }\href {https://doi.org/10.1103/PhysRevD.22.2227}
  {\bibfield  {journal} {\bibinfo  {journal} {Phys. Rev. D}\ }\textbf {\bibinfo
  {volume} {22}},\ \bibinfo {pages} {2227} (\bibinfo {year}
  {1980})}\BibitemShut {NoStop}%
\bibitem [{\citenamefont {Mohapatra}\ and\ \citenamefont
  {Senjanovic}(1981)}]{Mohapatra:1980yp}%
  \BibitemOpen
  \bibfield  {author} {\bibinfo {author} {\bibfnamefont {R.~N.}\ \bibnamefont
  {Mohapatra}}\ and\ \bibinfo {author} {\bibfnamefont {G.}~\bibnamefont
  {Senjanovic}},\ }\bibfield  {title} {\bibinfo {title} {{Neutrino Masses and
  Mixings in Gauge Models with Spontaneous Parity Violation}},\ }\href
  {https://doi.org/10.1103/PhysRevD.23.165} {\bibfield  {journal} {\bibinfo
  {journal} {Phys. Rev. D}\ }\textbf {\bibinfo {volume} {23}},\ \bibinfo
  {pages} {165} (\bibinfo {year} {1981})}\BibitemShut {NoStop}%
\bibitem [{\citenamefont {Magg}\ and\ \citenamefont
  {Wetterich}(1980)}]{Magg:1980ut}%
  \BibitemOpen
  \bibfield  {author} {\bibinfo {author} {\bibfnamefont {M.}~\bibnamefont
  {Magg}}\ and\ \bibinfo {author} {\bibfnamefont {C.}~\bibnamefont
  {Wetterich}},\ }\bibfield  {title} {\bibinfo {title} {{Neutrino Mass Problem
  and Gauge Hierarchy}},\ }\href {https://doi.org/10.1016/0370-2693(80)90825-4}
  {\bibfield  {journal} {\bibinfo  {journal} {Phys. Lett. B}\ }\textbf
  {\bibinfo {volume} {94}},\ \bibinfo {pages} {61} (\bibinfo {year}
  {1980})}\BibitemShut {NoStop}%
\bibitem [{\citenamefont {Huitu}\ \emph {et~al.}(1997)\citenamefont {Huitu},
  \citenamefont {Maalampi}, \citenamefont {Pietila},\ and\ \citenamefont
  {Raidal}}]{Huitu:1996su}%
  \BibitemOpen
  \bibfield  {author} {\bibinfo {author} {\bibfnamefont {K.}~\bibnamefont
  {Huitu}}, \bibinfo {author} {\bibfnamefont {J.}~\bibnamefont {Maalampi}},
  \bibinfo {author} {\bibfnamefont {A.}~\bibnamefont {Pietila}},\ and\ \bibinfo
  {author} {\bibfnamefont {M.}~\bibnamefont {Raidal}},\ }\bibfield  {title}
  {\bibinfo {title} {{Doubly charged Higgs at LHC}},\ }\href
  {https://doi.org/10.1016/S0550-3213(97)87466-4} {\bibfield  {journal}
  {\bibinfo  {journal} {Nucl. Phys. B}\ }\textbf {\bibinfo {volume} {487}},\
  \bibinfo {pages} {27} (\bibinfo {year} {1997})},\ \Eprint
  {https://arxiv.org/abs/hep-ph/9606311} {arXiv:hep-ph/9606311} \BibitemShut
  {NoStop}%
\bibitem [{\citenamefont {Gunion}\ \emph {et~al.}(1996)\citenamefont {Gunion},
  \citenamefont {Loomis},\ and\ \citenamefont {Pitts}}]{Gunion:1996pq}%
  \BibitemOpen
  \bibfield  {author} {\bibinfo {author} {\bibfnamefont {J.~F.}\ \bibnamefont
  {Gunion}}, \bibinfo {author} {\bibfnamefont {C.}~\bibnamefont {Loomis}},\
  and\ \bibinfo {author} {\bibfnamefont {K.~T.}\ \bibnamefont {Pitts}},\
  }\bibfield  {title} {\bibinfo {title} {{Searching for doubly charged Higgs
  bosons at future colliders}},\ }\href@noop {} {\bibfield  {journal} {\bibinfo
   {journal} {eConf}\ }\textbf {\bibinfo {volume} {C960625}},\ \bibinfo {pages}
  {LTH096} (\bibinfo {year} {1996})},\ \Eprint
  {https://arxiv.org/abs/hep-ph/9610237} {arXiv:hep-ph/9610237} \BibitemShut
  {NoStop}%
\bibitem [{\citenamefont {Chakrabarti}\ \emph {et~al.}(1998)\citenamefont
  {Chakrabarti}, \citenamefont {Choudhury}, \citenamefont {Godbole},\ and\
  \citenamefont {Mukhopadhyaya}}]{Chakrabarti:1998qy}%
  \BibitemOpen
  \bibfield  {author} {\bibinfo {author} {\bibfnamefont {S.}~\bibnamefont
  {Chakrabarti}}, \bibinfo {author} {\bibfnamefont {D.}~\bibnamefont
  {Choudhury}}, \bibinfo {author} {\bibfnamefont {R.~M.}\ \bibnamefont
  {Godbole}},\ and\ \bibinfo {author} {\bibfnamefont {B.}~\bibnamefont
  {Mukhopadhyaya}},\ }\bibfield  {title} {\bibinfo {title} {{Observing doubly
  charged Higgs bosons in photon-photon collisions}},\ }\href
  {https://doi.org/10.1016/S0370-2693(98)00743-6} {\bibfield  {journal}
  {\bibinfo  {journal} {Phys. Lett. B}\ }\textbf {\bibinfo {volume} {434}},\
  \bibinfo {pages} {347} (\bibinfo {year} {1998})},\ \Eprint
  {https://arxiv.org/abs/hep-ph/9804297} {arXiv:hep-ph/9804297} \BibitemShut
  {NoStop}%
\bibitem [{\citenamefont {Muhlleitner}\ and\ \citenamefont
  {Spira}(2003)}]{Muhlleitner:2003me}%
  \BibitemOpen
  \bibfield  {author} {\bibinfo {author} {\bibfnamefont {M.}~\bibnamefont
  {Muhlleitner}}\ and\ \bibinfo {author} {\bibfnamefont {M.}~\bibnamefont
  {Spira}},\ }\bibfield  {title} {\bibinfo {title} {{A Note on doubly charged
  Higgs pair production at hadron colliders}},\ }\href
  {https://doi.org/10.1103/PhysRevD.68.117701} {\bibfield  {journal} {\bibinfo
  {journal} {Phys. Rev. D}\ }\textbf {\bibinfo {volume} {68}},\ \bibinfo
  {pages} {117701} (\bibinfo {year} {2003})},\ \Eprint
  {https://arxiv.org/abs/hep-ph/0305288} {arXiv:hep-ph/0305288} \BibitemShut
  {NoStop}%
\bibitem [{\citenamefont {Akeroyd}\ and\ \citenamefont
  {Aoki}(2005)}]{Akeroyd:2005gt}%
  \BibitemOpen
  \bibfield  {author} {\bibinfo {author} {\bibfnamefont {A.~G.}\ \bibnamefont
  {Akeroyd}}\ and\ \bibinfo {author} {\bibfnamefont {M.}~\bibnamefont {Aoki}},\
  }\bibfield  {title} {\bibinfo {title} {{Single and pair production of doubly
  charged Higgs bosons at hadron colliders}},\ }\href
  {https://doi.org/10.1103/PhysRevD.72.035011} {\bibfield  {journal} {\bibinfo
  {journal} {Phys. Rev. D}\ }\textbf {\bibinfo {volume} {72}},\ \bibinfo
  {pages} {035011} (\bibinfo {year} {2005})},\ \Eprint
  {https://arxiv.org/abs/hep-ph/0506176} {arXiv:hep-ph/0506176} \BibitemShut
  {NoStop}%
\bibitem [{\citenamefont {Akeroyd}\ \emph {et~al.}(2008)\citenamefont
  {Akeroyd}, \citenamefont {Aoki},\ and\ \citenamefont
  {Sugiyama}}]{Akeroyd:2007zv}%
  \BibitemOpen
  \bibfield  {author} {\bibinfo {author} {\bibfnamefont {A.~G.}\ \bibnamefont
  {Akeroyd}}, \bibinfo {author} {\bibfnamefont {M.}~\bibnamefont {Aoki}},\ and\
  \bibinfo {author} {\bibfnamefont {H.}~\bibnamefont {Sugiyama}},\ }\bibfield
  {title} {\bibinfo {title} {{Probing Majorana Phases and Neutrino Mass
  Spectrum in the Higgs Triplet Model at the CERN LHC}},\ }\href
  {https://doi.org/10.1103/PhysRevD.77.075010} {\bibfield  {journal} {\bibinfo
  {journal} {Phys. Rev. D}\ }\textbf {\bibinfo {volume} {77}},\ \bibinfo
  {pages} {075010} (\bibinfo {year} {2008})},\ \Eprint
  {https://arxiv.org/abs/0712.4019} {arXiv:0712.4019 [hep-ph]} \BibitemShut
  {NoStop}%
\bibitem [{\citenamefont {Garayoa}\ and\ \citenamefont
  {Schwetz}(2008)}]{Garayoa:2007fw}%
  \BibitemOpen
  \bibfield  {author} {\bibinfo {author} {\bibfnamefont {J.}~\bibnamefont
  {Garayoa}}\ and\ \bibinfo {author} {\bibfnamefont {T.}~\bibnamefont
  {Schwetz}},\ }\bibfield  {title} {\bibinfo {title} {{Neutrino mass hierarchy
  and Majorana CP phases within the Higgs triplet model at the LHC}},\ }\href
  {https://doi.org/10.1088/1126-6708/2008/03/009} {\bibfield  {journal}
  {\bibinfo  {journal} {JHEP}\ }\textbf {\bibinfo {volume} {03}},\ \bibinfo
  {pages} {009}},\ \Eprint {https://arxiv.org/abs/0712.1453} {arXiv:0712.1453
  [hep-ph]} \BibitemShut {NoStop}%
\bibitem [{\citenamefont {Han}\ \emph {et~al.}(2007)\citenamefont {Han},
  \citenamefont {Mukhopadhyaya}, \citenamefont {Si},\ and\ \citenamefont
  {Wang}}]{Han:2007bk}%
  \BibitemOpen
  \bibfield  {author} {\bibinfo {author} {\bibfnamefont {T.}~\bibnamefont
  {Han}}, \bibinfo {author} {\bibfnamefont {B.}~\bibnamefont {Mukhopadhyaya}},
  \bibinfo {author} {\bibfnamefont {Z.}~\bibnamefont {Si}},\ and\ \bibinfo
  {author} {\bibfnamefont {K.}~\bibnamefont {Wang}},\ }\bibfield  {title}
  {\bibinfo {title} {{Pair production of doubly-charged scalars: Neutrino mass
  constraints and signals at the LHC}},\ }\href
  {https://doi.org/10.1103/PhysRevD.76.075013} {\bibfield  {journal} {\bibinfo
  {journal} {Phys. Rev. D}\ }\textbf {\bibinfo {volume} {76}},\ \bibinfo
  {pages} {075013} (\bibinfo {year} {2007})},\ \Eprint
  {https://arxiv.org/abs/0706.0441} {arXiv:0706.0441 [hep-ph]} \BibitemShut
  {NoStop}%
\bibitem [{\citenamefont {del Aguila}\ and\ \citenamefont
  {Aguilar-Saavedra}(2009)}]{delAguila:2008cj}%
  \BibitemOpen
  \bibfield  {author} {\bibinfo {author} {\bibfnamefont {F.}~\bibnamefont {del
  Aguila}}\ and\ \bibinfo {author} {\bibfnamefont {J.~A.}\ \bibnamefont
  {Aguilar-Saavedra}},\ }\bibfield  {title} {\bibinfo {title} {{Distinguishing
  seesaw models at LHC with multi-lepton signals}},\ }\href
  {https://doi.org/10.1016/j.nuclphysb.2008.12.029} {\bibfield  {journal}
  {\bibinfo  {journal} {Nucl. Phys. B}\ }\textbf {\bibinfo {volume} {813}},\
  \bibinfo {pages} {22} (\bibinfo {year} {2009})},\ \Eprint
  {https://arxiv.org/abs/0808.2468} {arXiv:0808.2468 [hep-ph]} \BibitemShut
  {NoStop}%
\bibitem [{\citenamefont {Fileviez~Perez}\ \emph
  {et~al.}(2008{\natexlab{a}})\citenamefont {Fileviez~Perez}, \citenamefont
  {Han}, \citenamefont {Huang}, \citenamefont {Li},\ and\ \citenamefont
  {Wang}}]{FileviezPerez:2008jbu}%
  \BibitemOpen
  \bibfield  {author} {\bibinfo {author} {\bibfnamefont {P.}~\bibnamefont
  {Fileviez~Perez}}, \bibinfo {author} {\bibfnamefont {T.}~\bibnamefont {Han}},
  \bibinfo {author} {\bibfnamefont {G.-y.}\ \bibnamefont {Huang}}, \bibinfo
  {author} {\bibfnamefont {T.}~\bibnamefont {Li}},\ and\ \bibinfo {author}
  {\bibfnamefont {K.}~\bibnamefont {Wang}},\ }\bibfield  {title} {\bibinfo
  {title} {{Neutrino Masses and the CERN LHC: Testing Type II Seesaw}},\ }\href
  {https://doi.org/10.1103/PhysRevD.78.015018} {\bibfield  {journal} {\bibinfo
  {journal} {Phys. Rev. D}\ }\textbf {\bibinfo {volume} {78}},\ \bibinfo
  {pages} {015018} (\bibinfo {year} {2008}{\natexlab{a}})},\ \Eprint
  {https://arxiv.org/abs/0805.3536} {arXiv:0805.3536 [hep-ph]} \BibitemShut
  {NoStop}%
\bibitem [{\citenamefont {Fileviez~Perez}\ \emph
  {et~al.}(2008{\natexlab{b}})\citenamefont {Fileviez~Perez}, \citenamefont
  {Han}, \citenamefont {Huang}, \citenamefont {Li},\ and\ \citenamefont
  {Wang}}]{Perez:2008ha}%
  \BibitemOpen
  \bibfield  {author} {\bibinfo {author} {\bibfnamefont {P.}~\bibnamefont
  {Fileviez~Perez}}, \bibinfo {author} {\bibfnamefont {T.}~\bibnamefont {Han}},
  \bibinfo {author} {\bibfnamefont {G.-y.}\ \bibnamefont {Huang}}, \bibinfo
  {author} {\bibfnamefont {T.}~\bibnamefont {Li}},\ and\ \bibinfo {author}
  {\bibfnamefont {K.}~\bibnamefont {Wang}},\ }\bibfield  {title} {\bibinfo
  {title} {{Neutrino Masses and the CERN LHC: Testing Type II Seesaw}},\ }\href
  {https://doi.org/10.1103/PhysRevD.78.015018} {\bibfield  {journal} {\bibinfo
  {journal} {Phys. Rev. D}\ }\textbf {\bibinfo {volume} {78}},\ \bibinfo
  {pages} {015018} (\bibinfo {year} {2008}{\natexlab{b}})},\ \Eprint
  {https://arxiv.org/abs/0805.3536} {arXiv:0805.3536 [hep-ph]} \BibitemShut
  {NoStop}%
\bibitem [{\citenamefont {Akeroyd}\ and\ \citenamefont
  {Chiang}(2009)}]{Akeroyd:2009hb}%
  \BibitemOpen
  \bibfield  {author} {\bibinfo {author} {\bibfnamefont {A.~G.}\ \bibnamefont
  {Akeroyd}}\ and\ \bibinfo {author} {\bibfnamefont {C.-W.}\ \bibnamefont
  {Chiang}},\ }\bibfield  {title} {\bibinfo {title} {{Doubly charged Higgs
  bosons and three-lepton signatures in the Higgs Triplet Model}},\ }\href
  {https://doi.org/10.1103/PhysRevD.80.113010} {\bibfield  {journal} {\bibinfo
  {journal} {Phys. Rev. D}\ }\textbf {\bibinfo {volume} {80}},\ \bibinfo
  {pages} {113010} (\bibinfo {year} {2009})},\ \Eprint
  {https://arxiv.org/abs/0909.4419} {arXiv:0909.4419 [hep-ph]} \BibitemShut
  {NoStop}%
\bibitem [{\citenamefont {Akeroyd}\ \emph {et~al.}(2010)\citenamefont
  {Akeroyd}, \citenamefont {Chiang},\ and\ \citenamefont
  {Gaur}}]{Akeroyd:2010ip}%
  \BibitemOpen
  \bibfield  {author} {\bibinfo {author} {\bibfnamefont {A.~G.}\ \bibnamefont
  {Akeroyd}}, \bibinfo {author} {\bibfnamefont {C.-W.}\ \bibnamefont
  {Chiang}},\ and\ \bibinfo {author} {\bibfnamefont {N.}~\bibnamefont {Gaur}},\
  }\bibfield  {title} {\bibinfo {title} {{Leptonic signatures of doubly charged
  Higgs boson production at the LHC}},\ }\href
  {https://doi.org/10.1007/JHEP11(2010)005} {\bibfield  {journal} {\bibinfo
  {journal} {JHEP}\ }\textbf {\bibinfo {volume} {11}},\ \bibinfo {pages}
  {005}},\ \Eprint {https://arxiv.org/abs/1009.2780} {arXiv:1009.2780 [hep-ph]}
  \BibitemShut {NoStop}%
\bibitem [{\citenamefont {Melfo}\ \emph {et~al.}(2012)\citenamefont {Melfo},
  \citenamefont {Nemevsek}, \citenamefont {Nesti}, \citenamefont {Senjanovic},\
  and\ \citenamefont {Zhang}}]{Melfo:2011nx}%
  \BibitemOpen
  \bibfield  {author} {\bibinfo {author} {\bibfnamefont {A.}~\bibnamefont
  {Melfo}}, \bibinfo {author} {\bibfnamefont {M.}~\bibnamefont {Nemevsek}},
  \bibinfo {author} {\bibfnamefont {F.}~\bibnamefont {Nesti}}, \bibinfo
  {author} {\bibfnamefont {G.}~\bibnamefont {Senjanovic}},\ and\ \bibinfo
  {author} {\bibfnamefont {Y.}~\bibnamefont {Zhang}},\ }\bibfield  {title}
  {\bibinfo {title} {{Type II Seesaw at LHC: The Roadmap}},\ }\href
  {https://doi.org/10.1103/PhysRevD.85.055018} {\bibfield  {journal} {\bibinfo
  {journal} {Phys. Rev. D}\ }\textbf {\bibinfo {volume} {85}},\ \bibinfo
  {pages} {055018} (\bibinfo {year} {2012})},\ \Eprint
  {https://arxiv.org/abs/1108.4416} {arXiv:1108.4416 [hep-ph]} \BibitemShut
  {NoStop}%
\bibitem [{\citenamefont {Aoki}\ \emph {et~al.}(2012)\citenamefont {Aoki},
  \citenamefont {Kanemura},\ and\ \citenamefont {Yagyu}}]{Aoki:2011pz}%
  \BibitemOpen
  \bibfield  {author} {\bibinfo {author} {\bibfnamefont {M.}~\bibnamefont
  {Aoki}}, \bibinfo {author} {\bibfnamefont {S.}~\bibnamefont {Kanemura}},\
  and\ \bibinfo {author} {\bibfnamefont {K.}~\bibnamefont {Yagyu}},\ }\bibfield
   {title} {\bibinfo {title} {{Testing the Higgs triplet model with the mass
  difference at the LHC}},\ }\href {https://doi.org/10.1103/PhysRevD.85.055007}
  {\bibfield  {journal} {\bibinfo  {journal} {Phys. Rev. D}\ }\textbf {\bibinfo
  {volume} {85}},\ \bibinfo {pages} {055007} (\bibinfo {year} {2012})},\
  \Eprint {https://arxiv.org/abs/1110.4625} {arXiv:1110.4625 [hep-ph]}
  \BibitemShut {NoStop}%
\bibitem [{\citenamefont {Akeroyd}\ and\ \citenamefont
  {Sugiyama}(2011)}]{Akeroyd:2011zza}%
  \BibitemOpen
  \bibfield  {author} {\bibinfo {author} {\bibfnamefont {A.~G.}\ \bibnamefont
  {Akeroyd}}\ and\ \bibinfo {author} {\bibfnamefont {H.}~\bibnamefont
  {Sugiyama}},\ }\bibfield  {title} {\bibinfo {title} {{Production of doubly
  charged scalars from the decay of singly charged scalars in the Higgs Triplet
  Model}},\ }\href {https://doi.org/10.1103/PhysRevD.84.035010} {\bibfield
  {journal} {\bibinfo  {journal} {Phys. Rev. D}\ }\textbf {\bibinfo {volume}
  {84}},\ \bibinfo {pages} {035010} (\bibinfo {year} {2011})},\ \Eprint
  {https://arxiv.org/abs/1105.2209} {arXiv:1105.2209 [hep-ph]} \BibitemShut
  {NoStop}%
\bibitem [{\citenamefont {Arbabifar}\ \emph {et~al.}(2013)\citenamefont
  {Arbabifar}, \citenamefont {Bahrami},\ and\ \citenamefont
  {Frank}}]{Arbabifar:2012bd}%
  \BibitemOpen
  \bibfield  {author} {\bibinfo {author} {\bibfnamefont {F.}~\bibnamefont
  {Arbabifar}}, \bibinfo {author} {\bibfnamefont {S.}~\bibnamefont {Bahrami}},\
  and\ \bibinfo {author} {\bibfnamefont {M.}~\bibnamefont {Frank}},\ }\bibfield
   {title} {\bibinfo {title} {{Neutral Higgs Bosons in the Higgs Triplet Model
  with nontrivial mixing}},\ }\href
  {https://doi.org/10.1103/PhysRevD.87.015020} {\bibfield  {journal} {\bibinfo
  {journal} {Phys. Rev. D}\ }\textbf {\bibinfo {volume} {87}},\ \bibinfo
  {pages} {015020} (\bibinfo {year} {2013})},\ \Eprint
  {https://arxiv.org/abs/1211.6797} {arXiv:1211.6797 [hep-ph]} \BibitemShut
  {NoStop}%
\bibitem [{\citenamefont {Chiang}\ \emph {et~al.}(2012)\citenamefont {Chiang},
  \citenamefont {Nomura},\ and\ \citenamefont {Tsumura}}]{Chiang:2012dk}%
  \BibitemOpen
  \bibfield  {author} {\bibinfo {author} {\bibfnamefont {C.-W.}\ \bibnamefont
  {Chiang}}, \bibinfo {author} {\bibfnamefont {T.}~\bibnamefont {Nomura}},\
  and\ \bibinfo {author} {\bibfnamefont {K.}~\bibnamefont {Tsumura}},\
  }\bibfield  {title} {\bibinfo {title} {{Search for doubly charged Higgs
  bosons using the same-sign diboson mode at the LHC}},\ }\href
  {https://doi.org/10.1103/PhysRevD.85.095023} {\bibfield  {journal} {\bibinfo
  {journal} {Phys. Rev. D}\ }\textbf {\bibinfo {volume} {85}},\ \bibinfo
  {pages} {095023} (\bibinfo {year} {2012})},\ \Eprint
  {https://arxiv.org/abs/1202.2014} {arXiv:1202.2014 [hep-ph]} \BibitemShut
  {NoStop}%
\bibitem [{\citenamefont {Akeroyd}\ \emph {et~al.}(2012)\citenamefont
  {Akeroyd}, \citenamefont {Moretti},\ and\ \citenamefont
  {Sugiyama}}]{Akeroyd:2012nd}%
  \BibitemOpen
  \bibfield  {author} {\bibinfo {author} {\bibfnamefont {A.~G.}\ \bibnamefont
  {Akeroyd}}, \bibinfo {author} {\bibfnamefont {S.}~\bibnamefont {Moretti}},\
  and\ \bibinfo {author} {\bibfnamefont {H.}~\bibnamefont {Sugiyama}},\
  }\bibfield  {title} {\bibinfo {title} {{Five-lepton and six-lepton signatures
  from production of neutral triplet scalars in the Higgs Triplet Model}},\
  }\href {https://doi.org/10.1103/PhysRevD.85.055026} {\bibfield  {journal}
  {\bibinfo  {journal} {Phys. Rev. D}\ }\textbf {\bibinfo {volume} {85}},\
  \bibinfo {pages} {055026} (\bibinfo {year} {2012})},\ \Eprint
  {https://arxiv.org/abs/1201.5047} {arXiv:1201.5047 [hep-ph]} \BibitemShut
  {NoStop}%
\bibitem [{\citenamefont {Chun}\ and\ \citenamefont
  {Sharma}(2012)}]{Chun:2012zu}%
  \BibitemOpen
  \bibfield  {author} {\bibinfo {author} {\bibfnamefont {E.~J.}\ \bibnamefont
  {Chun}}\ and\ \bibinfo {author} {\bibfnamefont {P.}~\bibnamefont {Sharma}},\
  }\bibfield  {title} {\bibinfo {title} {{Same-Sign Tetra-Leptons from Type II
  Seesaw}},\ }\href {https://doi.org/10.1007/JHEP08(2012)162} {\bibfield
  {journal} {\bibinfo  {journal} {JHEP}\ }\textbf {\bibinfo {volume} {08}},\
  \bibinfo {pages} {162}},\ \Eprint {https://arxiv.org/abs/1206.6278}
  {arXiv:1206.6278 [hep-ph]} \BibitemShut {NoStop}%
\bibitem [{\citenamefont {del \'Aguila}\ and\ \citenamefont
  {Chala}(2014)}]{delAguila:2013mia}%
  \BibitemOpen
  \bibfield  {author} {\bibinfo {author} {\bibfnamefont {F.}~\bibnamefont {del
  \'Aguila}}\ and\ \bibinfo {author} {\bibfnamefont {M.}~\bibnamefont
  {Chala}},\ }\bibfield  {title} {\bibinfo {title} {{LHC bounds on Lepton
  Number Violation mediated by doubly and singly-charged scalars}},\ }\href
  {https://doi.org/10.1007/JHEP03(2014)027} {\bibfield  {journal} {\bibinfo
  {journal} {JHEP}\ }\textbf {\bibinfo {volume} {03}},\ \bibinfo {pages}
  {027}},\ \Eprint {https://arxiv.org/abs/1311.1510} {arXiv:1311.1510 [hep-ph]}
  \BibitemShut {NoStop}%
\bibitem [{\citenamefont {Chun}\ and\ \citenamefont
  {Sharma}(2014)}]{Chun:2013vma}%
  \BibitemOpen
  \bibfield  {author} {\bibinfo {author} {\bibfnamefont {E.~J.}\ \bibnamefont
  {Chun}}\ and\ \bibinfo {author} {\bibfnamefont {P.}~\bibnamefont {Sharma}},\
  }\bibfield  {title} {\bibinfo {title} {{Search for a doubly-charged boson in
  four lepton final states in type II seesaw}},\ }\href
  {https://doi.org/10.1016/j.physletb.2013.11.056} {\bibfield  {journal}
  {\bibinfo  {journal} {Phys. Lett. B}\ }\textbf {\bibinfo {volume} {728}},\
  \bibinfo {pages} {256} (\bibinfo {year} {2014})},\ \Eprint
  {https://arxiv.org/abs/1309.6888} {arXiv:1309.6888 [hep-ph]} \BibitemShut
  {NoStop}%
\bibitem [{\citenamefont {Kanemura}\ \emph {et~al.}(2013)\citenamefont
  {Kanemura}, \citenamefont {Yagyu},\ and\ \citenamefont
  {Yokoya}}]{Kanemura:2013vxa}%
  \BibitemOpen
  \bibfield  {author} {\bibinfo {author} {\bibfnamefont {S.}~\bibnamefont
  {Kanemura}}, \bibinfo {author} {\bibfnamefont {K.}~\bibnamefont {Yagyu}},\
  and\ \bibinfo {author} {\bibfnamefont {H.}~\bibnamefont {Yokoya}},\
  }\bibfield  {title} {\bibinfo {title} {{First constraint on the mass of
  doubly-charged Higgs bosons in the same-sign diboson decay scenario at the
  LHC}},\ }\href {https://doi.org/10.1016/j.physletb.2013.08.054} {\bibfield
  {journal} {\bibinfo  {journal} {Phys. Lett. B}\ }\textbf {\bibinfo {volume}
  {726}},\ \bibinfo {pages} {316} (\bibinfo {year} {2013})},\ \Eprint
  {https://arxiv.org/abs/1305.2383} {arXiv:1305.2383 [hep-ph]} \BibitemShut
  {NoStop}%
\bibitem [{\citenamefont {Bhupal~Dev}\ \emph {et~al.}(2013)\citenamefont
  {Bhupal~Dev}, \citenamefont {Ghosh}, \citenamefont {Okada},\ and\
  \citenamefont {Saha}}]{Dev:2013ff}%
  \BibitemOpen
  \bibfield  {author} {\bibinfo {author} {\bibfnamefont {P.~S.}\ \bibnamefont
  {Bhupal~Dev}}, \bibinfo {author} {\bibfnamefont {D.~K.}\ \bibnamefont
  {Ghosh}}, \bibinfo {author} {\bibfnamefont {N.}~\bibnamefont {Okada}},\ and\
  \bibinfo {author} {\bibfnamefont {I.}~\bibnamefont {Saha}},\ }\bibfield
  {title} {\bibinfo {title} {{125 GeV Higgs Boson and the Type-II Seesaw
  Model}},\ }\href {https://doi.org/10.1007/JHEP03(2013)150} {\bibfield
  {journal} {\bibinfo  {journal} {JHEP}\ }\textbf {\bibinfo {volume} {03}},\
  \bibinfo {pages} {150}},\ \bibinfo {note} {[Erratum: JHEP 05, 049 (2013)]},\
  \Eprint {https://arxiv.org/abs/1301.3453} {arXiv:1301.3453 [hep-ph]}
  \BibitemShut {NoStop}%
\bibitem [{\citenamefont {Kanemura}\ \emph {et~al.}(2014)\citenamefont
  {Kanemura}, \citenamefont {Kikuchi}, \citenamefont {Yagyu},\ and\
  \citenamefont {Yokoya}}]{Kanemura:2014goa}%
  \BibitemOpen
  \bibfield  {author} {\bibinfo {author} {\bibfnamefont {S.}~\bibnamefont
  {Kanemura}}, \bibinfo {author} {\bibfnamefont {M.}~\bibnamefont {Kikuchi}},
  \bibinfo {author} {\bibfnamefont {K.}~\bibnamefont {Yagyu}},\ and\ \bibinfo
  {author} {\bibfnamefont {H.}~\bibnamefont {Yokoya}},\ }\bibfield  {title}
  {\bibinfo {title} {{Bounds on the mass of doubly-charged Higgs bosons in the
  same-sign diboson decay scenario}},\ }\href
  {https://doi.org/10.1103/PhysRevD.90.115018} {\bibfield  {journal} {\bibinfo
  {journal} {Phys. Rev. D}\ }\textbf {\bibinfo {volume} {90}},\ \bibinfo
  {pages} {115018} (\bibinfo {year} {2014})},\ \Eprint
  {https://arxiv.org/abs/1407.6547} {arXiv:1407.6547 [hep-ph]} \BibitemShut
  {NoStop}%
\bibitem [{\citenamefont {Kanemura}\ \emph {et~al.}(2015)\citenamefont
  {Kanemura}, \citenamefont {Kikuchi}, \citenamefont {Yokoya},\ and\
  \citenamefont {Yagyu}}]{Kanemura:2014ipa}%
  \BibitemOpen
  \bibfield  {author} {\bibinfo {author} {\bibfnamefont {S.}~\bibnamefont
  {Kanemura}}, \bibinfo {author} {\bibfnamefont {M.}~\bibnamefont {Kikuchi}},
  \bibinfo {author} {\bibfnamefont {H.}~\bibnamefont {Yokoya}},\ and\ \bibinfo
  {author} {\bibfnamefont {K.}~\bibnamefont {Yagyu}},\ }\bibfield  {title}
  {\bibinfo {title} {{LHC Run-I constraint on the mass of doubly charged Higgs
  bosons in the same-sign diboson decay scenario}},\ }\href
  {https://doi.org/10.1093/ptep/ptv071} {\bibfield  {journal} {\bibinfo
  {journal} {PTEP}\ }\textbf {\bibinfo {volume} {2015}},\ \bibinfo {pages}
  {051B02} (\bibinfo {year} {2015})},\ \Eprint
  {https://arxiv.org/abs/1412.7603} {arXiv:1412.7603 [hep-ph]} \BibitemShut
  {NoStop}%
\bibitem [{\citenamefont {Kang}\ \emph {et~al.}(2015)\citenamefont {Kang},
  \citenamefont {Li}, \citenamefont {Li}, \citenamefont {Liu},\ and\
  \citenamefont {Ning}}]{kang:2014jia}%
  \BibitemOpen
  \bibfield  {author} {\bibinfo {author} {\bibfnamefont {Z.}~\bibnamefont
  {Kang}}, \bibinfo {author} {\bibfnamefont {J.}~\bibnamefont {Li}}, \bibinfo
  {author} {\bibfnamefont {T.}~\bibnamefont {Li}}, \bibinfo {author}
  {\bibfnamefont {Y.}~\bibnamefont {Liu}},\ and\ \bibinfo {author}
  {\bibfnamefont {G.-Z.}\ \bibnamefont {Ning}},\ }\bibfield  {title} {\bibinfo
  {title} {{Light Doubly Charged Higgs Boson via the $WW^*$ Channel at LHC}},\
  }\href {https://doi.org/10.1140/epjc/s10052-015-3774-1} {\bibfield  {journal}
  {\bibinfo  {journal} {Eur. Phys. J. C}\ }\textbf {\bibinfo {volume} {75}},\
  \bibinfo {pages} {574} (\bibinfo {year} {2015})},\ \Eprint
  {https://arxiv.org/abs/1404.5207} {arXiv:1404.5207 [hep-ph]} \BibitemShut
  {NoStop}%
\bibitem [{\citenamefont {Han}\ \emph {et~al.}(2015{\natexlab{a}})\citenamefont
  {Han}, \citenamefont {Ding},\ and\ \citenamefont {Liao}}]{Han:2015hba}%
  \BibitemOpen
  \bibfield  {author} {\bibinfo {author} {\bibfnamefont {Z.-L.}\ \bibnamefont
  {Han}}, \bibinfo {author} {\bibfnamefont {R.}~\bibnamefont {Ding}},\ and\
  \bibinfo {author} {\bibfnamefont {Y.}~\bibnamefont {Liao}},\ }\bibfield
  {title} {\bibinfo {title} {{LHC Phenomenology of Type II Seesaw:
  Nondegenerate Case}},\ }\href {https://doi.org/10.1103/PhysRevD.91.093006}
  {\bibfield  {journal} {\bibinfo  {journal} {Phys. Rev. D}\ }\textbf {\bibinfo
  {volume} {91}},\ \bibinfo {pages} {093006} (\bibinfo {year}
  {2015}{\natexlab{a}})},\ \Eprint {https://arxiv.org/abs/1502.05242}
  {arXiv:1502.05242 [hep-ph]} \BibitemShut {NoStop}%
\bibitem [{\citenamefont {Han}\ \emph {et~al.}(2015{\natexlab{b}})\citenamefont
  {Han}, \citenamefont {Ding},\ and\ \citenamefont {Liao}}]{Han:2015sca}%
  \BibitemOpen
  \bibfield  {author} {\bibinfo {author} {\bibfnamefont {Z.-L.}\ \bibnamefont
  {Han}}, \bibinfo {author} {\bibfnamefont {R.}~\bibnamefont {Ding}},\ and\
  \bibinfo {author} {\bibfnamefont {Y.}~\bibnamefont {Liao}},\ }\bibfield
  {title} {\bibinfo {title} {{LHC phenomenology of the type II seesaw
  mechanism: Observability of neutral scalars in the nondegenerate case}},\
  }\href {https://doi.org/10.1103/PhysRevD.92.033014} {\bibfield  {journal}
  {\bibinfo  {journal} {Phys. Rev. D}\ }\textbf {\bibinfo {volume} {92}},\
  \bibinfo {pages} {033014} (\bibinfo {year} {2015}{\natexlab{b}})},\ \Eprint
  {https://arxiv.org/abs/1506.08996} {arXiv:1506.08996 [hep-ph]} \BibitemShut
  {NoStop}%
\bibitem [{\citenamefont {Mitra}\ \emph {et~al.}(2017)\citenamefont {Mitra},
  \citenamefont {Niyogi},\ and\ \citenamefont {Spannowsky}}]{Mitra:2016wpr}%
  \BibitemOpen
  \bibfield  {author} {\bibinfo {author} {\bibfnamefont {M.}~\bibnamefont
  {Mitra}}, \bibinfo {author} {\bibfnamefont {S.}~\bibnamefont {Niyogi}},\ and\
  \bibinfo {author} {\bibfnamefont {M.}~\bibnamefont {Spannowsky}},\ }\bibfield
   {title} {\bibinfo {title} {{Type-II Seesaw Model and Multilepton Signatures
  at Hadron Colliders}},\ }\href {https://doi.org/10.1103/PhysRevD.95.035042}
  {\bibfield  {journal} {\bibinfo  {journal} {Phys. Rev. D}\ }\textbf {\bibinfo
  {volume} {95}},\ \bibinfo {pages} {035042} (\bibinfo {year} {2017})},\
  \Eprint {https://arxiv.org/abs/1611.09594} {arXiv:1611.09594 [hep-ph]}
  \BibitemShut {NoStop}%
\bibitem [{\citenamefont {Ghosh}\ \emph {et~al.}(2018)\citenamefont {Ghosh},
  \citenamefont {Ghosh}, \citenamefont {Saha},\ and\ \citenamefont
  {Shaw}}]{Ghosh:2017pxl}%
  \BibitemOpen
  \bibfield  {author} {\bibinfo {author} {\bibfnamefont {D.~K.}\ \bibnamefont
  {Ghosh}}, \bibinfo {author} {\bibfnamefont {N.}~\bibnamefont {Ghosh}},
  \bibinfo {author} {\bibfnamefont {I.}~\bibnamefont {Saha}},\ and\ \bibinfo
  {author} {\bibfnamefont {A.}~\bibnamefont {Shaw}},\ }\bibfield  {title}
  {\bibinfo {title} {{Revisiting the high-scale validity of the type II seesaw
  model with novel LHC signature}},\ }\href
  {https://doi.org/10.1103/PhysRevD.97.115022} {\bibfield  {journal} {\bibinfo
  {journal} {Phys. Rev. D}\ }\textbf {\bibinfo {volume} {97}},\ \bibinfo
  {pages} {115022} (\bibinfo {year} {2018})},\ \Eprint
  {https://arxiv.org/abs/1711.06062} {arXiv:1711.06062 [hep-ph]} \BibitemShut
  {NoStop}%
\bibitem [{\citenamefont {Antusch}\ \emph {et~al.}(2019)\citenamefont
  {Antusch}, \citenamefont {Fischer}, \citenamefont {Hammad},\ and\
  \citenamefont {Scherb}}]{Antusch:2018svb}%
  \BibitemOpen
  \bibfield  {author} {\bibinfo {author} {\bibfnamefont {S.}~\bibnamefont
  {Antusch}}, \bibinfo {author} {\bibfnamefont {O.}~\bibnamefont {Fischer}},
  \bibinfo {author} {\bibfnamefont {A.}~\bibnamefont {Hammad}},\ and\ \bibinfo
  {author} {\bibfnamefont {C.}~\bibnamefont {Scherb}},\ }\bibfield  {title}
  {\bibinfo {title} {{Low scale type II seesaw: Present constraints and
  prospects for displaced vertex searches}},\ }\href
  {https://doi.org/10.1007/JHEP02(2019)157} {\bibfield  {journal} {\bibinfo
  {journal} {JHEP}\ }\textbf {\bibinfo {volume} {02}},\ \bibinfo {pages}
  {157}},\ \Eprint {https://arxiv.org/abs/1811.03476} {arXiv:1811.03476
  [hep-ph]} \BibitemShut {NoStop}%
\bibitem [{\citenamefont {Bhupal~Dev}\ and\ \citenamefont
  {Zhang}(2018)}]{BhupalDev:2018tox}%
  \BibitemOpen
  \bibfield  {author} {\bibinfo {author} {\bibfnamefont {P.~S.}\ \bibnamefont
  {Bhupal~Dev}}\ and\ \bibinfo {author} {\bibfnamefont {Y.}~\bibnamefont
  {Zhang}},\ }\bibfield  {title} {\bibinfo {title} {{Displaced vertex
  signatures of doubly charged scalars in the type-II seesaw and its left-right
  extensions}},\ }\href {https://doi.org/10.1007/JHEP10(2018)199} {\bibfield
  {journal} {\bibinfo  {journal} {JHEP}\ }\textbf {\bibinfo {volume} {10}},\
  \bibinfo {pages} {199}},\ \Eprint {https://arxiv.org/abs/1808.00943}
  {arXiv:1808.00943 [hep-ph]} \BibitemShut {NoStop}%
\bibitem [{\citenamefont {de~Melo}\ \emph {et~al.}(2019)\citenamefont
  {de~Melo}, \citenamefont {Queiroz},\ and\ \citenamefont
  {Villamizar}}]{deMelo:2019asm}%
  \BibitemOpen
  \bibfield  {author} {\bibinfo {author} {\bibfnamefont {T.~B.}\ \bibnamefont
  {de~Melo}}, \bibinfo {author} {\bibfnamefont {F.~S.}\ \bibnamefont
  {Queiroz}},\ and\ \bibinfo {author} {\bibfnamefont {Y.}~\bibnamefont
  {Villamizar}},\ }\bibfield  {title} {\bibinfo {title} {{Doubly Charged Scalar
  at the High-Luminosity and High-Energy LHC}},\ }\href
  {https://doi.org/10.1142/S0217751X19501574} {\bibfield  {journal} {\bibinfo
  {journal} {Int. J. Mod. Phys. A}\ }\textbf {\bibinfo {volume} {34}},\
  \bibinfo {pages} {1950157} (\bibinfo {year} {2019})},\ \Eprint
  {https://arxiv.org/abs/1909.07429} {arXiv:1909.07429 [hep-ph]} \BibitemShut
  {NoStop}%
\bibitem [{\citenamefont {Primulando}\ \emph {et~al.}(2019)\citenamefont
  {Primulando}, \citenamefont {Julio},\ and\ \citenamefont
  {Uttayarat}}]{Primulando:2019evb}%
  \BibitemOpen
  \bibfield  {author} {\bibinfo {author} {\bibfnamefont {R.}~\bibnamefont
  {Primulando}}, \bibinfo {author} {\bibfnamefont {J.}~\bibnamefont {Julio}},\
  and\ \bibinfo {author} {\bibfnamefont {P.}~\bibnamefont {Uttayarat}},\
  }\bibfield  {title} {\bibinfo {title} {{Scalar phenomenology in type-II
  seesaw model}},\ }\href {https://doi.org/10.1007/JHEP08(2019)024} {\bibfield
  {journal} {\bibinfo  {journal} {JHEP}\ }\textbf {\bibinfo {volume} {08}},\
  \bibinfo {pages} {024}},\ \Eprint {https://arxiv.org/abs/1903.02493}
  {arXiv:1903.02493 [hep-ph]} \BibitemShut {NoStop}%
\bibitem [{\citenamefont {Chun}\ \emph {et~al.}(2020)\citenamefont {Chun},
  \citenamefont {Khan}, \citenamefont {Mandal}, \citenamefont {Mitra},\ and\
  \citenamefont {Shil}}]{Chun:2019hce}%
  \BibitemOpen
  \bibfield  {author} {\bibinfo {author} {\bibfnamefont {E.~J.}\ \bibnamefont
  {Chun}}, \bibinfo {author} {\bibfnamefont {S.}~\bibnamefont {Khan}}, \bibinfo
  {author} {\bibfnamefont {S.}~\bibnamefont {Mandal}}, \bibinfo {author}
  {\bibfnamefont {M.}~\bibnamefont {Mitra}},\ and\ \bibinfo {author}
  {\bibfnamefont {S.}~\bibnamefont {Shil}},\ }\bibfield  {title} {\bibinfo
  {title} {{Same-sign tetralepton signature at the Large Hadron Collider and a
  future $pp$ collider}},\ }\href {https://doi.org/10.1103/PhysRevD.101.075008}
  {\bibfield  {journal} {\bibinfo  {journal} {Phys. Rev. D}\ }\textbf {\bibinfo
  {volume} {101}},\ \bibinfo {pages} {075008} (\bibinfo {year} {2020})},\
  \Eprint {https://arxiv.org/abs/1911.00971} {arXiv:1911.00971 [hep-ph]}
  \BibitemShut {NoStop}%
\bibitem [{\citenamefont {Padhan}\ \emph {et~al.}(2020)\citenamefont {Padhan},
  \citenamefont {Das}, \citenamefont {Mitra},\ and\ \citenamefont
  {Kumar~Nayak}}]{Padhan:2019jlc}%
  \BibitemOpen
  \bibfield  {author} {\bibinfo {author} {\bibfnamefont {R.}~\bibnamefont
  {Padhan}}, \bibinfo {author} {\bibfnamefont {D.}~\bibnamefont {Das}},
  \bibinfo {author} {\bibfnamefont {M.}~\bibnamefont {Mitra}},\ and\ \bibinfo
  {author} {\bibfnamefont {A.}~\bibnamefont {Kumar~Nayak}},\ }\bibfield
  {title} {\bibinfo {title} {{Probing doubly and singly charged Higgs bosons at
  the $pp$ collider HE-LHC}},\ }\href
  {https://doi.org/10.1103/PhysRevD.101.075050} {\bibfield  {journal} {\bibinfo
   {journal} {Phys. Rev. D}\ }\textbf {\bibinfo {volume} {101}},\ \bibinfo
  {pages} {075050} (\bibinfo {year} {2020})},\ \Eprint
  {https://arxiv.org/abs/1909.10495} {arXiv:1909.10495 [hep-ph]} \BibitemShut
  {NoStop}%
\bibitem [{\citenamefont {Gluza}\ \emph {et~al.}(2021)\citenamefont {Gluza},
  \citenamefont {Kordiaczynska},\ and\ \citenamefont
  {Srivastava}}]{Gluza:2020qrt}%
  \BibitemOpen
  \bibfield  {author} {\bibinfo {author} {\bibfnamefont {J.}~\bibnamefont
  {Gluza}}, \bibinfo {author} {\bibfnamefont {M.}~\bibnamefont
  {Kordiaczynska}},\ and\ \bibinfo {author} {\bibfnamefont {T.}~\bibnamefont
  {Srivastava}},\ }\bibfield  {title} {\bibinfo {title} {{Discriminating the
  HTM and MLRSM models in collider studies via doubly charged Higgs boson pair
  production and the subsequent leptonic decays}},\ }\href
  {https://doi.org/10.1088/1674-1137/abfe51} {\bibfield  {journal} {\bibinfo
  {journal} {Chin. Phys. C}\ }\textbf {\bibinfo {volume} {45}},\ \bibinfo
  {pages} {073113} (\bibinfo {year} {2021})},\ \Eprint
  {https://arxiv.org/abs/2006.04610} {arXiv:2006.04610 [hep-ph]} \BibitemShut
  {NoStop}%
\bibitem [{\citenamefont {Ashanujjaman}\ and\ \citenamefont
  {Ghosh}(2022)}]{Ashanujjaman:2021txz}%
  \BibitemOpen
  \bibfield  {author} {\bibinfo {author} {\bibfnamefont {S.}~\bibnamefont
  {Ashanujjaman}}\ and\ \bibinfo {author} {\bibfnamefont {K.}~\bibnamefont
  {Ghosh}},\ }\bibfield  {title} {\bibinfo {title} {{Revisiting type-II
  see-saw: present limits and future prospects at LHC}},\ }\href
  {https://doi.org/10.1007/JHEP03(2022)195} {\bibfield  {journal} {\bibinfo
  {journal} {JHEP}\ }\textbf {\bibinfo {volume} {03}},\ \bibinfo {pages}
  {195}},\ \Eprint {https://arxiv.org/abs/2108.10952} {arXiv:2108.10952
  [hep-ph]} \BibitemShut {NoStop}%
\bibitem [{\citenamefont {Ashanujjaman}\ \emph {et~al.}(2023)\citenamefont
  {Ashanujjaman}, \citenamefont {Ghosh},\ and\ \citenamefont
  {Sahu}}]{Ashanujjaman:2022ofg}%
  \BibitemOpen
  \bibfield  {author} {\bibinfo {author} {\bibfnamefont {S.}~\bibnamefont
  {Ashanujjaman}}, \bibinfo {author} {\bibfnamefont {K.}~\bibnamefont
  {Ghosh}},\ and\ \bibinfo {author} {\bibfnamefont {R.}~\bibnamefont {Sahu}},\
  }\bibfield  {title} {\bibinfo {title} {{Low-mass doubly charged Higgs bosons
  at the LHC}},\ }\href {https://doi.org/10.1103/PhysRevD.107.015018}
  {\bibfield  {journal} {\bibinfo  {journal} {Phys. Rev. D}\ }\textbf {\bibinfo
  {volume} {107}},\ \bibinfo {pages} {015018} (\bibinfo {year} {2023})},\
  \Eprint {https://arxiv.org/abs/2211.00632} {arXiv:2211.00632 [hep-ph]}
  \BibitemShut {NoStop}%
\bibitem [{\citenamefont {Rodejohann}\ and\ \citenamefont
  {Zhang}(2011)}]{Rodejohann:2010bv}%
  \BibitemOpen
  \bibfield  {author} {\bibinfo {author} {\bibfnamefont {W.}~\bibnamefont
  {Rodejohann}}\ and\ \bibinfo {author} {\bibfnamefont {H.}~\bibnamefont
  {Zhang}},\ }\bibfield  {title} {\bibinfo {title} {{Higgs triplets at
  like-sign linear colliders and neutrino mixing}},\ }\href
  {https://doi.org/10.1103/PhysRevD.83.073005} {\bibfield  {journal} {\bibinfo
  {journal} {Phys. Rev. D}\ }\textbf {\bibinfo {volume} {83}},\ \bibinfo
  {pages} {073005} (\bibinfo {year} {2011})},\ \Eprint
  {https://arxiv.org/abs/1011.3606} {arXiv:1011.3606 [hep-ph]} \BibitemShut
  {NoStop}%
\bibitem [{\citenamefont {Blunier}\ \emph {et~al.}(2017)\citenamefont
  {Blunier}, \citenamefont {Cottin}, \citenamefont {D\'\i{}az},\ and\
  \citenamefont {Koch}}]{Blunier:2016peh}%
  \BibitemOpen
  \bibfield  {author} {\bibinfo {author} {\bibfnamefont {S.}~\bibnamefont
  {Blunier}}, \bibinfo {author} {\bibfnamefont {G.}~\bibnamefont {Cottin}},
  \bibinfo {author} {\bibfnamefont {M.~A.}\ \bibnamefont {D\'\i{}az}},\ and\
  \bibinfo {author} {\bibfnamefont {B.}~\bibnamefont {Koch}},\ }\bibfield
  {title} {\bibinfo {title} {{Phenomenology of a Higgs triplet model at future
  $e^{+}e^{-}$ colliders}},\ }\href
  {https://doi.org/10.1103/PhysRevD.95.075038} {\bibfield  {journal} {\bibinfo
  {journal} {Phys. Rev. D}\ }\textbf {\bibinfo {volume} {95}},\ \bibinfo
  {pages} {075038} (\bibinfo {year} {2017})},\ \Eprint
  {https://arxiv.org/abs/1611.07896} {arXiv:1611.07896 [hep-ph]} \BibitemShut
  {NoStop}%
\bibitem [{\citenamefont {Nomura}\ \emph {et~al.}(2018)\citenamefont {Nomura},
  \citenamefont {Okada},\ and\ \citenamefont {Yokoya}}]{Nomura:2017abh}%
  \BibitemOpen
  \bibfield  {author} {\bibinfo {author} {\bibfnamefont {T.}~\bibnamefont
  {Nomura}}, \bibinfo {author} {\bibfnamefont {H.}~\bibnamefont {Okada}},\ and\
  \bibinfo {author} {\bibfnamefont {H.}~\bibnamefont {Yokoya}},\ }\bibfield
  {title} {\bibinfo {title} {{Discriminating leptonic Yukawa interactions with
  doubly charged scalar at the ILC}},\ }\href
  {https://doi.org/10.1016/j.nuclphysb.2018.02.011} {\bibfield  {journal}
  {\bibinfo  {journal} {Nucl. Phys. B}\ }\textbf {\bibinfo {volume} {929}},\
  \bibinfo {pages} {193} (\bibinfo {year} {2018})},\ \Eprint
  {https://arxiv.org/abs/1702.03396} {arXiv:1702.03396 [hep-ph]} \BibitemShut
  {NoStop}%
\bibitem [{\citenamefont {Crivellin}\ \emph {et~al.}(2019)\citenamefont
  {Crivellin}, \citenamefont {Ghezzi}, \citenamefont {Panizzi}, \citenamefont
  {Pruna},\ and\ \citenamefont {Signer}}]{Crivellin:2018ahj}%
  \BibitemOpen
  \bibfield  {author} {\bibinfo {author} {\bibfnamefont {A.}~\bibnamefont
  {Crivellin}}, \bibinfo {author} {\bibfnamefont {M.}~\bibnamefont {Ghezzi}},
  \bibinfo {author} {\bibfnamefont {L.}~\bibnamefont {Panizzi}}, \bibinfo
  {author} {\bibfnamefont {G.~M.}\ \bibnamefont {Pruna}},\ and\ \bibinfo
  {author} {\bibfnamefont {A.}~\bibnamefont {Signer}},\ }\bibfield  {title}
  {\bibinfo {title} {{Low- and high-energy phenomenology of a doubly charged
  scalar}},\ }\href {https://doi.org/10.1103/PhysRevD.99.035004} {\bibfield
  {journal} {\bibinfo  {journal} {Phys. Rev. D}\ }\textbf {\bibinfo {volume}
  {99}},\ \bibinfo {pages} {035004} (\bibinfo {year} {2019})},\ \Eprint
  {https://arxiv.org/abs/1807.10224} {arXiv:1807.10224 [hep-ph]} \BibitemShut
  {NoStop}%
\bibitem [{\citenamefont {Agrawal}\ \emph {et~al.}(2018)\citenamefont
  {Agrawal}, \citenamefont {Mitra}, \citenamefont {Niyogi}, \citenamefont
  {Shil},\ and\ \citenamefont {Spannowsky}}]{Agrawal:2018pci}%
  \BibitemOpen
  \bibfield  {author} {\bibinfo {author} {\bibfnamefont {P.}~\bibnamefont
  {Agrawal}}, \bibinfo {author} {\bibfnamefont {M.}~\bibnamefont {Mitra}},
  \bibinfo {author} {\bibfnamefont {S.}~\bibnamefont {Niyogi}}, \bibinfo
  {author} {\bibfnamefont {S.}~\bibnamefont {Shil}},\ and\ \bibinfo {author}
  {\bibfnamefont {M.}~\bibnamefont {Spannowsky}},\ }\bibfield  {title}
  {\bibinfo {title} {{Probing the Type-II Seesaw Mechanism through the
  Production of Higgs Bosons at a Lepton Collider}},\ }\href
  {https://doi.org/10.1103/PhysRevD.98.015024} {\bibfield  {journal} {\bibinfo
  {journal} {Phys. Rev. D}\ }\textbf {\bibinfo {volume} {98}},\ \bibinfo
  {pages} {015024} (\bibinfo {year} {2018})},\ \Eprint
  {https://arxiv.org/abs/1803.00677} {arXiv:1803.00677 [hep-ph]} \BibitemShut
  {NoStop}%
\bibitem [{\citenamefont {Rahili}\ \emph {et~al.}(2019)\citenamefont {Rahili},
  \citenamefont {Arhrib},\ and\ \citenamefont {Benbrik}}]{Rahili:2019ixf}%
  \BibitemOpen
  \bibfield  {author} {\bibinfo {author} {\bibfnamefont {L.}~\bibnamefont
  {Rahili}}, \bibinfo {author} {\bibfnamefont {A.}~\bibnamefont {Arhrib}},\
  and\ \bibinfo {author} {\bibfnamefont {R.}~\bibnamefont {Benbrik}},\
  }\bibfield  {title} {\bibinfo {title} {{Associated production of SM Higgs
  with a photon in type-II seesaw models at the ILC}},\ }\href
  {https://doi.org/10.1140/epjc/s10052-019-7471-3} {\bibfield  {journal}
  {\bibinfo  {journal} {Eur. Phys. J. C}\ }\textbf {\bibinfo {volume} {79}},\
  \bibinfo {pages} {940} (\bibinfo {year} {2019})},\ \Eprint
  {https://arxiv.org/abs/1909.07793} {arXiv:1909.07793 [hep-ph]} \BibitemShut
  {NoStop}%
\bibitem [{\citenamefont {Bandyopadhyay}\ \emph {et~al.}(2020)\citenamefont
  {Bandyopadhyay}, \citenamefont {Karan},\ and\ \citenamefont
  {Sen}}]{Bandyopadhyay:2020mnp}%
  \BibitemOpen
  \bibfield  {author} {\bibinfo {author} {\bibfnamefont {P.}~\bibnamefont
  {Bandyopadhyay}}, \bibinfo {author} {\bibfnamefont {A.}~\bibnamefont
  {Karan}},\ and\ \bibinfo {author} {\bibfnamefont {C.}~\bibnamefont {Sen}},\
  }\bibfield  {title} {\bibinfo {title} {{Discerning Signatures of Seesaw
  Models and Complementarity of Leptonic Colliders}},\ }\href@noop {} {\
  (\bibinfo {year} {2020})},\ \Eprint {https://arxiv.org/abs/2011.04191}
  {arXiv:2011.04191 [hep-ph]} \BibitemShut {NoStop}%
\bibitem [{\citenamefont {Ashanujjaman}\ \emph {et~al.}(2022)\citenamefont
  {Ashanujjaman}, \citenamefont {Ghosh},\ and\ \citenamefont
  {Huitu}}]{Ashanujjaman:2022tdn}%
  \BibitemOpen
  \bibfield  {author} {\bibinfo {author} {\bibfnamefont {S.}~\bibnamefont
  {Ashanujjaman}}, \bibinfo {author} {\bibfnamefont {K.}~\bibnamefont
  {Ghosh}},\ and\ \bibinfo {author} {\bibfnamefont {K.}~\bibnamefont {Huitu}},\
  }\bibfield  {title} {\bibinfo {title} {{Type-II see-saw: searching the LHC
  elusive low-mass triplet-like Higgses at $e^-e^+$ colliders}},\ }\href
  {https://doi.org/10.1103/PhysRevD.106.075028} {\bibfield  {journal} {\bibinfo
   {journal} {Phys. Rev. D}\ }\textbf {\bibinfo {volume} {106}},\ \bibinfo
  {pages} {075028} (\bibinfo {year} {2022})},\ \Eprint
  {https://arxiv.org/abs/2205.14983} {arXiv:2205.14983 [hep-ph]} \BibitemShut
  {NoStop}%
\bibitem [{\citenamefont {Rodejohann}(2010)}]{Rodejohann:2010jh}%
  \BibitemOpen
  \bibfield  {author} {\bibinfo {author} {\bibfnamefont {W.}~\bibnamefont
  {Rodejohann}},\ }\bibfield  {title} {\bibinfo {title} {{Inverse Neutrino-less
  Double Beta Decay Revisited: Neutrinos, Higgs Triplets and a Muon
  Collider}},\ }\href {https://doi.org/10.1103/PhysRevD.81.114001} {\bibfield
  {journal} {\bibinfo  {journal} {Phys. Rev. D}\ }\textbf {\bibinfo {volume}
  {81}},\ \bibinfo {pages} {114001} (\bibinfo {year} {2010})},\ \Eprint
  {https://arxiv.org/abs/1005.2854} {arXiv:1005.2854 [hep-ph]} \BibitemShut
  {NoStop}%
\bibitem [{\citenamefont {Li}\ \emph {et~al.}(2023)\citenamefont {Li},
  \citenamefont {Yao},\ and\ \citenamefont {Yuan}}]{Li:2023ksw}%
  \BibitemOpen
  \bibfield  {author} {\bibinfo {author} {\bibfnamefont {T.}~\bibnamefont
  {Li}}, \bibinfo {author} {\bibfnamefont {C.-Y.}\ \bibnamefont {Yao}},\ and\
  \bibinfo {author} {\bibfnamefont {M.}~\bibnamefont {Yuan}},\ }\bibfield
  {title} {\bibinfo {title} {{Revealing the origin of neutrino masses through
  the Type II Seesaw mechanism at high-energy muon colliders}},\ }\href
  {https://doi.org/10.1007/JHEP03(2023)137} {\bibfield  {journal} {\bibinfo
  {journal} {JHEP}\ }\textbf {\bibinfo {volume} {03}},\ \bibinfo {pages}
  {137}},\ \Eprint {https://arxiv.org/abs/2301.07274} {arXiv:2301.07274
  [hep-ph]} \BibitemShut {NoStop}%
\bibitem [{\citenamefont {Maharathy}\ and\ \citenamefont
  {Mitra}(2023)}]{Maharathy:2023dtp}%
  \BibitemOpen
  \bibfield  {author} {\bibinfo {author} {\bibfnamefont {S.~P.}\ \bibnamefont
  {Maharathy}}\ and\ \bibinfo {author} {\bibfnamefont {M.}~\bibnamefont
  {Mitra}},\ }\bibfield  {title} {\bibinfo {title} {{Type-II see-saw at
  $\mu^+$$\mu^-$ collider}},\ }\href@noop {} {\  (\bibinfo {year} {2023})},\
  \Eprint {https://arxiv.org/abs/2304.08732} {arXiv:2304.08732 [hep-ph]}
  \BibitemShut {NoStop}%
\bibitem [{\citenamefont {Fridell}\ \emph {et~al.}(2023)\citenamefont
  {Fridell}, \citenamefont {Kitano},\ and\ \citenamefont
  {Takai}}]{Fridell:2023gjx}%
  \BibitemOpen
  \bibfield  {author} {\bibinfo {author} {\bibfnamefont {K.}~\bibnamefont
  {Fridell}}, \bibinfo {author} {\bibfnamefont {R.}~\bibnamefont {Kitano}},\
  and\ \bibinfo {author} {\bibfnamefont {R.}~\bibnamefont {Takai}},\ }\bibfield
   {title} {\bibinfo {title} {{Lepton flavor physics at $\mu^+ \mu^+$
  colliders}},\ }\href@noop {} {\  (\bibinfo {year} {2023})},\ \Eprint
  {https://arxiv.org/abs/2304.14020} {arXiv:2304.14020 [hep-ph]} \BibitemShut
  {NoStop}%
\bibitem [{\citenamefont {Dev}\ \emph {et~al.}(2019)\citenamefont {Dev},
  \citenamefont {Khan}, \citenamefont {Mitra},\ and\ \citenamefont
  {Rai}}]{Dev:2019hev}%
  \BibitemOpen
  \bibfield  {author} {\bibinfo {author} {\bibfnamefont {P.~S.~B.}\
  \bibnamefont {Dev}}, \bibinfo {author} {\bibfnamefont {S.}~\bibnamefont
  {Khan}}, \bibinfo {author} {\bibfnamefont {M.}~\bibnamefont {Mitra}},\ and\
  \bibinfo {author} {\bibfnamefont {S.~K.}\ \bibnamefont {Rai}},\ }\bibfield
  {title} {\bibinfo {title} {{Doubly-charged Higgs boson at a future
  electron-proton collider}},\ }\href
  {https://doi.org/10.1103/PhysRevD.99.115015} {\bibfield  {journal} {\bibinfo
  {journal} {Phys. Rev. D}\ }\textbf {\bibinfo {volume} {99}},\ \bibinfo
  {pages} {115015} (\bibinfo {year} {2019})},\ \Eprint
  {https://arxiv.org/abs/1903.01431} {arXiv:1903.01431 [hep-ph]} \BibitemShut
  {NoStop}%
\bibitem [{\citenamefont {Yang}\ and\ \citenamefont
  {Yang}(2022)}]{Yang:2021skb}%
  \BibitemOpen
  \bibfield  {author} {\bibinfo {author} {\bibfnamefont {X.-H.}\ \bibnamefont
  {Yang}}\ and\ \bibinfo {author} {\bibfnamefont {Z.-J.}\ \bibnamefont
  {Yang}},\ }\bibfield  {title} {\bibinfo {title} {{Doubly charged Higgs
  production at future $ep$ colliders}},\ }\href
  {https://doi.org/10.1088/1674-1137/ac581b} {\bibfield  {journal} {\bibinfo
  {journal} {Chin. Phys. C}\ }\textbf {\bibinfo {volume} {46}},\ \bibinfo
  {pages} {063107} (\bibinfo {year} {2022})},\ \Eprint
  {https://arxiv.org/abs/2103.11412} {arXiv:2103.11412 [hep-ph]} \BibitemShut
  {NoStop}%
\bibitem [{\citenamefont {Deppisch}\ \emph {et~al.}(2015)\citenamefont
  {Deppisch}, \citenamefont {Bhupal~Dev},\ and\ \citenamefont
  {Pilaftsis}}]{Deppisch:2015qwa}%
  \BibitemOpen
  \bibfield  {author} {\bibinfo {author} {\bibfnamefont {F.~F.}\ \bibnamefont
  {Deppisch}}, \bibinfo {author} {\bibfnamefont {P.~S.}\ \bibnamefont
  {Bhupal~Dev}},\ and\ \bibinfo {author} {\bibfnamefont {A.}~\bibnamefont
  {Pilaftsis}},\ }\bibfield  {title} {\bibinfo {title} {{Neutrinos and Collider
  Physics}},\ }\href {https://doi.org/10.1088/1367-2630/17/7/075019} {\bibfield
   {journal} {\bibinfo  {journal} {New J. Phys.}\ }\textbf {\bibinfo {volume}
  {17}},\ \bibinfo {pages} {075019} (\bibinfo {year} {2015})},\ \Eprint
  {https://arxiv.org/abs/1502.06541} {arXiv:1502.06541 [hep-ph]} \BibitemShut
  {NoStop}%
\bibitem [{\citenamefont {Cai}\ \emph {et~al.}(2018)\citenamefont {Cai},
  \citenamefont {Han}, \citenamefont {Li},\ and\ \citenamefont
  {Ruiz}}]{Cai:2017mow}%
  \BibitemOpen
  \bibfield  {author} {\bibinfo {author} {\bibfnamefont {Y.}~\bibnamefont
  {Cai}}, \bibinfo {author} {\bibfnamefont {T.}~\bibnamefont {Han}}, \bibinfo
  {author} {\bibfnamefont {T.}~\bibnamefont {Li}},\ and\ \bibinfo {author}
  {\bibfnamefont {R.}~\bibnamefont {Ruiz}},\ }\bibfield  {title} {\bibinfo
  {title} {{Lepton Number Violation: Seesaw Models and Their Collider Tests}},\
  }\href {https://doi.org/10.3389/fphy.2018.00040} {\bibfield  {journal}
  {\bibinfo  {journal} {Front. in Phys.}\ }\textbf {\bibinfo {volume} {6}},\
  \bibinfo {pages} {40} (\bibinfo {year} {2018})},\ \Eprint
  {https://arxiv.org/abs/1711.02180} {arXiv:1711.02180 [hep-ph]} \BibitemShut
  {NoStop}%
\bibitem [{\citenamefont {Chun}\ \emph {et~al.}(2003)\citenamefont {Chun},
  \citenamefont {Lee},\ and\ \citenamefont {Park}}]{Chun:2003ej}%
  \BibitemOpen
  \bibfield  {author} {\bibinfo {author} {\bibfnamefont {E.~J.}\ \bibnamefont
  {Chun}}, \bibinfo {author} {\bibfnamefont {K.~Y.}\ \bibnamefont {Lee}},\ and\
  \bibinfo {author} {\bibfnamefont {S.~C.}\ \bibnamefont {Park}},\ }\bibfield
  {title} {\bibinfo {title} {{Testing Higgs triplet model and neutrino mass
  patterns}},\ }\href {https://doi.org/10.1016/S0370-2693(03)00770-6}
  {\bibfield  {journal} {\bibinfo  {journal} {Phys. Lett. B}\ }\textbf
  {\bibinfo {volume} {566}},\ \bibinfo {pages} {142} (\bibinfo {year}
  {2003})},\ \Eprint {https://arxiv.org/abs/hep-ph/0304069}
  {arXiv:hep-ph/0304069} \BibitemShut {NoStop}%
\bibitem [{\citenamefont {Kadastik}\ \emph {et~al.}(2008)\citenamefont
  {Kadastik}, \citenamefont {Raidal},\ and\ \citenamefont
  {Rebane}}]{Kadastik:2007yd}%
  \BibitemOpen
  \bibfield  {author} {\bibinfo {author} {\bibfnamefont {M.}~\bibnamefont
  {Kadastik}}, \bibinfo {author} {\bibfnamefont {M.}~\bibnamefont {Raidal}},\
  and\ \bibinfo {author} {\bibfnamefont {L.}~\bibnamefont {Rebane}},\
  }\bibfield  {title} {\bibinfo {title} {{Direct determination of neutrino mass
  parameters at future colliders}},\ }\href
  {https://doi.org/10.1103/PhysRevD.77.115023} {\bibfield  {journal} {\bibinfo
  {journal} {Phys. Rev. D}\ }\textbf {\bibinfo {volume} {77}},\ \bibinfo
  {pages} {115023} (\bibinfo {year} {2008})},\ \Eprint
  {https://arxiv.org/abs/0712.3912} {arXiv:0712.3912 [hep-ph]} \BibitemShut
  {NoStop}%
\bibitem [{\citenamefont {Arhrib}\ \emph {et~al.}(2011)\citenamefont {Arhrib},
  \citenamefont {Benbrik}, \citenamefont {Chabab}, \citenamefont {Moultaka},
  \citenamefont {Peyranere}, \citenamefont {Rahili},\ and\ \citenamefont
  {Ramadan}}]{Arhrib:2011uy}%
  \BibitemOpen
  \bibfield  {author} {\bibinfo {author} {\bibfnamefont {A.}~\bibnamefont
  {Arhrib}}, \bibinfo {author} {\bibfnamefont {R.}~\bibnamefont {Benbrik}},
  \bibinfo {author} {\bibfnamefont {M.}~\bibnamefont {Chabab}}, \bibinfo
  {author} {\bibfnamefont {G.}~\bibnamefont {Moultaka}}, \bibinfo {author}
  {\bibfnamefont {M.~C.}\ \bibnamefont {Peyranere}}, \bibinfo {author}
  {\bibfnamefont {L.}~\bibnamefont {Rahili}},\ and\ \bibinfo {author}
  {\bibfnamefont {J.}~\bibnamefont {Ramadan}},\ }\bibfield  {title} {\bibinfo
  {title} {{The Higgs Potential in the Type II Seesaw Model}},\ }\href
  {https://doi.org/10.1103/PhysRevD.84.095005} {\bibfield  {journal} {\bibinfo
  {journal} {Phys. Rev. D}\ }\textbf {\bibinfo {volume} {84}},\ \bibinfo
  {pages} {095005} (\bibinfo {year} {2011})},\ \Eprint
  {https://arxiv.org/abs/1105.1925} {arXiv:1105.1925 [hep-ph]} \BibitemShut
  {NoStop}%
\bibitem [{\citenamefont {Chun}\ \emph {et~al.}(2012)\citenamefont {Chun},
  \citenamefont {Lee},\ and\ \citenamefont {Sharma}}]{Chun:2012jw}%
  \BibitemOpen
  \bibfield  {author} {\bibinfo {author} {\bibfnamefont {E.~J.}\ \bibnamefont
  {Chun}}, \bibinfo {author} {\bibfnamefont {H.~M.}\ \bibnamefont {Lee}},\ and\
  \bibinfo {author} {\bibfnamefont {P.}~\bibnamefont {Sharma}},\ }\bibfield
  {title} {\bibinfo {title} {{Vacuum Stability, Perturbativity, EWPD and
  Higgs-to-diphoton rate in Type II Seesaw Models}},\ }\href
  {https://doi.org/10.1007/JHEP11(2012)106} {\bibfield  {journal} {\bibinfo
  {journal} {JHEP}\ }\textbf {\bibinfo {volume} {11}},\ \bibinfo {pages}
  {106}},\ \Eprint {https://arxiv.org/abs/1209.1303} {arXiv:1209.1303 [hep-ph]}
  \BibitemShut {NoStop}%
\bibitem [{\citenamefont {Das}\ and\ \citenamefont
  {Santamaria}(2016)}]{Das:2016bir}%
  \BibitemOpen
  \bibfield  {author} {\bibinfo {author} {\bibfnamefont {D.}~\bibnamefont
  {Das}}\ and\ \bibinfo {author} {\bibfnamefont {A.}~\bibnamefont
  {Santamaria}},\ }\bibfield  {title} {\bibinfo {title} {{Updated scalar sector
  constraints in the Higgs triplet model}},\ }\href
  {https://doi.org/10.1103/PhysRevD.94.015015} {\bibfield  {journal} {\bibinfo
  {journal} {Phys. Rev. D}\ }\textbf {\bibinfo {volume} {94}},\ \bibinfo
  {pages} {015015} (\bibinfo {year} {2016})},\ \Eprint
  {https://arxiv.org/abs/1604.08099} {arXiv:1604.08099 [hep-ph]} \BibitemShut
  {NoStop}%
\bibitem [{\citenamefont {Aad}\ \emph {et~al.}(2012)\citenamefont {Aad} \emph
  {et~al.}}]{ATLAS:2012hi}%
  \BibitemOpen
  \bibfield  {author} {\bibinfo {author} {\bibfnamefont {G.}~\bibnamefont
  {Aad}} \emph {et~al.} (\bibinfo {collaboration} {ATLAS}),\ }\bibfield
  {title} {\bibinfo {title} {{Search for doubly-charged Higgs bosons in
  like-sign dilepton final states at $\sqrt{s}=7$ TeV with the ATLAS
  detector}},\ }\href {https://doi.org/10.1140/epjc/s10052-012-2244-2}
  {\bibfield  {journal} {\bibinfo  {journal} {Eur. Phys. J. C}\ }\textbf
  {\bibinfo {volume} {72}},\ \bibinfo {pages} {2244} (\bibinfo {year}
  {2012})},\ \Eprint {https://arxiv.org/abs/1210.5070} {arXiv:1210.5070
  [hep-ex]} \BibitemShut {NoStop}%
\bibitem [{\citenamefont {Chatrchyan}\ \emph {et~al.}(2012)\citenamefont
  {Chatrchyan} \emph {et~al.}}]{Chatrchyan:2012ya}%
  \BibitemOpen
  \bibfield  {author} {\bibinfo {author} {\bibfnamefont {S.}~\bibnamefont
  {Chatrchyan}} \emph {et~al.} (\bibinfo {collaboration} {CMS}),\ }\bibfield
  {title} {\bibinfo {title} {{A Search for a Doubly-Charged Higgs Boson in $pp$
  Collisions at $\sqrt{s}=7$ TeV}},\ }\href
  {https://doi.org/10.1140/epjc/s10052-012-2189-5} {\bibfield  {journal}
  {\bibinfo  {journal} {Eur. Phys. J. C}\ }\textbf {\bibinfo {volume} {72}},\
  \bibinfo {pages} {2189} (\bibinfo {year} {2012})},\ \Eprint
  {https://arxiv.org/abs/1207.2666} {arXiv:1207.2666 [hep-ex]} \BibitemShut
  {NoStop}%
\bibitem [{\citenamefont {Aad}\ \emph {et~al.}(2015)\citenamefont {Aad} \emph
  {et~al.}}]{ATLAS:2014kca}%
  \BibitemOpen
  \bibfield  {author} {\bibinfo {author} {\bibfnamefont {G.}~\bibnamefont
  {Aad}} \emph {et~al.} (\bibinfo {collaboration} {ATLAS}),\ }\bibfield
  {title} {\bibinfo {title} {{Search for anomalous production of prompt
  same-sign lepton pairs and pair-produced doubly charged Higgs bosons with $
  \sqrt{s}=8 $ TeV $pp$ collisions using the ATLAS detector}},\ }\href
  {https://doi.org/10.1007/JHEP03(2015)041} {\bibfield  {journal} {\bibinfo
  {journal} {JHEP}\ }\textbf {\bibinfo {volume} {03}},\ \bibinfo {pages}
  {041}},\ \Eprint {https://arxiv.org/abs/1412.0237} {arXiv:1412.0237 [hep-ex]}
  \BibitemShut {NoStop}%
\bibitem [{\citenamefont {Khachatryan}\ \emph {et~al.}(2015)\citenamefont
  {Khachatryan} \emph {et~al.}}]{Khachatryan:2014sta}%
  \BibitemOpen
  \bibfield  {author} {\bibinfo {author} {\bibfnamefont {V.}~\bibnamefont
  {Khachatryan}} \emph {et~al.} (\bibinfo {collaboration} {CMS}),\ }\bibfield
  {title} {\bibinfo {title} {{Study of vector boson scattering and search for
  new physics in events with two same-sign leptons and two jets}},\ }\href
  {https://doi.org/10.1103/PhysRevLett.114.051801} {\bibfield  {journal}
  {\bibinfo  {journal} {Phys. Rev. Lett.}\ }\textbf {\bibinfo {volume} {114}},\
  \bibinfo {pages} {051801} (\bibinfo {year} {2015})},\ \Eprint
  {https://arxiv.org/abs/1410.6315} {arXiv:1410.6315 [hep-ex]} \BibitemShut
  {NoStop}%
\bibitem [{CMS(2016)}]{CMS:2016cpz}%
  \BibitemOpen
  \bibfield  {title} {\bibinfo {title} {{Search for a doubly-charged Higgs
  boson with $\sqrt{s}=8~\mathrm{TeV}$ $pp$ collisions at the CMS
  experiment}},\ }\href@noop {} {\  (\bibinfo {year} {2016})}\BibitemShut
  {NoStop}%
\bibitem [{CMS(2017)}]{CMS:2017pet}%
  \BibitemOpen
  \bibfield  {title} {\bibinfo {title} {{A search for doubly-charged Higgs
  boson production in three and four lepton final states at
  $\sqrt{s}=13~\mathrm{TeV}$}},\ }\href@noop {} {\  (\bibinfo {year}
  {2017})}\BibitemShut {NoStop}%
\bibitem [{\citenamefont {Aaboud}\ \emph
  {et~al.}(2018{\natexlab{a}})\citenamefont {Aaboud} \emph
  {et~al.}}]{Aaboud:2017qph}%
  \BibitemOpen
  \bibfield  {author} {\bibinfo {author} {\bibfnamefont {M.}~\bibnamefont
  {Aaboud}} \emph {et~al.} (\bibinfo {collaboration} {ATLAS}),\ }\bibfield
  {title} {\bibinfo {title} {{Search for doubly charged Higgs boson production
  in multi-lepton final states with the ATLAS detector using
  proton\textendash{}proton collisions at $\sqrt{s}=13\,\text {TeV}$}},\ }\href
  {https://doi.org/10.1140/epjc/s10052-018-5661-z} {\bibfield  {journal}
  {\bibinfo  {journal} {Eur. Phys. J. C}\ }\textbf {\bibinfo {volume} {78}},\
  \bibinfo {pages} {199} (\bibinfo {year} {2018}{\natexlab{a}})},\ \Eprint
  {https://arxiv.org/abs/1710.09748} {arXiv:1710.09748 [hep-ex]} \BibitemShut
  {NoStop}%
\bibitem [{\citenamefont {Sirunyan}\ \emph {et~al.}(2018)\citenamefont
  {Sirunyan} \emph {et~al.}}]{CMS:2017fhs}%
  \BibitemOpen
  \bibfield  {author} {\bibinfo {author} {\bibfnamefont {A.~M.}\ \bibnamefont
  {Sirunyan}} \emph {et~al.} (\bibinfo {collaboration} {CMS}),\ }\bibfield
  {title} {\bibinfo {title} {{Observation of electroweak production of
  same-sign W boson pairs in the two jet and two same-sign lepton final state
  in proton-proton collisions at $\sqrt{s} = $ 13 TeV}},\ }\href
  {https://doi.org/10.1103/PhysRevLett.120.081801} {\bibfield  {journal}
  {\bibinfo  {journal} {Phys. Rev. Lett.}\ }\textbf {\bibinfo {volume} {120}},\
  \bibinfo {pages} {081801} (\bibinfo {year} {2018})},\ \Eprint
  {https://arxiv.org/abs/1709.05822} {arXiv:1709.05822 [hep-ex]} \BibitemShut
  {NoStop}%
\bibitem [{\citenamefont {Aaboud}\ \emph {et~al.}(2019)\citenamefont {Aaboud}
  \emph {et~al.}}]{Aaboud:2018qcu}%
  \BibitemOpen
  \bibfield  {author} {\bibinfo {author} {\bibfnamefont {M.}~\bibnamefont
  {Aaboud}} \emph {et~al.} (\bibinfo {collaboration} {ATLAS}),\ }\bibfield
  {title} {\bibinfo {title} {{Search for doubly charged scalar bosons decaying
  into same-sign $W$ boson pairs with the ATLAS detector}},\ }\href
  {https://doi.org/10.1140/epjc/s10052-018-6500-y} {\bibfield  {journal}
  {\bibinfo  {journal} {Eur. Phys. J. C}\ }\textbf {\bibinfo {volume} {79}},\
  \bibinfo {pages} {58} (\bibinfo {year} {2019})},\ \Eprint
  {https://arxiv.org/abs/1808.01899} {arXiv:1808.01899 [hep-ex]} \BibitemShut
  {NoStop}%
\bibitem [{\citenamefont {Aad}\ \emph {et~al.}(2021{\natexlab{a}})\citenamefont
  {Aad} \emph {et~al.}}]{Aad:2021lzu}%
  \BibitemOpen
  \bibfield  {author} {\bibinfo {author} {\bibfnamefont {G.}~\bibnamefont
  {Aad}} \emph {et~al.} (\bibinfo {collaboration} {ATLAS}),\ }\bibfield
  {title} {\bibinfo {title} {{Search for doubly and singly charged Higgs bosons
  decaying into vector bosons in multi-lepton final states with the ATLAS
  detector using proton-proton collisions at $ \sqrt{\mathrm{s}} $ = 13 TeV}},\
  }\href {https://doi.org/10.1007/JHEP06(2021)146} {\bibfield  {journal}
  {\bibinfo  {journal} {JHEP}\ }\textbf {\bibinfo {volume} {06}},\ \bibinfo
  {pages} {146}},\ \Eprint {https://arxiv.org/abs/2101.11961} {arXiv:2101.11961
  [hep-ex]} \BibitemShut {NoStop}%
\bibitem [{\citenamefont {Aad}\ \emph {et~al.}(2021{\natexlab{b}})\citenamefont
  {Aad} \emph {et~al.}}]{ATLAS:2021jol}%
  \BibitemOpen
  \bibfield  {author} {\bibinfo {author} {\bibfnamefont {G.}~\bibnamefont
  {Aad}} \emph {et~al.} (\bibinfo {collaboration} {ATLAS}),\ }\bibfield
  {title} {\bibinfo {title} {{Search for doubly and singly charged Higgs bosons
  decaying into vector bosons in multi-lepton final states with the ATLAS
  detector using proton-proton collisions at $ \sqrt{\mathrm{s}} $ = 13 TeV}},\
  }\href {https://doi.org/10.1007/JHEP06(2021)146} {\bibfield  {journal}
  {\bibinfo  {journal} {JHEP}\ }\textbf {\bibinfo {volume} {06}},\ \bibinfo
  {pages} {146}},\ \Eprint {https://arxiv.org/abs/2101.11961} {arXiv:2101.11961
  [hep-ex]} \BibitemShut {NoStop}%
\bibitem [{ATL(2022)}]{ATLAS:2022yzd}%
  \BibitemOpen
  \bibfield  {title} {\bibinfo {title} {{Search for doubly charged Higgs boson
  production in multi-lepton final states using $139\,\text{fb}^{-1}$ of
  proton--proton collisions at $\sqrt{s}= 13\,\text{TeV}$ with the ATLAS
  detector}},\ }\href@noop {} {\  (\bibinfo {year} {2022})}\BibitemShut
  {NoStop}%
\bibitem [{\citenamefont {Aad}\ \emph {et~al.}(2021{\natexlab{c}})\citenamefont
  {Aad} \emph {et~al.}}]{ATLAS:2021kxv}%
  \BibitemOpen
  \bibfield  {author} {\bibinfo {author} {\bibfnamefont {G.}~\bibnamefont
  {Aad}} \emph {et~al.} (\bibinfo {collaboration} {ATLAS}),\ }\bibfield
  {title} {\bibinfo {title} {{Search for new phenomena in events with an
  energetic jet and missing transverse momentum in $pp$ collisions at $\sqrt
  {s}$ =13 TeV with the ATLAS detector}},\ }\href
  {https://doi.org/10.1103/PhysRevD.103.112006} {\bibfield  {journal} {\bibinfo
   {journal} {Phys. Rev. D}\ }\textbf {\bibinfo {volume} {103}},\ \bibinfo
  {pages} {112006} (\bibinfo {year} {2021}{\natexlab{c}})},\ \Eprint
  {https://arxiv.org/abs/2102.10874} {arXiv:2102.10874 [hep-ex]} \BibitemShut
  {NoStop}%
\bibitem [{\citenamefont {Tumasyan}\ \emph {et~al.}(2021)\citenamefont
  {Tumasyan} \emph {et~al.}}]{CMS:2021far}%
  \BibitemOpen
  \bibfield  {author} {\bibinfo {author} {\bibfnamefont {A.}~\bibnamefont
  {Tumasyan}} \emph {et~al.} (\bibinfo {collaboration} {CMS}),\ }\bibfield
  {title} {\bibinfo {title} {{Search for new particles in events with energetic
  jets and large missing transverse momentum in proton-proton collisions at $
  \sqrt{s} $ = 13 TeV}},\ }\href {https://doi.org/10.1007/JHEP11(2021)153}
  {\bibfield  {journal} {\bibinfo  {journal} {JHEP}\ }\textbf {\bibinfo
  {volume} {11}},\ \bibinfo {pages} {153}},\ \Eprint
  {https://arxiv.org/abs/2107.13021} {arXiv:2107.13021 [hep-ex]} \BibitemShut
  {NoStop}%
\bibitem [{\citenamefont {Aad}\ \emph {et~al.}(2020{\natexlab{a}})\citenamefont
  {Aad} \emph {et~al.}}]{ATLAS:2019lng}%
  \BibitemOpen
  \bibfield  {author} {\bibinfo {author} {\bibfnamefont {G.}~\bibnamefont
  {Aad}} \emph {et~al.} (\bibinfo {collaboration} {ATLAS}),\ }\bibfield
  {title} {\bibinfo {title} {{Searches for electroweak production of
  supersymmetric particles with compressed mass spectra in $\sqrt{s}=$ 13 TeV
  $pp$ collisions with the ATLAS detector}},\ }\href
  {https://doi.org/10.1103/PhysRevD.101.052005} {\bibfield  {journal} {\bibinfo
   {journal} {Phys. Rev. D}\ }\textbf {\bibinfo {volume} {101}},\ \bibinfo
  {pages} {052005} (\bibinfo {year} {2020}{\natexlab{a}})},\ \Eprint
  {https://arxiv.org/abs/1911.12606} {arXiv:1911.12606 [hep-ex]} \BibitemShut
  {NoStop}%
\bibitem [{\citenamefont {Tumasyan}\ \emph {et~al.}(2022)\citenamefont
  {Tumasyan} \emph {et~al.}}]{CMS:2021edw}%
  \BibitemOpen
  \bibfield  {author} {\bibinfo {author} {\bibfnamefont {A.}~\bibnamefont
  {Tumasyan}} \emph {et~al.} (\bibinfo {collaboration} {CMS}),\ }\bibfield
  {title} {\bibinfo {title} {{Search for supersymmetry in final states with two
  or three soft leptons and missing transverse momentum in proton-proton
  collisions at $ \sqrt{s} $ = 13 TeV}},\ }\href
  {https://doi.org/10.1007/JHEP04(2022)091} {\bibfield  {journal} {\bibinfo
  {journal} {JHEP}\ }\textbf {\bibinfo {volume} {04}},\ \bibinfo {pages}
  {091}},\ \Eprint {https://arxiv.org/abs/2111.06296} {arXiv:2111.06296
  [hep-ex]} \BibitemShut {NoStop}%
\bibitem [{\citenamefont {Staub}(2014)}]{Staub:2013tta}%
  \BibitemOpen
  \bibfield  {author} {\bibinfo {author} {\bibfnamefont {F.}~\bibnamefont
  {Staub}},\ }\bibfield  {title} {\bibinfo {title} {{SARAH 4 : A tool for (not
  only SUSY) model builders}},\ }\href
  {https://doi.org/10.1016/j.cpc.2014.02.018} {\bibfield  {journal} {\bibinfo
  {journal} {Comput. Phys. Commun.}\ }\textbf {\bibinfo {volume} {185}},\
  \bibinfo {pages} {1773} (\bibinfo {year} {2014})},\ \Eprint
  {https://arxiv.org/abs/1309.7223} {arXiv:1309.7223 [hep-ph]} \BibitemShut
  {NoStop}%
\bibitem [{\citenamefont {Staub}(2015)}]{Staub:2015kfa}%
  \BibitemOpen
  \bibfield  {author} {\bibinfo {author} {\bibfnamefont {F.}~\bibnamefont
  {Staub}},\ }\bibfield  {title} {\bibinfo {title} {{Exploring new models in
  all detail with SARAH}},\ }\href {https://doi.org/10.1155/2015/840780}
  {\bibfield  {journal} {\bibinfo  {journal} {Adv. High Energy Phys.}\ }\textbf
  {\bibinfo {volume} {2015}},\ \bibinfo {pages} {840780} (\bibinfo {year}
  {2015})},\ \Eprint {https://arxiv.org/abs/1503.04200} {arXiv:1503.04200
  [hep-ph]} \BibitemShut {NoStop}%
\bibitem [{\citenamefont {Degrande}\ \emph {et~al.}(2012)\citenamefont
  {Degrande}, \citenamefont {Duhr}, \citenamefont {Fuks}, \citenamefont
  {Grellscheid}, \citenamefont {Mattelaer},\ and\ \citenamefont
  {Reiter}}]{Degrande:2011ua}%
  \BibitemOpen
  \bibfield  {author} {\bibinfo {author} {\bibfnamefont {C.}~\bibnamefont
  {Degrande}}, \bibinfo {author} {\bibfnamefont {C.}~\bibnamefont {Duhr}},
  \bibinfo {author} {\bibfnamefont {B.}~\bibnamefont {Fuks}}, \bibinfo {author}
  {\bibfnamefont {D.}~\bibnamefont {Grellscheid}}, \bibinfo {author}
  {\bibfnamefont {O.}~\bibnamefont {Mattelaer}},\ and\ \bibinfo {author}
  {\bibfnamefont {T.}~\bibnamefont {Reiter}},\ }\bibfield  {title} {\bibinfo
  {title} {{UFO - The Universal FeynRules Output}},\ }\href
  {https://doi.org/10.1016/j.cpc.2012.01.022} {\bibfield  {journal} {\bibinfo
  {journal} {Comput. Phys. Commun.}\ }\textbf {\bibinfo {volume} {183}},\
  \bibinfo {pages} {1201} (\bibinfo {year} {2012})},\ \Eprint
  {https://arxiv.org/abs/1108.2040} {arXiv:1108.2040 [hep-ph]} \BibitemShut
  {NoStop}%
\bibitem [{\citenamefont {Alwall}\ \emph {et~al.}(2011)\citenamefont {Alwall},
  \citenamefont {Herquet}, \citenamefont {Maltoni}, \citenamefont {Mattelaer},\
  and\ \citenamefont {Stelzer}}]{Alwall:2011uj}%
  \BibitemOpen
  \bibfield  {author} {\bibinfo {author} {\bibfnamefont {J.}~\bibnamefont
  {Alwall}}, \bibinfo {author} {\bibfnamefont {M.}~\bibnamefont {Herquet}},
  \bibinfo {author} {\bibfnamefont {F.}~\bibnamefont {Maltoni}}, \bibinfo
  {author} {\bibfnamefont {O.}~\bibnamefont {Mattelaer}},\ and\ \bibinfo
  {author} {\bibfnamefont {T.}~\bibnamefont {Stelzer}},\ }\bibfield  {title}
  {\bibinfo {title} {{MadGraph 5 : Going Beyond}},\ }\href
  {https://doi.org/10.1007/JHEP06(2011)128} {\bibfield  {journal} {\bibinfo
  {journal} {JHEP}\ }\textbf {\bibinfo {volume} {06}},\ \bibinfo {pages}
  {128}},\ \Eprint {https://arxiv.org/abs/1106.0522} {arXiv:1106.0522 [hep-ph]}
  \BibitemShut {NoStop}%
\bibitem [{\citenamefont {Alwall}\ \emph {et~al.}(2014)\citenamefont {Alwall},
  \citenamefont {Frederix}, \citenamefont {Frixione}, \citenamefont {Hirschi},
  \citenamefont {Maltoni}, \citenamefont {Mattelaer}, \citenamefont {Shao},
  \citenamefont {Stelzer}, \citenamefont {Torrielli},\ and\ \citenamefont
  {Zaro}}]{Alwall:2014hca}%
  \BibitemOpen
  \bibfield  {author} {\bibinfo {author} {\bibfnamefont {J.}~\bibnamefont
  {Alwall}}, \bibinfo {author} {\bibfnamefont {R.}~\bibnamefont {Frederix}},
  \bibinfo {author} {\bibfnamefont {S.}~\bibnamefont {Frixione}}, \bibinfo
  {author} {\bibfnamefont {V.}~\bibnamefont {Hirschi}}, \bibinfo {author}
  {\bibfnamefont {F.}~\bibnamefont {Maltoni}}, \bibinfo {author} {\bibfnamefont
  {O.}~\bibnamefont {Mattelaer}}, \bibinfo {author} {\bibfnamefont {H.~S.}\
  \bibnamefont {Shao}}, \bibinfo {author} {\bibfnamefont {T.}~\bibnamefont
  {Stelzer}}, \bibinfo {author} {\bibfnamefont {P.}~\bibnamefont {Torrielli}},\
  and\ \bibinfo {author} {\bibfnamefont {M.}~\bibnamefont {Zaro}},\ }\bibfield
  {title} {\bibinfo {title} {{The automated computation of tree-level and
  next-to-leading order differential cross sections, and their matching to
  parton shower simulations}},\ }\href
  {https://doi.org/10.1007/JHEP07(2014)079} {\bibfield  {journal} {\bibinfo
  {journal} {JHEP}\ }\textbf {\bibinfo {volume} {07}},\ \bibinfo {pages}
  {079}},\ \Eprint {https://arxiv.org/abs/1405.0301} {arXiv:1405.0301 [hep-ph]}
  \BibitemShut {NoStop}%
\bibitem [{\citenamefont {Ball}\ \emph {et~al.}(2013)\citenamefont {Ball},
  \citenamefont {Bertone}, \citenamefont {Carrazza}, \citenamefont
  {Del~Debbio}, \citenamefont {Forte}, \citenamefont {Guffanti}, \citenamefont
  {Hartland},\ and\ \citenamefont {Rojo}}]{Ball:2013hta}%
  \BibitemOpen
  \bibfield  {author} {\bibinfo {author} {\bibfnamefont {R.~D.}\ \bibnamefont
  {Ball}}, \bibinfo {author} {\bibfnamefont {V.}~\bibnamefont {Bertone}},
  \bibinfo {author} {\bibfnamefont {S.}~\bibnamefont {Carrazza}}, \bibinfo
  {author} {\bibfnamefont {L.}~\bibnamefont {Del~Debbio}}, \bibinfo {author}
  {\bibfnamefont {S.}~\bibnamefont {Forte}}, \bibinfo {author} {\bibfnamefont
  {A.}~\bibnamefont {Guffanti}}, \bibinfo {author} {\bibfnamefont {N.~P.}\
  \bibnamefont {Hartland}},\ and\ \bibinfo {author} {\bibfnamefont
  {J.}~\bibnamefont {Rojo}} (\bibinfo {collaboration} {NNPDF}),\ }\bibfield
  {title} {\bibinfo {title} {{Parton distributions with QED corrections}},\
  }\href {https://doi.org/10.1016/j.nuclphysb.2013.10.010} {\bibfield
  {journal} {\bibinfo  {journal} {Nucl. Phys. B}\ }\textbf {\bibinfo {volume}
  {877}},\ \bibinfo {pages} {290} (\bibinfo {year} {2013})},\ \Eprint
  {https://arxiv.org/abs/1308.0598} {arXiv:1308.0598 [hep-ph]} \BibitemShut
  {NoStop}%
\bibitem [{\citenamefont {Ball}\ \emph {et~al.}(2015)\citenamefont {Ball} \emph
  {et~al.}}]{NNPDF:2014otw}%
  \BibitemOpen
  \bibfield  {author} {\bibinfo {author} {\bibfnamefont {R.~D.}\ \bibnamefont
  {Ball}} \emph {et~al.} (\bibinfo {collaboration} {NNPDF}),\ }\bibfield
  {title} {\bibinfo {title} {{Parton distributions for the LHC Run II}},\
  }\href {https://doi.org/10.1007/JHEP04(2015)040} {\bibfield  {journal}
  {\bibinfo  {journal} {JHEP}\ }\textbf {\bibinfo {volume} {04}},\ \bibinfo
  {pages} {040}},\ \Eprint {https://arxiv.org/abs/1410.8849} {arXiv:1410.8849
  [hep-ph]} \BibitemShut {NoStop}%
\bibitem [{\citenamefont {Fuks}\ \emph {et~al.}(2020)\citenamefont {Fuks},
  \citenamefont {Nemev\v{s}ek},\ and\ \citenamefont {Ruiz}}]{Fuks:2019clu}%
  \BibitemOpen
  \bibfield  {author} {\bibinfo {author} {\bibfnamefont {B.}~\bibnamefont
  {Fuks}}, \bibinfo {author} {\bibfnamefont {M.}~\bibnamefont {Nemev\v{s}ek}},\
  and\ \bibinfo {author} {\bibfnamefont {R.}~\bibnamefont {Ruiz}},\ }\bibfield
  {title} {\bibinfo {title} {{Doubly Charged Higgs Boson Production at Hadron
  Colliders}},\ }\href {https://doi.org/10.1103/PhysRevD.101.075022} {\bibfield
   {journal} {\bibinfo  {journal} {Phys. Rev. D}\ }\textbf {\bibinfo {volume}
  {101}},\ \bibinfo {pages} {075022} (\bibinfo {year} {2020})},\ \Eprint
  {https://arxiv.org/abs/1912.08975} {arXiv:1912.08975 [hep-ph]} \BibitemShut
  {NoStop}%
\bibitem [{\citenamefont {Sj\"ostrand}\ \emph {et~al.}(2015)\citenamefont
  {Sj\"ostrand}, \citenamefont {Ask}, \citenamefont {Christiansen},
  \citenamefont {Corke}, \citenamefont {Desai}, \citenamefont {Ilten},
  \citenamefont {Mrenna}, \citenamefont {Prestel}, \citenamefont {Rasmussen},\
  and\ \citenamefont {Skands}}]{Sjostrand:2014zea}%
  \BibitemOpen
  \bibfield  {author} {\bibinfo {author} {\bibfnamefont {T.}~\bibnamefont
  {Sj\"ostrand}}, \bibinfo {author} {\bibfnamefont {S.}~\bibnamefont {Ask}},
  \bibinfo {author} {\bibfnamefont {J.~R.}\ \bibnamefont {Christiansen}},
  \bibinfo {author} {\bibfnamefont {R.}~\bibnamefont {Corke}}, \bibinfo
  {author} {\bibfnamefont {N.}~\bibnamefont {Desai}}, \bibinfo {author}
  {\bibfnamefont {P.}~\bibnamefont {Ilten}}, \bibinfo {author} {\bibfnamefont
  {S.}~\bibnamefont {Mrenna}}, \bibinfo {author} {\bibfnamefont
  {S.}~\bibnamefont {Prestel}}, \bibinfo {author} {\bibfnamefont {C.~O.}\
  \bibnamefont {Rasmussen}},\ and\ \bibinfo {author} {\bibfnamefont {P.~Z.}\
  \bibnamefont {Skands}},\ }\bibfield  {title} {\bibinfo {title} {{An
  introduction to PYTHIA 8.2}},\ }\href
  {https://doi.org/10.1016/j.cpc.2015.01.024} {\bibfield  {journal} {\bibinfo
  {journal} {Comput. Phys. Commun.}\ }\textbf {\bibinfo {volume} {191}},\
  \bibinfo {pages} {159} (\bibinfo {year} {2015})},\ \Eprint
  {https://arxiv.org/abs/1410.3012} {arXiv:1410.3012 [hep-ph]} \BibitemShut
  {NoStop}%
\bibitem [{\citenamefont {Catani}\ \emph {et~al.}(2009)\citenamefont {Catani},
  \citenamefont {Cieri}, \citenamefont {Ferrera}, \citenamefont {de~Florian},\
  and\ \citenamefont {Grazzini}}]{Catani:2009sm}%
  \BibitemOpen
  \bibfield  {author} {\bibinfo {author} {\bibfnamefont {S.}~\bibnamefont
  {Catani}}, \bibinfo {author} {\bibfnamefont {L.}~\bibnamefont {Cieri}},
  \bibinfo {author} {\bibfnamefont {G.}~\bibnamefont {Ferrera}}, \bibinfo
  {author} {\bibfnamefont {D.}~\bibnamefont {de~Florian}},\ and\ \bibinfo
  {author} {\bibfnamefont {M.}~\bibnamefont {Grazzini}},\ }\bibfield  {title}
  {\bibinfo {title} {{Vector boson production at hadron colliders: a fully
  exclusive QCD calculation at NNLO}},\ }\href
  {https://doi.org/10.1103/PhysRevLett.103.082001} {\bibfield  {journal}
  {\bibinfo  {journal} {Phys. Rev. Lett.}\ }\textbf {\bibinfo {volume} {103}},\
  \bibinfo {pages} {082001} (\bibinfo {year} {2009})},\ \Eprint
  {https://arxiv.org/abs/0903.2120} {arXiv:0903.2120 [hep-ph]} \BibitemShut
  {NoStop}%
\bibitem [{\citenamefont {Balossini}\ \emph {et~al.}(2010)\citenamefont
  {Balossini}, \citenamefont {Montagna}, \citenamefont {Carloni~Calame},
  \citenamefont {Moretti}, \citenamefont {Nicrosini}, \citenamefont
  {Piccinini}, \citenamefont {Treccani},\ and\ \citenamefont
  {Vicini}}]{Balossini:2009sa}%
  \BibitemOpen
  \bibfield  {author} {\bibinfo {author} {\bibfnamefont {G.}~\bibnamefont
  {Balossini}}, \bibinfo {author} {\bibfnamefont {G.}~\bibnamefont {Montagna}},
  \bibinfo {author} {\bibfnamefont {C.~M.}\ \bibnamefont {Carloni~Calame}},
  \bibinfo {author} {\bibfnamefont {M.}~\bibnamefont {Moretti}}, \bibinfo
  {author} {\bibfnamefont {O.}~\bibnamefont {Nicrosini}}, \bibinfo {author}
  {\bibfnamefont {F.}~\bibnamefont {Piccinini}}, \bibinfo {author}
  {\bibfnamefont {M.}~\bibnamefont {Treccani}},\ and\ \bibinfo {author}
  {\bibfnamefont {A.}~\bibnamefont {Vicini}},\ }\bibfield  {title} {\bibinfo
  {title} {{Combination of electroweak and QCD corrections to single W
  production at the Fermilab Tevatron and the CERN LHC}},\ }\href
  {https://doi.org/10.1007/JHEP01(2010)013} {\bibfield  {journal} {\bibinfo
  {journal} {JHEP}\ }\textbf {\bibinfo {volume} {01}},\ \bibinfo {pages}
  {013}},\ \Eprint {https://arxiv.org/abs/0907.0276} {arXiv:0907.0276 [hep-ph]}
  \BibitemShut {NoStop}%
\bibitem [{\citenamefont {Campbell}\ \emph {et~al.}(2011)\citenamefont
  {Campbell}, \citenamefont {Ellis},\ and\ \citenamefont
  {Williams}}]{Campbell:2011bn}%
  \BibitemOpen
  \bibfield  {author} {\bibinfo {author} {\bibfnamefont {J.~M.}\ \bibnamefont
  {Campbell}}, \bibinfo {author} {\bibfnamefont {R.~K.}\ \bibnamefont
  {Ellis}},\ and\ \bibinfo {author} {\bibfnamefont {C.}~\bibnamefont
  {Williams}},\ }\bibfield  {title} {\bibinfo {title} {{Vector boson pair
  production at the LHC}},\ }\href {https://doi.org/10.1007/JHEP07(2011)018}
  {\bibfield  {journal} {\bibinfo  {journal} {JHEP}\ }\textbf {\bibinfo
  {volume} {07}},\ \bibinfo {pages} {018}},\ \Eprint
  {https://arxiv.org/abs/1105.0020} {arXiv:1105.0020 [hep-ph]} \BibitemShut
  {NoStop}%
\bibitem [{\citenamefont {Cascioli}\ \emph {et~al.}(2014)\citenamefont
  {Cascioli}, \citenamefont {Gehrmann}, \citenamefont {Grazzini}, \citenamefont
  {Kallweit}, \citenamefont {Maierh\"ofer}, \citenamefont {von Manteuffel},
  \citenamefont {Pozzorini}, \citenamefont {Rathlev}, \citenamefont
  {Tancredi},\ and\ \citenamefont {Weihs}}]{Cascioli:2014yka}%
  \BibitemOpen
  \bibfield  {author} {\bibinfo {author} {\bibfnamefont {F.}~\bibnamefont
  {Cascioli}}, \bibinfo {author} {\bibfnamefont {T.}~\bibnamefont {Gehrmann}},
  \bibinfo {author} {\bibfnamefont {M.}~\bibnamefont {Grazzini}}, \bibinfo
  {author} {\bibfnamefont {S.}~\bibnamefont {Kallweit}}, \bibinfo {author}
  {\bibfnamefont {P.}~\bibnamefont {Maierh\"ofer}}, \bibinfo {author}
  {\bibfnamefont {A.}~\bibnamefont {von Manteuffel}}, \bibinfo {author}
  {\bibfnamefont {S.}~\bibnamefont {Pozzorini}}, \bibinfo {author}
  {\bibfnamefont {D.}~\bibnamefont {Rathlev}}, \bibinfo {author} {\bibfnamefont
  {L.}~\bibnamefont {Tancredi}},\ and\ \bibinfo {author} {\bibfnamefont
  {E.}~\bibnamefont {Weihs}},\ }\bibfield  {title} {\bibinfo {title} {{ZZ
  production at hadron colliders in NNLO QCD}},\ }\href
  {https://doi.org/10.1016/j.physletb.2014.06.056} {\bibfield  {journal}
  {\bibinfo  {journal} {Phys. Lett. B}\ }\textbf {\bibinfo {volume} {735}},\
  \bibinfo {pages} {311} (\bibinfo {year} {2014})},\ \Eprint
  {https://arxiv.org/abs/1405.2219} {arXiv:1405.2219 [hep-ph]} \BibitemShut
  {NoStop}%
\bibitem [{\citenamefont {Campbell}\ \emph {et~al.}(2016)\citenamefont
  {Campbell}, \citenamefont {Ellis},\ and\ \citenamefont
  {Williams}}]{Campbell:2016jau}%
  \BibitemOpen
  \bibfield  {author} {\bibinfo {author} {\bibfnamefont {J.~M.}\ \bibnamefont
  {Campbell}}, \bibinfo {author} {\bibfnamefont {R.~K.}\ \bibnamefont
  {Ellis}},\ and\ \bibinfo {author} {\bibfnamefont {C.}~\bibnamefont
  {Williams}},\ }\bibfield  {title} {\bibinfo {title} {{Associated production
  of a Higgs boson at NNLO}},\ }\href {https://doi.org/10.1007/JHEP06(2016)179}
  {\bibfield  {journal} {\bibinfo  {journal} {JHEP}\ }\textbf {\bibinfo
  {volume} {06}},\ \bibinfo {pages} {179}},\ \Eprint
  {https://arxiv.org/abs/1601.00658} {arXiv:1601.00658 [hep-ph]} \BibitemShut
  {NoStop}%
\bibitem [{\citenamefont {de~Florian}\ \emph {et~al.}(2016)\citenamefont
  {de~Florian} \emph {et~al.}}]{LHCHiggsCrossSectionWorkingGroup:2016ypw}%
  \BibitemOpen
  \bibfield  {author} {\bibinfo {author} {\bibfnamefont {D.}~\bibnamefont
  {de~Florian}} \emph {et~al.} (\bibinfo {collaboration} {LHC Higgs Cross
  Section Working Group}),\ }\bibfield  {title} {\bibinfo {title} {{Handbook of
  LHC Higgs Cross Sections: 4. Deciphering the Nature of the Higgs Sector}}\
  }\textbf {\bibinfo {volume} {2/2017}},\ \href
  {https://doi.org/10.23731/CYRM-2017-002} {10.23731/CYRM-2017-002} (\bibinfo
  {year} {2016}),\ \Eprint {https://arxiv.org/abs/1610.07922} {arXiv:1610.07922
  [hep-ph]} \BibitemShut {NoStop}%
\bibitem [{\citenamefont {Shen}\ \emph {et~al.}(2017)\citenamefont {Shen},
  \citenamefont {Zhang}, \citenamefont {Ma}, \citenamefont {Li},\ and\
  \citenamefont {Guo}}]{Shen:2016ape}%
  \BibitemOpen
  \bibfield  {author} {\bibinfo {author} {\bibfnamefont {Y.-B.}\ \bibnamefont
  {Shen}}, \bibinfo {author} {\bibfnamefont {R.-Y.}\ \bibnamefont {Zhang}},
  \bibinfo {author} {\bibfnamefont {W.-G.}\ \bibnamefont {Ma}}, \bibinfo
  {author} {\bibfnamefont {X.-Z.}\ \bibnamefont {Li}},\ and\ \bibinfo {author}
  {\bibfnamefont {L.}~\bibnamefont {Guo}},\ }\bibfield  {title} {\bibinfo
  {title} {{NLO QCD and electroweak corrections to WWW production at the
  LHC}},\ }\href {https://doi.org/10.1103/PhysRevD.95.073005} {\bibfield
  {journal} {\bibinfo  {journal} {Phys. Rev. D}\ }\textbf {\bibinfo {volume}
  {95}},\ \bibinfo {pages} {073005} (\bibinfo {year} {2017})},\ \Eprint
  {https://arxiv.org/abs/1605.00554} {arXiv:1605.00554 [hep-ph]} \BibitemShut
  {NoStop}%
\bibitem [{\citenamefont {Nhung}\ \emph {et~al.}(2013)\citenamefont {Nhung},
  \citenamefont {Ninh},\ and\ \citenamefont {Weber}}]{Nhung:2013jta}%
  \BibitemOpen
  \bibfield  {author} {\bibinfo {author} {\bibfnamefont {D.~T.}\ \bibnamefont
  {Nhung}}, \bibinfo {author} {\bibfnamefont {L.~D.}\ \bibnamefont {Ninh}},\
  and\ \bibinfo {author} {\bibfnamefont {M.~M.}\ \bibnamefont {Weber}},\
  }\bibfield  {title} {\bibinfo {title} {{NLO corrections to WWZ production at
  the LHC}},\ }\href {https://doi.org/10.1007/JHEP12(2013)096} {\bibfield
  {journal} {\bibinfo  {journal} {JHEP}\ }\textbf {\bibinfo {volume} {12}},\
  \bibinfo {pages} {096}},\ \Eprint {https://arxiv.org/abs/1307.7403}
  {arXiv:1307.7403 [hep-ph]} \BibitemShut {NoStop}%
\bibitem [{\citenamefont {Shen}\ \emph {et~al.}(2015)\citenamefont {Shen},
  \citenamefont {Zhang}, \citenamefont {Ma}, \citenamefont {Li}, \citenamefont
  {Zhang},\ and\ \citenamefont {Guo}}]{Shen:2015cwj}%
  \BibitemOpen
  \bibfield  {author} {\bibinfo {author} {\bibfnamefont {Y.-B.}\ \bibnamefont
  {Shen}}, \bibinfo {author} {\bibfnamefont {R.-Y.}\ \bibnamefont {Zhang}},
  \bibinfo {author} {\bibfnamefont {W.-G.}\ \bibnamefont {Ma}}, \bibinfo
  {author} {\bibfnamefont {X.-Z.}\ \bibnamefont {Li}}, \bibinfo {author}
  {\bibfnamefont {Y.}~\bibnamefont {Zhang}},\ and\ \bibinfo {author}
  {\bibfnamefont {L.}~\bibnamefont {Guo}},\ }\bibfield  {title} {\bibinfo
  {title} {{NLO QCD + NLO EW corrections to $WZZ$ productions with leptonic
  decays at the LHC}},\ }\href {https://doi.org/10.1007/JHEP10(2015)186}
  {\bibfield  {journal} {\bibinfo  {journal} {JHEP}\ }\textbf {\bibinfo
  {volume} {10}},\ \bibinfo {pages} {186}},\ \bibinfo {note} {[Erratum: JHEP
  10, 156 (2016)]},\ \Eprint {https://arxiv.org/abs/1507.03693}
  {arXiv:1507.03693 [hep-ph]} \BibitemShut {NoStop}%
\bibitem [{\citenamefont {Wang}\ \emph {et~al.}(2016)\citenamefont {Wang},
  \citenamefont {Zhang}, \citenamefont {Ma}, \citenamefont {Guo}, \citenamefont
  {Li},\ and\ \citenamefont {Wang}}]{Wang:2016fvj}%
  \BibitemOpen
  \bibfield  {author} {\bibinfo {author} {\bibfnamefont {H.}~\bibnamefont
  {Wang}}, \bibinfo {author} {\bibfnamefont {R.-Y.}\ \bibnamefont {Zhang}},
  \bibinfo {author} {\bibfnamefont {W.-G.}\ \bibnamefont {Ma}}, \bibinfo
  {author} {\bibfnamefont {L.}~\bibnamefont {Guo}}, \bibinfo {author}
  {\bibfnamefont {X.-Z.}\ \bibnamefont {Li}},\ and\ \bibinfo {author}
  {\bibfnamefont {S.-M.}\ \bibnamefont {Wang}},\ }\bibfield  {title} {\bibinfo
  {title} {{NLO QCD + EW corrections to ZZZ production with subsequent leptonic
  decays at the LHC}},\ }\href {https://doi.org/10.1088/0954-3899/43/11/115001}
  {\bibfield  {journal} {\bibinfo  {journal} {J. Phys. G}\ }\textbf {\bibinfo
  {volume} {43}},\ \bibinfo {pages} {115001} (\bibinfo {year} {2016})},\
  \Eprint {https://arxiv.org/abs/1610.05876} {arXiv:1610.05876 [hep-ph]}
  \BibitemShut {NoStop}%
\bibitem [{\citenamefont {Frederix}\ \emph {et~al.}(2014)\citenamefont
  {Frederix}, \citenamefont {Frixione}, \citenamefont {Hirschi}, \citenamefont
  {Maltoni}, \citenamefont {Mattelaer}, \citenamefont {Torrielli},
  \citenamefont {Vryonidou},\ and\ \citenamefont {Zaro}}]{Frederix:2014hta}%
  \BibitemOpen
  \bibfield  {author} {\bibinfo {author} {\bibfnamefont {R.}~\bibnamefont
  {Frederix}}, \bibinfo {author} {\bibfnamefont {S.}~\bibnamefont {Frixione}},
  \bibinfo {author} {\bibfnamefont {V.}~\bibnamefont {Hirschi}}, \bibinfo
  {author} {\bibfnamefont {F.}~\bibnamefont {Maltoni}}, \bibinfo {author}
  {\bibfnamefont {O.}~\bibnamefont {Mattelaer}}, \bibinfo {author}
  {\bibfnamefont {P.}~\bibnamefont {Torrielli}}, \bibinfo {author}
  {\bibfnamefont {E.}~\bibnamefont {Vryonidou}},\ and\ \bibinfo {author}
  {\bibfnamefont {M.}~\bibnamefont {Zaro}},\ }\bibfield  {title} {\bibinfo
  {title} {{Higgs pair production at the LHC with NLO and parton-shower
  effects}},\ }\href {https://doi.org/10.1016/j.physletb.2014.03.026}
  {\bibfield  {journal} {\bibinfo  {journal} {Phys. Lett. B}\ }\textbf
  {\bibinfo {volume} {732}},\ \bibinfo {pages} {142} (\bibinfo {year}
  {2014})},\ \Eprint {https://arxiv.org/abs/1401.7340} {arXiv:1401.7340
  [hep-ph]} \BibitemShut {NoStop}%
\bibitem [{\citenamefont {Kidonakis}(2015)}]{Kidonakis:2015nna}%
  \BibitemOpen
  \bibfield  {author} {\bibinfo {author} {\bibfnamefont {N.}~\bibnamefont
  {Kidonakis}},\ }\bibfield  {title} {\bibinfo {title} {{Theoretical results
  for electroweak-boson and single-top production}},\ }\href
  {https://doi.org/10.22323/1.247.0170} {\bibfield  {journal} {\bibinfo
  {journal} {PoS}\ }\textbf {\bibinfo {volume} {DIS2015}},\ \bibinfo {pages}
  {170} (\bibinfo {year} {2015})},\ \Eprint {https://arxiv.org/abs/1506.04072}
  {arXiv:1506.04072 [hep-ph]} \BibitemShut {NoStop}%
\bibitem [{\citenamefont {Muselli}\ \emph {et~al.}(2015)\citenamefont
  {Muselli}, \citenamefont {Bonvini}, \citenamefont {Forte}, \citenamefont
  {Marzani},\ and\ \citenamefont {Ridolfi}}]{Muselli:2015kba}%
  \BibitemOpen
  \bibfield  {author} {\bibinfo {author} {\bibfnamefont {C.}~\bibnamefont
  {Muselli}}, \bibinfo {author} {\bibfnamefont {M.}~\bibnamefont {Bonvini}},
  \bibinfo {author} {\bibfnamefont {S.}~\bibnamefont {Forte}}, \bibinfo
  {author} {\bibfnamefont {S.}~\bibnamefont {Marzani}},\ and\ \bibinfo {author}
  {\bibfnamefont {G.}~\bibnamefont {Ridolfi}},\ }\bibfield  {title} {\bibinfo
  {title} {{Top Quark Pair Production beyond NNLO}},\ }\href
  {https://doi.org/10.1007/JHEP08(2015)076} {\bibfield  {journal} {\bibinfo
  {journal} {JHEP}\ }\textbf {\bibinfo {volume} {08}},\ \bibinfo {pages}
  {076}},\ \Eprint {https://arxiv.org/abs/1505.02006} {arXiv:1505.02006
  [hep-ph]} \BibitemShut {NoStop}%
\bibitem [{\citenamefont {Broggio}\ \emph {et~al.}(2019)\citenamefont
  {Broggio}, \citenamefont {Ferroglia}, \citenamefont {Frederix}, \citenamefont
  {Pagani}, \citenamefont {Pecjak},\ and\ \citenamefont
  {Tsinikos}}]{Broggio:2019ewu}%
  \BibitemOpen
  \bibfield  {author} {\bibinfo {author} {\bibfnamefont {A.}~\bibnamefont
  {Broggio}}, \bibinfo {author} {\bibfnamefont {A.}~\bibnamefont {Ferroglia}},
  \bibinfo {author} {\bibfnamefont {R.}~\bibnamefont {Frederix}}, \bibinfo
  {author} {\bibfnamefont {D.}~\bibnamefont {Pagani}}, \bibinfo {author}
  {\bibfnamefont {B.~D.}\ \bibnamefont {Pecjak}},\ and\ \bibinfo {author}
  {\bibfnamefont {I.}~\bibnamefont {Tsinikos}},\ }\bibfield  {title} {\bibinfo
  {title} {{Top-quark pair hadroproduction in association with a heavy boson at
  NLO+NNLL including EW corrections}},\ }\href
  {https://doi.org/10.1007/JHEP08(2019)039} {\bibfield  {journal} {\bibinfo
  {journal} {JHEP}\ }\textbf {\bibinfo {volume} {08}},\ \bibinfo {pages}
  {039}},\ \Eprint {https://arxiv.org/abs/1907.04343} {arXiv:1907.04343
  [hep-ph]} \BibitemShut {NoStop}%
\bibitem [{\citenamefont {Frederix}\ \emph {et~al.}(2018)\citenamefont
  {Frederix}, \citenamefont {Pagani},\ and\ \citenamefont
  {Zaro}}]{Frederix:2017wme}%
  \BibitemOpen
  \bibfield  {author} {\bibinfo {author} {\bibfnamefont {R.}~\bibnamefont
  {Frederix}}, \bibinfo {author} {\bibfnamefont {D.}~\bibnamefont {Pagani}},\
  and\ \bibinfo {author} {\bibfnamefont {M.}~\bibnamefont {Zaro}},\ }\bibfield
  {title} {\bibinfo {title} {{Large NLO corrections in $t\bar{t}W^{\pm}$ and
  $t\bar{t}t\bar{t}$ hadroproduction from supposedly subleading EW
  contributions}},\ }\href {https://doi.org/10.1007/JHEP02(2018)031} {\bibfield
   {journal} {\bibinfo  {journal} {JHEP}\ }\textbf {\bibinfo {volume} {02}},\
  \bibinfo {pages} {031}},\ \Eprint {https://arxiv.org/abs/1711.02116}
  {arXiv:1711.02116 [hep-ph]} \BibitemShut {NoStop}%
\bibitem [{ATL(2016)}]{ATLAS:2016iqc}%
  \BibitemOpen
  \bibfield  {title} {\bibinfo {title} {{Electron efficiency measurements with
  the ATLAS detector using the 2015 LHC proton-proton collision data}},\
  }\href@noop {} {\  (\bibinfo {year} {2016})}\BibitemShut {NoStop}%
\bibitem [{\citenamefont {Aaboud}\ \emph
  {et~al.}(2018{\natexlab{b}})\citenamefont {Aaboud} \emph
  {et~al.}}]{ATLAS:2017xqs}%
  \BibitemOpen
  \bibfield  {author} {\bibinfo {author} {\bibfnamefont {M.}~\bibnamefont
  {Aaboud}} \emph {et~al.} (\bibinfo {collaboration} {ATLAS}),\ }\bibfield
  {title} {\bibinfo {title} {{Search for doubly charged Higgs boson production
  in multi-lepton final states with the ATLAS detector using
  proton\textendash{}proton collisions at $\sqrt{s}=13\,\text {TeV}$}},\ }\href
  {https://doi.org/10.1140/epjc/s10052-018-5661-z} {\bibfield  {journal}
  {\bibinfo  {journal} {Eur. Phys. J. C}\ }\textbf {\bibinfo {volume} {78}},\
  \bibinfo {pages} {199} (\bibinfo {year} {2018}{\natexlab{b}})},\ \Eprint
  {https://arxiv.org/abs/1710.09748} {arXiv:1710.09748 [hep-ex]} \BibitemShut
  {NoStop}%
\bibitem [{\citenamefont {de~Favereau}\ \emph {et~al.}(2014)\citenamefont
  {de~Favereau}, \citenamefont {Delaere}, \citenamefont {Demin}, \citenamefont
  {Giammanco}, \citenamefont {Lema\^\i{}tre}, \citenamefont {Mertens},\ and\
  \citenamefont {Selvaggi}}]{deFavereau:2013fsa}%
  \BibitemOpen
  \bibfield  {author} {\bibinfo {author} {\bibfnamefont {J.}~\bibnamefont
  {de~Favereau}}, \bibinfo {author} {\bibfnamefont {C.}~\bibnamefont
  {Delaere}}, \bibinfo {author} {\bibfnamefont {P.}~\bibnamefont {Demin}},
  \bibinfo {author} {\bibfnamefont {A.}~\bibnamefont {Giammanco}}, \bibinfo
  {author} {\bibfnamefont {V.}~\bibnamefont {Lema\^\i{}tre}}, \bibinfo {author}
  {\bibfnamefont {A.}~\bibnamefont {Mertens}},\ and\ \bibinfo {author}
  {\bibfnamefont {M.}~\bibnamefont {Selvaggi}} (\bibinfo {collaboration}
  {DELPHES 3}),\ }\bibfield  {title} {\bibinfo {title} {{DELPHES 3, A modular
  framework for fast simulation of a generic collider experiment}},\ }\href
  {https://doi.org/10.1007/JHEP02(2014)057} {\bibfield  {journal} {\bibinfo
  {journal} {JHEP}\ }\textbf {\bibinfo {volume} {02}},\ \bibinfo {pages}
  {057}},\ \Eprint {https://arxiv.org/abs/1307.6346} {arXiv:1307.6346 [hep-ex]}
  \BibitemShut {NoStop}%
\bibitem [{\citenamefont {Cacciari}\ \emph {et~al.}(2008)\citenamefont
  {Cacciari}, \citenamefont {Salam},\ and\ \citenamefont
  {Soyez}}]{Cacciari:2008gp}%
  \BibitemOpen
  \bibfield  {author} {\bibinfo {author} {\bibfnamefont {M.}~\bibnamefont
  {Cacciari}}, \bibinfo {author} {\bibfnamefont {G.~P.}\ \bibnamefont
  {Salam}},\ and\ \bibinfo {author} {\bibfnamefont {G.}~\bibnamefont {Soyez}},\
  }\bibfield  {title} {\bibinfo {title} {{The anti-$k_t$ jet clustering
  algorithm}},\ }\href {https://doi.org/10.1088/1126-6708/2008/04/063}
  {\bibfield  {journal} {\bibinfo  {journal} {JHEP}\ }\textbf {\bibinfo
  {volume} {04}},\ \bibinfo {pages} {063}},\ \Eprint
  {https://arxiv.org/abs/0802.1189} {arXiv:0802.1189 [hep-ph]} \BibitemShut
  {NoStop}%
\bibitem [{\citenamefont {Cacciari}\ \emph {et~al.}(2012)\citenamefont
  {Cacciari}, \citenamefont {Salam},\ and\ \citenamefont
  {Soyez}}]{Cacciari:2011ma}%
  \BibitemOpen
  \bibfield  {author} {\bibinfo {author} {\bibfnamefont {M.}~\bibnamefont
  {Cacciari}}, \bibinfo {author} {\bibfnamefont {G.~P.}\ \bibnamefont
  {Salam}},\ and\ \bibinfo {author} {\bibfnamefont {G.}~\bibnamefont {Soyez}},\
  }\bibfield  {title} {\bibinfo {title} {{FastJet User Manual}},\ }\href
  {https://doi.org/10.1140/epjc/s10052-012-1896-2} {\bibfield  {journal}
  {\bibinfo  {journal} {Eur. Phys. J. C}\ }\textbf {\bibinfo {volume} {72}},\
  \bibinfo {pages} {1896} (\bibinfo {year} {2012})},\ \Eprint
  {https://arxiv.org/abs/1111.6097} {arXiv:1111.6097 [hep-ph]} \BibitemShut
  {NoStop}%
\bibitem [{\citenamefont {Aaboud}\ \emph {et~al.}(2017)\citenamefont {Aaboud}
  \emph {et~al.}}]{ATLAS:2016wtr}%
  \BibitemOpen
  \bibfield  {author} {\bibinfo {author} {\bibfnamefont {M.}~\bibnamefont
  {Aaboud}} \emph {et~al.} (\bibinfo {collaboration} {ATLAS}),\ }\bibfield
  {title} {\bibinfo {title} {{Performance of the ATLAS Trigger System in
  2015}},\ }\href {https://doi.org/10.1140/epjc/s10052-017-4852-3} {\bibfield
  {journal} {\bibinfo  {journal} {Eur. Phys. J. C}\ }\textbf {\bibinfo {volume}
  {77}},\ \bibinfo {pages} {317} (\bibinfo {year} {2017})},\ \Eprint
  {https://arxiv.org/abs/1611.09661} {arXiv:1611.09661 [hep-ex]} \BibitemShut
  {NoStop}%
\bibitem [{\citenamefont {Aad}\ \emph {et~al.}(2020{\natexlab{b}})\citenamefont
  {Aad} \emph {et~al.}}]{ATLAS:2019dpa}%
  \BibitemOpen
  \bibfield  {author} {\bibinfo {author} {\bibfnamefont {G.}~\bibnamefont
  {Aad}} \emph {et~al.} (\bibinfo {collaboration} {ATLAS}),\ }\bibfield
  {title} {\bibinfo {title} {{Performance of electron and photon triggers in
  ATLAS during LHC Run 2}},\ }\href
  {https://doi.org/10.1140/epjc/s10052-019-7500-2} {\bibfield  {journal}
  {\bibinfo  {journal} {Eur. Phys. J. C}\ }\textbf {\bibinfo {volume} {80}},\
  \bibinfo {pages} {47} (\bibinfo {year} {2020}{\natexlab{b}})},\ \Eprint
  {https://arxiv.org/abs/1909.00761} {arXiv:1909.00761 [hep-ex]} \BibitemShut
  {NoStop}%
\bibitem [{\citenamefont {Sirunyan}\ \emph {et~al.}(2020)\citenamefont
  {Sirunyan} \emph {et~al.}}]{CMS:2020cmk}%
  \BibitemOpen
  \bibfield  {author} {\bibinfo {author} {\bibfnamefont {A.~M.}\ \bibnamefont
  {Sirunyan}} \emph {et~al.} (\bibinfo {collaboration} {CMS}),\ }\bibfield
  {title} {\bibinfo {title} {{Performance of the CMS Level-1 trigger in
  proton-proton collisions at $\sqrt{s} =$ 13 TeV}},\ }\href
  {https://doi.org/10.1088/1748-0221/15/10/P10017} {\bibfield  {journal}
  {\bibinfo  {journal} {JINST}\ }\textbf {\bibinfo {volume} {15}}\bibfield
  {number} {\bibinfo  {number} { (10)},\ \bibinfo {pages} {P10017}},\ }\Eprint
  {https://arxiv.org/abs/2006.10165} {arXiv:2006.10165 [hep-ex]} \BibitemShut
  {NoStop}%
\bibitem [{\citenamefont {Hocker}\ \emph {et~al.}(2007)\citenamefont {Hocker}
  \emph {et~al.}}]{Hocker:2007ht}%
  \BibitemOpen
  \bibfield  {author} {\bibinfo {author} {\bibfnamefont {A.}~\bibnamefont
  {Hocker}} \emph {et~al.},\ }\bibfield  {title} {\bibinfo {title} {{TMVA -
  Toolkit for Multivariate Data Analysis}},\ }\href@noop {} {\  (\bibinfo
  {year} {2007})},\ \Eprint {https://arxiv.org/abs/physics/0703039}
  {arXiv:physics/0703039} \BibitemShut {NoStop}%
\bibitem [{\citenamefont {Brun}\ \emph {et~al.}(2000)\citenamefont {Brun},
  \citenamefont {Rademakers},\ and\ \citenamefont {Panacek}}]{Brun:2000es}%
  \BibitemOpen
  \bibfield  {author} {\bibinfo {author} {\bibfnamefont {R.}~\bibnamefont
  {Brun}}, \bibinfo {author} {\bibfnamefont {F.}~\bibnamefont {Rademakers}},\
  and\ \bibinfo {author} {\bibfnamefont {S.}~\bibnamefont {Panacek}},\
  }\bibfield  {title} {\bibinfo {title} {{ROOT, an object oriented data
  analysis framework}},\ }in\ \href {https://doi.org/10.5170/CERN-2000-013.11}
  {\emph {\bibinfo {booktitle} {{CERN School of Computing (CSC 2000)}}}}\
  (\bibinfo {year} {2000})\ pp.\ \bibinfo {pages} {11--42}\BibitemShut
  {NoStop}%
\bibitem [{\citenamefont {Cowan}\ \emph {et~al.}(2011)\citenamefont {Cowan},
  \citenamefont {Cranmer}, \citenamefont {Gross},\ and\ \citenamefont
  {Vitells}}]{Cowan:2010js}%
  \BibitemOpen
  \bibfield  {author} {\bibinfo {author} {\bibfnamefont {G.}~\bibnamefont
  {Cowan}}, \bibinfo {author} {\bibfnamefont {K.}~\bibnamefont {Cranmer}},
  \bibinfo {author} {\bibfnamefont {E.}~\bibnamefont {Gross}},\ and\ \bibinfo
  {author} {\bibfnamefont {O.}~\bibnamefont {Vitells}},\ }\bibfield  {title}
  {\bibinfo {title} {{Asymptotic formulae for likelihood-based tests of new
  physics}},\ }\href {https://doi.org/10.1140/epjc/s10052-011-1554-0}
  {\bibfield  {journal} {\bibinfo  {journal} {Eur. Phys. J. C}\ }\textbf
  {\bibinfo {volume} {71}},\ \bibinfo {pages} {1554} (\bibinfo {year}
  {2011})},\ \bibinfo {note} {[Erratum: Eur.Phys.J.C 73, 2501 (2013)]},\
  \Eprint {https://arxiv.org/abs/1007.1727} {arXiv:1007.1727 [physics.data-an]}
  \BibitemShut {NoStop}%
\bibitem [{\citenamefont {Li}\ and\ \citenamefont {Ma}(1983)}]{Li:1983fv}%
  \BibitemOpen
  \bibfield  {author} {\bibinfo {author} {\bibfnamefont {T.~P.}\ \bibnamefont
  {Li}}\ and\ \bibinfo {author} {\bibfnamefont {Y.~Q.}\ \bibnamefont {Ma}},\
  }\bibfield  {title} {\bibinfo {title} {{Analysis methods for results in
  gamma-ray astronomy}},\ }\href {https://doi.org/10.1086/161295} {\bibfield
  {journal} {\bibinfo  {journal} {Astrophys. J.}\ }\textbf {\bibinfo {volume}
  {272}},\ \bibinfo {pages} {317} (\bibinfo {year} {1983})}\BibitemShut
  {NoStop}%
\bibitem [{\citenamefont {Cousins}\ \emph {et~al.}(2008)\citenamefont
  {Cousins}, \citenamefont {Linnemann},\ and\ \citenamefont
  {Tucker}}]{Cousins:2007yta}%
  \BibitemOpen
  \bibfield  {author} {\bibinfo {author} {\bibfnamefont {R.~D.}\ \bibnamefont
  {Cousins}}, \bibinfo {author} {\bibfnamefont {J.~T.}\ \bibnamefont
  {Linnemann}},\ and\ \bibinfo {author} {\bibfnamefont {J.}~\bibnamefont
  {Tucker}},\ }\bibfield  {title} {\bibinfo {title} {{Evaluation of three
  methods for calculating statistical significance when incorporating a
  systematic uncertainty into a test of the background-only hypothesis for a
  Poisson process}},\ }\href {https://doi.org/10.1016/j.nima.2008.07.086}
  {\bibfield  {journal} {\bibinfo  {journal} {Nucl. Instrum. Meth. A}\ }\textbf
  {\bibinfo {volume} {595}},\ \bibinfo {pages} {480} (\bibinfo {year}
  {2008})},\ \Eprint {https://arxiv.org/abs/physics/0702156}
  {arXiv:physics/0702156} \BibitemShut {NoStop}%
\end{thebibliography}%

\end{document}